%% file: main.tex
\documentclass{article}
\usepackage{packages}

\title{Synchrotron-based pore-network modeling of two-phase flow in Nubian Sandstone and implications for capillary trapping of carbon dioxide}

\author[1,2]{Mahmoud Hefny\thanks{Corresponding author: M. Hefny\\E-mail address: \href{mailto:mhefny@ethz.ch}{\tt mhefny@ethz.ch},\\ Postal address: Department of Earth Science, Sonneggstrasse 05, 8092 Zürich, Switzerland}}
\author[3,4]{ChaoZhong Qin}
\author[1,5]{Martin O. Saar}
\author[1]{Anozie Ebigbo}

\affil[1]{\small{Geothermal Energy and Geofluids, Institute of Geophysics, ETH Zurich, Switzerland}}
\affil[2]{Geology Department, South Valley University, Egypt}
\affil[3]{State Key Laboratory of Coal Mine Disaster Dynamics and Control, Chongqing University, China}
\affil[4]{Mechanical Engineering Department, Technology University of Eindhoven, The Netherlands}
\affil[5]{Department of Earth and Environmental Sciences, University of Minnesota, Minneapolis, USA}

\date{}
\begin{document}

\maketitle
\thispagestyle{fancy}
%\input{Sections/pnm_abstract.tex}
% =========================================================
\begin{abstract}
\label{pnm_abstract}
Depleted oil fields in the Gulf of Suez (Egypt) can serve as geothermal reservoirs for power generation using a CO$_{\text2}$-Plume Geothermal (CPG) system, while geologically sequestering CO$_{\text2}$. This entails the injection of a substantial amount of CO$_{\text2}$ into the highly permeable brine-saturated Nubian Sandstone. Numerical models of two-phase flow processes are indispensable for predicting the CO$_{\text2}$-plume migration at a representative geological scale. Such models require reliable constitutive relationships, including relative permeability and capillary pressure curves. In this study, quasi-static pore-network modelling has been used to simulate the equilibrium positions of fluid--fluid interfaces, and thus determine the capillary pressure and relative permeability curves.

Three-dimensional images with a voxel size of 0.65 $\mathrm{\micro m^3}$ of a Nubian Sandstone rock sample have been obtained using Synchrotron Radiation X-ray Tomographic Microscopy. From the images, topological properties of pores/throats were constructed. Using a pore-network model, we performed a sequential primary drainage--main imbibition cycle of quasi-static invasion in order to quantify (1) the CO$_{\text2}$ and brine relative permeability curves, (2) the effect of initial wetting-phase saturation (i.e. the saturation at the point of reversal from drainage to imbibition) on the residual–trapping potential, and (3) study the relative permeability–saturation hysteresis. The results 
improve our understanding of the potential magnitude of capillary trapping in Nubian Sandstone, essential for future field-scale simulations. Further, an initial basin-scale assessment of CO$_{\text2}$ storage capacity, which incorporates capillary trapping, yields a range of 14--49\,GtCO$_{\text2}$ in Nubian Sandstone, Gulf of Suez Basin.  %  
\end{abstract}

{\bf Keywords:} Pore-Network Modelling, Carbon Capture and Storage, CO$_{\text2}$-Plume Geothermal, Nubian Sandstone (Egypt), Residual Trapping.\\

{\bf Highlights:}
\begin{itemize}
\small{
    \item 3D images of a Nubian Sandstone rock have been obtained using Synchrotron Radiation X-ray Tomographic Microscopy.
    \item Lab experiments and image analyses provide a pore-size distribution and $P_\text{c}\!-\!S_\text{w}$ relationship for the CO$_{\text2}$-brine system.
    \item Relative permeability curves have been obtained for Nubian sandstone using quasi-static pore-network modelling.
    \item The Land trapping coefficient for CO$_{\text2}$ in the Nubian Sandstone sample has been quantified as 1.2.
    \item The sensitivity study reflects the importance of the wettability and aspect ratio on the residual CO$_{\text2}$ trapping.}
\end{itemize}
% ========================================================

%\begin{linenumbers}
% ========================================================
%\input{Sections/pnm_introduction.tex}

\newpage
\section{Introduction}
\label{pnm_introduction}

%=========================================================
\subsection{Background}
\label{pnm_backgound}
The continuous rise in atmospheric carbon dioxide concentrations due to the combustion of fossil fuels could, if not abated, lead to irreversible climate consequences \citep{Hartfield2018State2017}. In fact, the Intergovernmental Panel on Climate Change (IPCC) concluded that the combustion of fossil fuels, to generate electricity and heat, is responsible for at least 25\% of the total amount of CO$_{\text2}$ that has been emitted to the atmosphere by human activities \citep{IPCC2014ClimateChange.}. To reduce CO$_{\text2}$ emissions to the atmosphere to the required net-zero value, several methods, called `stabilization wedges’ in \cite{Pacala2004StabilizationTechnologies.}, are required, including more wide-spread implementation of renewable energies and Carbon (or CO$_{\text2}$) Capture and geologic Storage (CCS). During CCS, CO$_{\text2}$ is captured at a point-source CO$_{\text2}$ emitter, such as a fossil-fueled power plant, a cement manufacturer, or other process industries, or directly from the air \citep{Fasihi2019Techno-economicPlants}, and is injected into suitable subsurface formations for permanent geologic storage \citep{IPCC2005CarbonStorage, Michael2010GeologicalOperations}. Geological storage of CO$_{\text2}$ can occur as structural/stratigraphic, capillary/residual, dissolution, and/or mineral trapping \citep{Benson2008COsub2/subFormations}, as discussed in more detail below. Ideally, the CO$_{\text2}$ storage formation (e.g. sandstone or carbonate rock) is overlain by more than one low- to practically zero-permeability caprock (consisting, for example, of shale). Such multiple confinements increase the CO$_{\text2}$ retention security, necessary to impede CO$_{\text2}$ leakage and upward migration.  

It has been suggested \cite[e.g.,][]{Randolph2011CombiningSequestration, Randolph2011, Adams2014OnSystems,Adams2015AConditions,Garapati2015BrineWell, Ezekiel2020CombiningGeneration} to combine CCS and/or CO$_{\text2}$-based enhanced gas/oil recovery with geothermal energy extraction by using the CO$_{\text2}$, that is structurally trapped in the geologic reservoir, as the geothermal working fluid to extract enthalpy, while still permanently sequestering 100\% of the CO$_{\text2}$ as a CO$_{\text2}$ plume, constituting a CO$_{\text2}$ Capture, Utilization, and Storage (CCUS) system. During such  CO$_{\text2}$-Plume Geothermal (CPG) power plant operations, a portion of the geothermally heated CO$_{\text2}$ rises buoyantly to the land surface, passes through a turbine, is cooled, and is reinjected into the CO$_{\text2}$ storage reservoir, along with the CO$_{\text2}$ coming from the CO$_{\text2}$ capture facility.

\begin{figure}[hbtp]
\begin{center}
{\small\includegraphics[width=0.82\textwidth]{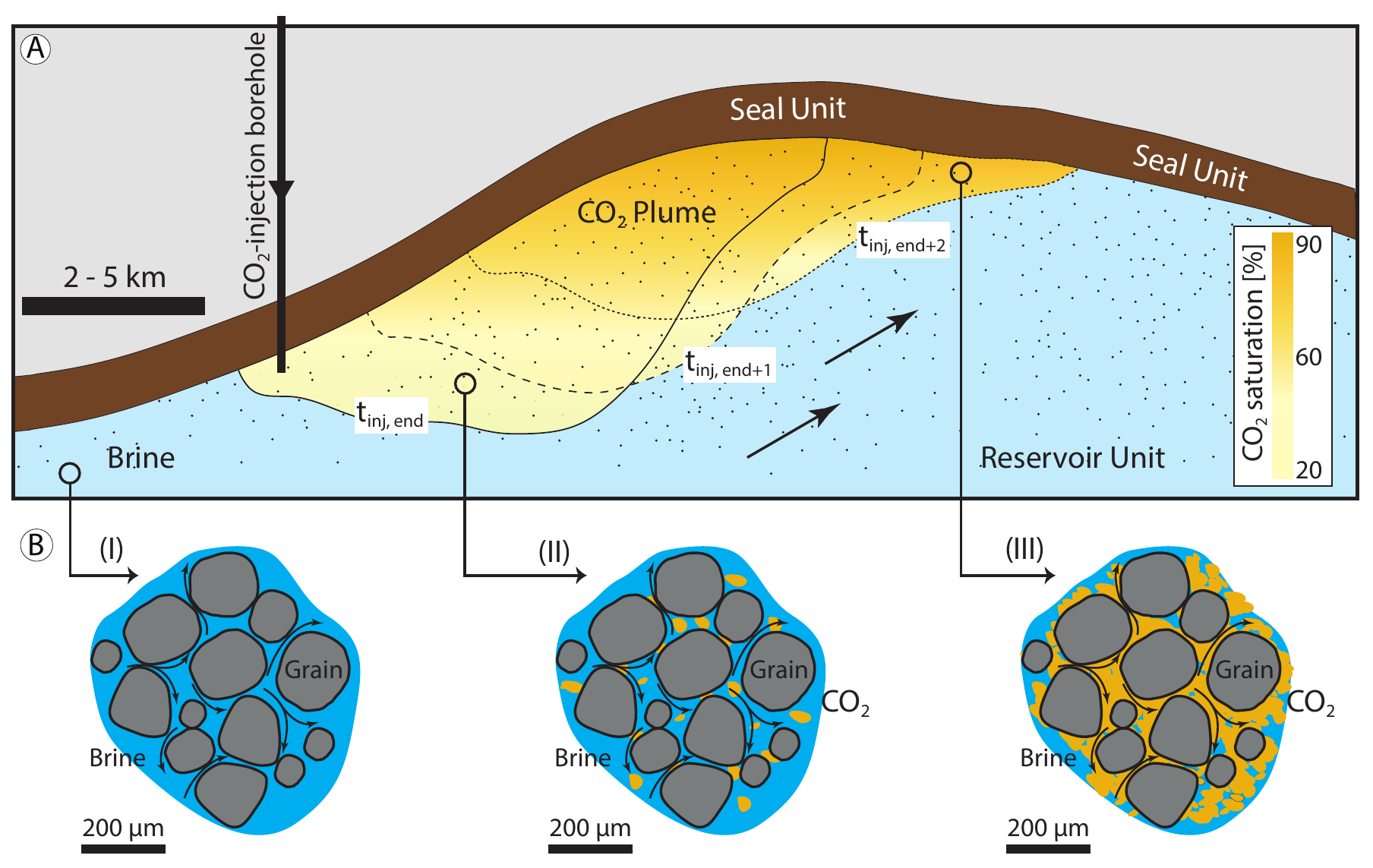}}
\caption{\small{$\left[\mathrm{A}\right]$ Schematic diagram of key processes of CO$_{\text2}$ sequestration in a saline aquifer. 
As caprocks are typically not perfectly horizontal, the injected CO$_{\text2}$ will tend to move updip and once the injection of CO$_{\text2}$ ends, the entire CO$_{\text2}$ plume will typically move updip.
At the trailing edge of the CO$_{\text2}$ plume, this leads to re-imbibition of native formation fluid (e.g., brine), reducing the CO$_{\text2}$ saturation in the pore space. When this imbibition process is completed, some amount of CO$_{\text2}$ remains trapped by capillary forces. This ``residual trapping” effectively immobilizes a large portion of the injected CO$_{\text2}$ and can, under favorable geologic conditions, remove the necessity of a sealing unit (caprock) to ensure permanent CO$_{\text2}$ storage underground. Depths are not to scale. $\left[\mathrm{B}\right]$ Typical pore-scale configurations of fluid distributions at various pore-space saturations.}} 
\label{fig:pnm_conceputal_model}
\end{center}
\end{figure}

The injection of CO$_{\text2}$ into a deep saline formation will lead to the displacement of some of the resident formation brine in the pore space (i.e. drainage).
Upon injection into the formation, the CO$_{\text2}$ tends to spread radially at the beginning, when the movement of the CO$_{\text2}$ is dominated by viscous forces. As the CO$_{\text2}$ plume continues to spread farther away from the injection location, the CO$_{\text2}$ will move upwards, due to buoyancy, approaching the overlying caprock, where it is prevented from further vertical movement and is thus essentially pooling upwards against the caprock (structural/stratigraphic trapping) (Figure~\ref{fig:pnm_conceputal_model}A). As caprocks are typically not perfectly horizontal, the CO$_{\text2}$ will tend to move updip and once the injection of CO$_{\text2}$ ends, possibly after many decades of CO$_{\text2}$ injection, the entire CO$_{\text2}$ plume will typically move updip. As shown in Figure~\ref{fig:pnm_conceputal_model}B, at the trailing edge of the CO$_{\text2}$ plume, brine can imbibe back into the pore space (i.e. imbibition), typically leading to the disconnection of small blobs (10s--100s of micrometers in size) of CO$_{\text2}$ (capillary/residual trapping). This phenomenon, at the continuum scale, can effectively immobilize a large portion of the injected CO$_{\text2}$ by capillary forces, slowing down the CO$_{\text2}$ plume migration, and increasing the CO$_{\text2}$ storage safety \citep{IPCC2005CarbonStorage,Krevor2015}. The amount of residual trapping depends on the pore-scale balance between viscous and capillary forces (quantifiable by the capillary number, $N_c$), wettability (i.e. contact angle), the saturation of the CO$_{\text2}$ (non-wetting) phase when drainage switches to imbibition, and pore-space geometry \citep{Zuo2014Process-dependentSandstone}.

Recent advances in core-flooding experiments and in-situ contact-angle measurements, using X-ray tomographic imaging, show that the CO$_{\text2}$-brine-quartz system is strongly water-wet, as contact angles of $\theta \mathrm{<50^{\circ}}$ were found \citep{Krevor2012RelativeConditions, Andrew2014Pore-scaleMicrotomography,Zuo2014Process-dependentSandstone}. The difficulties inherent in the prediction of CO$_{\text2}$-brine flow in deep reservoirs require analyzing capillary interactions at the pore scale in order to obtain the constitutive relationships between capillary pressure, $(P_\mathrm{c})$, relative permeability, $(k_\mathrm{r})$, and phase saturation, $(S_\mathrm{\kappa})$. Such multiphase fluid flow is controlled by a variety of interdependent physical factors, including the actual geometry of the porous medium, the fluid properties (viscosity and surface tension), and the interaction between the solid and the fluids (contact angle) \citep{Valvatne2004PredictiveMedia, Andrew2015TheConditions}. Henceforth, the brine and the CO$_{\text2}$ phases will be referred to as the wetting and the non-wetting phases, respectively.

In the following, we focus on the conditions, where capillary forces dominate over viscous forces (i.e. quasi-static conditions with low-capillary numbers of $\mathrm{\; 10^{-8} - 10^{-7}}$). These conditions are important after the CO$_{\text2}$ injection has stopped (Figure~\ref{fig:pnm_conceputal_model}). We apply the same fluid-displacement concepts as experimentally observed by \cite{Lenormand1983MechanismsDucts,  Lenormand1988NumericalMedia} and numerically predicted by \cite{Oren1998ExtendingModels} and \cite{Valvatne2004PredictiveMedia}. 

Numerous oil/gas reservoirs in the Gulf of Suez (Egypt), that are either in the depletion phase or have already been depleted, could be suitable for CCS/CPG (i.e. CCUS) development. \cite{Hefny} characterize the Nubian Sandstone, a common rock type in hydrocarbon reservoirs, found in the Gulf of Suez, and investigate, if it has potential to be used for CCS and, in particlar, for CPG. Here, we focus on investigating the pore-scale CO$_{\text2}$-trapping mechanisms and on estimating the CO$_{\text2}$-storage-capacity estimation of the Nubian Sandstone, using three-dimensional imaging via a fast Synchrotron Radiation X‐ray CT source.  
%
%=========================================================
\subsection{Pore-scale trapping mechanisms}
\label{pnm_capillarytrapping}
Numerical modeling of two-phase flow in a natural porous medium at the pore scale is often employed to obtain two-phase constitutive relations such as capillary pressure and relative permeability curves, and to understand pore-scale displacement mechanisms \cite[e.g.][and references therein]{Patzek2001VerificationImbibition,Blunt2013Pore-scaleModelling,Berg2016ConnectedImbibition}. The application of high-flux synchrotron radiation in the hard X-ray range provides improved time and spatial resolution, permitting the investigation of the internal geometry and topology of microstructures down to sub-micrometer levels \citep{Donoghue2006SynchrotronEmbryos}. The constant improvement in both imaging capacity and computational power allows usage of discretized images of the pore structure as a computational mesh to perform fluid-flow simulations directly \citep{Blunt2013Pore-scaleModelling, Gostick2017}. 

Pore-scale models can be broadly divided into direct numerical simulation models (such as volume-of-fluid and Lattice Boltzmann methods), and pore-network models (PNM). In addition to being computationally cheap, PNM can simulate flow in pore networks that preserve the pore connectivity expressed in pore-space images. Each (network) element is assigned a set of geometric properties (e.g. size and shape) that closely match the values measured directly from the images.

Using a high-performance computing cluster, the multiform geometrical and essential topological properties of the Nubian Sandstone pore network (PN) have been directly extracted from Synchrotron Radiation X‐ray Tomographic Microscopy (SRXTM) data. Further, we have performed quasi-static pore-network modeling,  based on the intricate structure of the Nubian Sandstone~PN. In this work, we are interested in the flow processes occurring between the two immiscible fluid phases at the trailing edge of the CO$_{\text2}$ plume and these processes' relevance for trapping CO$_{\text2}$ in Nubian Sandstone. 

To investigate imbibition processes at the trailing edge of the CO$_{\text2}$ plume, the following pore-scale processes are considered:
\begin{enumerate}
    \item Piston-like (\textit{pl}) advance is characterized by the displacement of the nonwetting phase from a narrow pore throat into an adjoining pore body that was initially filled with the wetting phase. It causes a drop in capillary pressure, but it does not result in the trapping of the nonwetting phase.
    \item Snap-off (\textit{so}) occurs when two wetting layers in a pore body swell, such that they touch and coalesce, disconnecting the non-wetting phase within the neighboring pore bodies from the main connected pathway. 
    \item Co-operative pore-body filling (\textit{cpf}) refers to the invasion of a non-wetting-phase-filled pore throat by arc menisci~(AM) or wetting layers initially present in corners, crevices and rough surfaces of the pore body. At a critical capillary pressure, the AMs fuse and the center of the throat spontaneously fills with the wetting phase. This can lead to the trapping of the non-wetting phase.
\end{enumerate}

%=========================================================
\subsection{Objectives}
\label{pnm_objectives}
The main objective of this paper is to characterize the two-phase properties of the Nubian Sandstone as it pertains to CCS and CPG by implementing a quasi-static PN model. Of particular interest is the quantification of the potential for immobilization of CO$_{\text2}$ by capillary forces and how this compares to other sandstones. In addition, the PN model is used to simulate sequential primary drainage/main imbibition cycles to obtain capillary pressure--saturation and relative permeability--saturation curves. Such constitutive relationships are important inputs for large-scale numerical simulations of CCS/CPG operations. These results complement rigorous laboratory experiments for understanding fluid-displacement mechanisms in porous media.

We have endeavored to replicate pressure, temperature, and salinity conditions that are similar to those found in the reservoir of interest. An important outcome of this study is the basin-scale assessment of CO$_{\text2}$ storage capacity for Nubian Sandstone in the Gulf of Suez basin in which capillary trapping is considered. %The main outcome of this study is the computation of capillary pressure and relative permeability curves for the CO$_{\text2}$-brine system in Nubian Sandstone.

% ========================================================
\section{Rock formation characterization}
\label{pnm_rock_characterization}
The Nubian Sandstone -- a major reservoir-rock formation is found in Egypt and extends regionally-- has the potential to serve as a deep geological formation for long-term CO$_{\text2}$ storage and related applications, such as CPG-based power generation, as discussed in the introduction. \cite{Hefny} presented a three-dimensional static reservoir model for Nubian Sandstone, which includes the formation's geology and rock physics properties. For the current study, Nubian Sandstone blocks have been retrieved from the central Gulf of Suez (cGOS). 

We conduct a routine optical microscopy analysis, employing an automated QEMSCAN (Quantitative Evaluation of Minerals by Scanning Electron Microscopy) Quanta~650F technique at the University of Geneva, Switzerland, providing quantitative mineralogical compositions at a scanning resolution of 5~$\mathrm{\micro m}$. Here, the mineral identification maps are based on back-scattered electron values, energy-dispersive X-ray spectra, and X-ray count rates. These tools are integrated to assess the mineralogical composition and rock type. 

\begin{figure}[hbtp]
\begin{center}
{\small\includegraphics[width=0.405\textwidth]{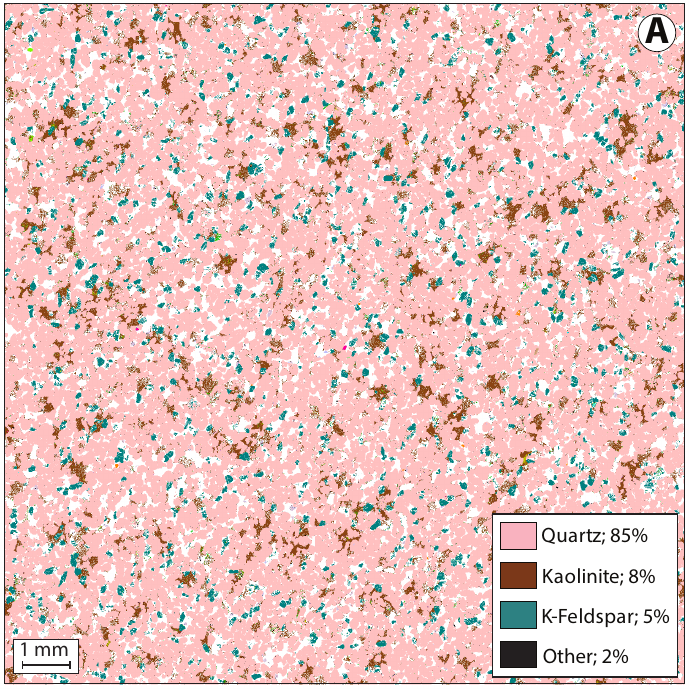}}
{\small\includegraphics[width=0.4\textwidth]{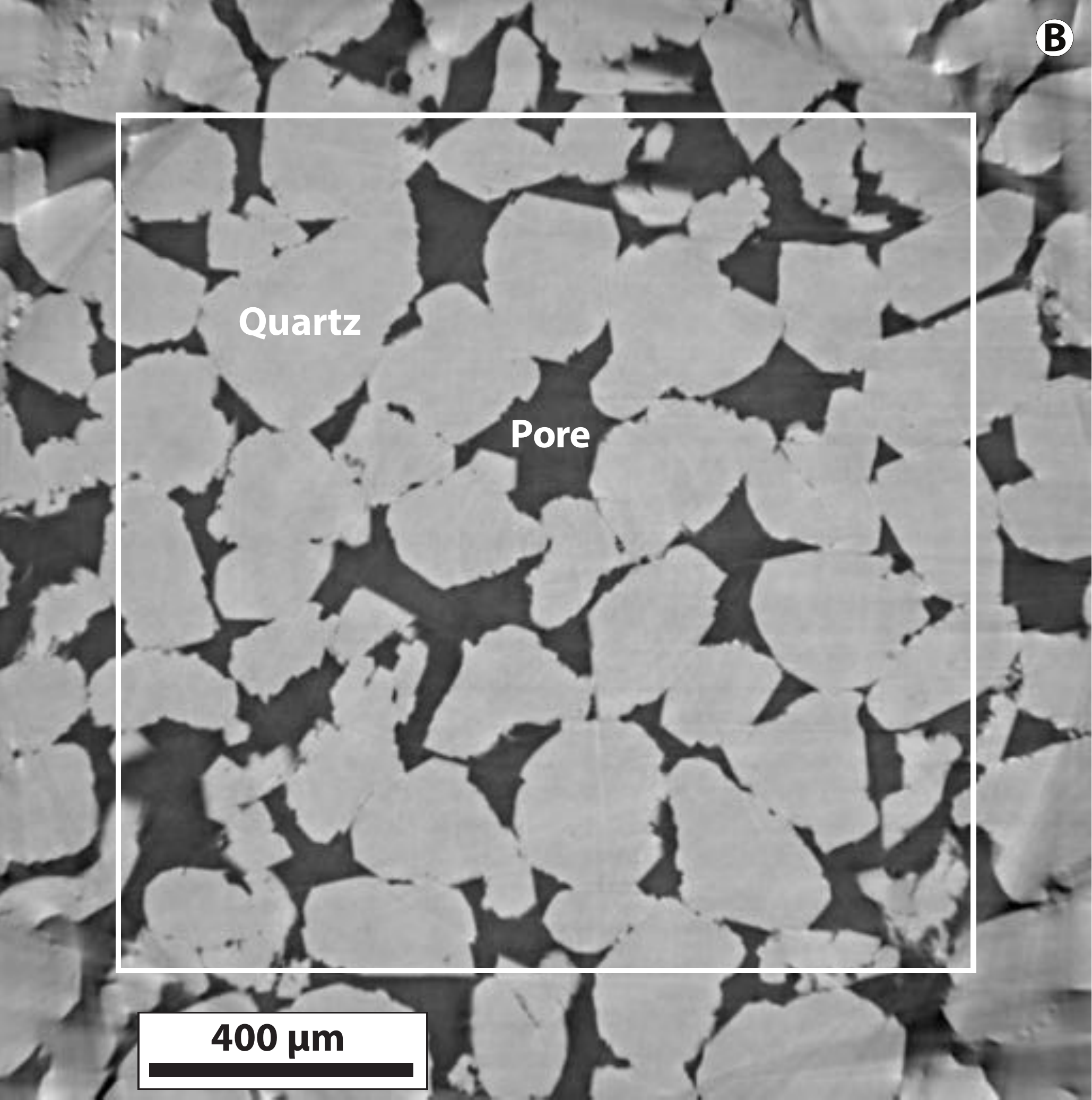}}
\caption{\small{[A] The QEMSCAN image shows the dominance of fine-grained quartz in the Nubian Sandstone. [B] A 2D slice through the SRXTM-volume raw dataset of the Nubian Sandstone (voxel size is $\mathrm{0.65} \;\mathrm{\micro m^3}$). Dark gray areas are pore space (air), while light gray areas represent the mineral grains (quartz). Due to the experiment setup, the data that are used for pore-network modeling are located only within the marked white square. The corresponding file name of this slice \texttt{`SRXTM\_Nubian\_Sandstone\_06\_rec.8bit\_0084.tif’} is given in the full raw dataset. The full SRXTM raw dataset ($\mathrm{2560\!\times\!2560\!\times\!4320}$ pixels and 8-bits) is provided as a supplementary to this publication.} }
\label{fig:pnm_nubian_sandstone_qemscan_srxtm}
\end{center}
\end{figure}

The QEMSCAN analysis shows that the majority of the distinguished mineral phases in the Nubian Sandstone are medium to fine quartz grains (85\%) with some kaolinite (8\%), 5\% k-feldspar, and 2\% other minerals, trapped within the pore system, as shown in Figure~\ref{fig:pnm_nubian_sandstone_qemscan_srxtm}A. The quartz grain types (which, in turn, are mainly monocrystalline with sharp edges; Figure~\ref{fig:pnm_nubian_sandstone_qemscan_srxtm}B) are cemented by silica. This Palaeozoic quartz arenite sandstone was deposited in continental to fluvial braided river systems. It represents the primary prolific conventional reservoir in the ~cGOS. Such a quartz-dominated system suggests that neither mineral dissolution/precipitation nor clay swelling require consideration during the quasi-static simulations.

For the laboratory rock physics experiments, three cylindrical plugs of  $\mathrm{25.5 \;mm}$ diameter and  $\mathrm{>\!30 \;mm}$ height were cored from the rock block parallel to each other to investigate the intrinsic rock properties and quantify possible anisotropies. A lathe was used to produce a high degree of parallelism between the upper and lower surfaces of the cylindrical plugs, with $\mathrm{\pm 5 \; \micro m}$ precision. The plugs were dried in an oven at $100\mathrm{^{\circ}C}$ for at least 48~hours to remove trapped pore fluid. Before performing further measurements, the plugs were stored in a desiccator with silica gel to prevent moisture from re-entering.

Using a gas-displacement-based apparatus (He-pycnometer Accupyc~II~1340, micromeritics), the matrix volumes of the cylindrical plugs were determined under ambient conditions in a range that keeps the standard deviation of the measurements below~5\%.  Beforehand, the plugs’ masses were measured, using a high-precision balance ($\mathrm{\pm 5.0\times \!10^{-7} kg \; tolerance}$). The bulk volume, matrix volume, dry mass, grain density, and effective porosity (i.e. the volume of interconnected pores) were calculated. 

A cylindrical subplug (with $ 6\;\mathrm{mm}$ diameter and $\mathrm{\sim \!10 \; mm}$ length, that was recovered from the main plug) was used to perform high-pressure Mercury Intrusion Porosimetry (MIP) to evaluate the pore radius, pore-size distribution (PSD), and porosity-related characteristics of Nubian Sandstone. The PSD observations are based on the behavior of a non-wetting liquid in a capillary, following the Young-Laplace relationship. The experiment was set up initially at low pressure ($\mathrm{40 \; MPa}$), followed by a stepwise increase in pressure up to~$\mathrm{400 \; MPa}$. At~$\mathrm{400 \; MPa}$, mercury can be forced to invade pores as small as $\mathrm{5 \;nm}$ in radius, constituting microporosity \citep{Karger2011FlowRock}. At high capillary pressures, the wetting phase (air) is pushed farther into the pore corners, as the radius of curvature of the interface decreases. 

\begin{figure}[hbtp]
\begin{center}
{\small\includegraphics[width=0.83\textwidth]{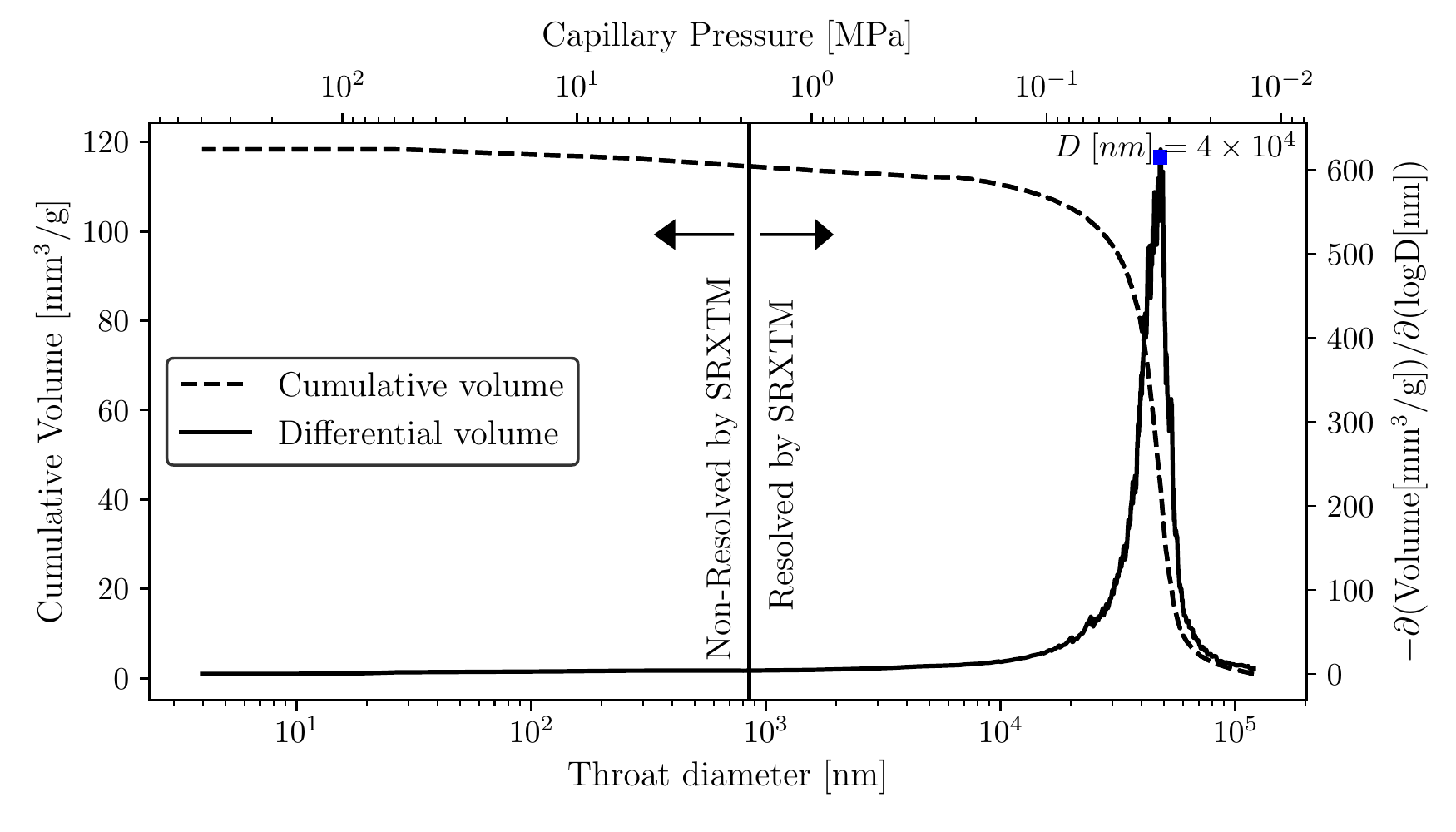}}
\caption{\small{Pore-size distribution of the Nubian Sandstone sample and cumulative mercury volume intruded as a function of pore-throat radius and capillary pressure. Both are determined by Mercury Intrusion Porosimetry (MIP). The majority of pores in the Nubian Sandstone are located in the zone resolved by SRXTM. The mean diameter of the pore elements (throats) is shown above the pore-size distribution curve $\mathrm{\left(\overline{D} =4\times\!10^4 \; nm\right)}$.} }
\label{fig:pnm_nubian_mip_psd}
\end{center}
\end{figure}

To validate the results of our PN quasi-static simulator, we compare the pore-volume density data and the scaled air--mercury $P_\mathrm{c}\!-\!S_\mathrm{w}$ experiment curve with the simulation results extracted from the 3D rock model. Mercury Intrusion Porosimetry (MIP) shows a lognormal pore-size distribution of the Nubian Sandstone samples. MIP also shows that the majority of the pore bodies fall within the resolvable range of our imaging capacity (Figure~\ref{fig:pnm_nubian_mip_psd}). 

% =========================================================
\section{Computerized tomography scanning and image analysis}
\label{pnm_computerized_tomographic_image_analysis}

%===========================================================
\subsection{Imaging technique}
\label{pnm_imaging_technique}
High‐resolution Synchrotron Radiation X‐ray Tomographic Microscopy (SRXTM) can be employed to investigate in a non-destructive and quantitative way the internal topology of micro-structures. The SRXTM measurements were conducted on a cylindrical subplug (with $ 6\;\mathrm{mm}$ diameter and $\mathrm{\sim \!10 \; mm}$ length that was recovered from the main plug). The parallelism of the ends of the cylindrically shaped subplug was achieved by gently grinding the cuttings, first on the side on a rock saw blade, then by hand, using sandpaper (grit 120).

We performed the imaging work at the TOmographic Microscopy and Coherent rAdiology experimenTs (TOMCAT) beamline station, at the third-generation synchrotron facilities, at the Swiss Light Source, Paul Scherrer Institute, Switzerland \citep[see][for technical specifications]{Marone2012RegriddingImaging.}. For the best-contrast phases (pore and solid), the sample was exposed to a parallel beam of monochromatic synchrotron X-ray radiation for $\mathrm{500\; ms}$ exposure time at $\mathrm{26 \; keV}$ photon energy and $\mathrm{401.25\; mA}$ ring current. This beam energy is optimized to provide a sufficient photon flux, capable of penetrating the $\mathrm{6\;mm}$ diameter subplug and ensuring the best image contrast (signal-to-noise ratio). The lower-energy X-rays that hit the cylindrical subplug and do not improve imaging were filtered out using $\mathrm{100\;\micro m \;Al}$, $\mathrm{10\;\micro m \;Cu}$, and $\mathrm{10\;\micro m \;Fe}$ sheets, while the remaining X-rays hit the cylindrical subplug. The field of view (FOV) covered $\mathrm{1.3\times1.4 \;mm^2}$ of the original size of the subplug. 

The transmitted X-rays were converted into visible light by a cerium-doped lutetium aluminium garnet $\mathrm{(LuAG\!: 20 \;\micro m)}$ scintillator and projected at 10$\times$ magnification onto a high-speed CMOS camera (PCO.Edge 5.5; PCO AG, Germany) with $\mathrm{2560\!\times\!2560}$ pixels, leading to an effective pixel width of $\mathrm{0.65 \;\micro m}$. A sample-to-scintillator distance of $\mathrm{36 \;mm}$ yielded a small amount of edge enhancement in the images. 

Each tomogram was computed from $1801$ projections ($\mathrm{500 \;ms}$ exposure time) over a $180\mathrm{^{\circ}}$ rotation by a gridded Fourier transform-based reconstruction algorithm with a Parzen filter using a filtered backprojection algorithm \citep{Marone2012RegriddingImaging.}. Projections were magnified by microscope optics and digitized by a high-resolution CCD camera, which results in 8-bit (256 gray values) tiff-format images. We provide, through the online access of this article, the SRXTM images data set along with a summary of the experiment conditions and the characteristic properties of the SRXTM images. 

The reconstruction center was found for the first and last image in the sequence and linearly interpolated between these two values for the others. The reconstructed volumes were filtered with a $\mathrm{3\!\times\!3\!\times\!3}$ median filter, segmented with local connectivity-based thresholding, and processed further as described in the next section. The image datasets were then cropped around the plug so each one consisted of around $\mathrm{2160\!\times\!2160\!\times\!4320}$ $\left(\mathrm{\sim\!2.02\times\!10^{10}}\right)$ pixels with a voxel size of $\mathrm{0.65 \;\micro m^3}$ at 10-fold optical magnification. A 3D volume with dimension $\mathrm{2160\!\times\!2160\!\times\!4320}$ voxels was cropped from the original 3D raw data set to avoid poor image quality around the edges of the scan volume.
%
%===========================================================
\subsection{Image processing and segmentation}
\label{pnm_image_processing_segmentation}
\begin{figure}[hbtp]
\begin{center}
{\small\includegraphics[width=0.97\textwidth]{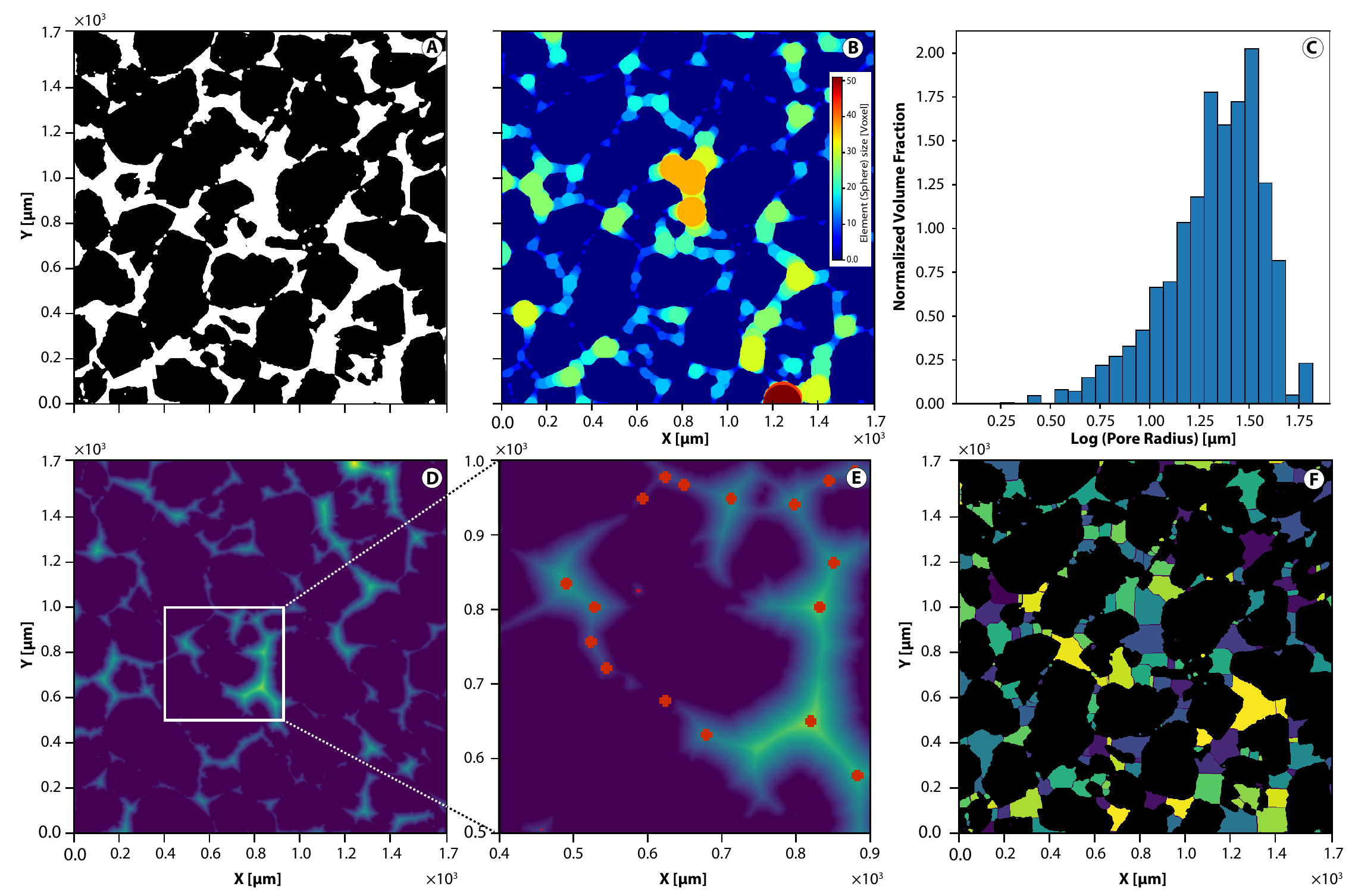}}
\caption{\small{Illustration of the marker-based watershed segmentation workflow on a 2D image. [A] Binarization of the same gray image shown in Figure \ref{fig:pnm_nubian_sandstone_qemscan_srxtm}B by implementing the multi-level Otsu method \citep{Otsu1979AHistograms}. [B] Application of the local Thickness filter with a spherical structuring element to get actual information about the pore-size distribution (i.e. determination of the radius of the largest sphere that fits inside the pore space; $R_\text{max}$) from the CT images. [C] The pore-size distributions determined using the simulated Local Thickness filter. The pore-volume distribution indicates that approximately 80\% of the total pore volume comprises pores with a radius smaller than $\mathrm{55 \; \micro m}$. The histogram of the pore distribution shows that the majority of the pore radii are in the range of $\mathrm{17 - 50 \; \micro m}$, closely matching the experiment values obtained from Mercury Intrusion Porosimetry; see Section \ref{pnm_rock_characterization} and Figure~\ref{fig:pnm_nubian_mip_psd}. [D] Distance transform of void space. [E] Peaks are properly located at the pore centers and marked with $5\times$ exaggerated red dots overlying the global distance transform map. [F] Segmented pore regions resulting from the marker-based watershed segmentation overlain by the mineral phase (black).}}
\label{fig:pnm_watershed_segmentation_workflow}
\end{center}
\end{figure}

The gray-level SRXTM images of the Nubian Sandstone have only one major identifiable mineral phase (Quartz; from Section \ref{pnm_rock_characterization}) with sharp edges (Figure \ref{fig:pnm_nubian_sandstone_qemscan_srxtm}B). The attenuation coefficient of the X-ray (grayscale intensities) in the mineral phase is higher than the attenuation coefficient in pore spaces (darker gray). A bimodal histogram for each image, therefore, was applied using Otsu’s method of automatic thresholding \citep{Otsu1979AHistograms}; Figure \ref{fig:pnm_watershed_segmentation_workflow}A. The threshold value was set for the local minimum to alleviate spurious creation of catchment basins.

Image segmentation is done by delineating and labeling sets of pixels within the domain of gray-level SRXTM images into a number of disjoint non-empty sets of objects depending on the gray-level histogram, neighborhood information, and prior knowledge of the image statistics \citep{Soille2014MorphologicalApplications}. The watershed transformation, which treats pixel values as local topography, was used to segment the images. However, watershed segmentation does not straightforwardly provide acceptable results due to over-segmentation. Alleviating this effect will be elaborated upon in the following. The Local Thickness filter replaces every voxel in the binary images with the radius of the largest spherical kernel that fits inside the void space (Figure~\ref{fig:pnm_watershed_segmentation_workflow}B). Analysis of the histogram of the voxel values provides information about the pore-size distribution as illustrated in Figure~\ref{fig:pnm_watershed_segmentation_workflow}C.

After thresholding, a labeling procedure was performed on the obtained binary images to isolate and quantify pore-space voxels from solid voxels using the Euclidian distance map. The distance transform of a bi-level image is defined by the shortest distance from every (pore-space) pixel to the nearest non-zero foreground-valued (solid) pixel and is employed to distinguish between pore bodies and pore throats. Every voxel is assigned to a pore, such that the distance map increases in the direction of the pore centre; (Figure~\ref{fig:pnm_watershed_segmentation_workflow}D). The algorithm also finds throat surfaces by looking for restrictions in the void space that bound pores: here the distance map increases on either side of the throat surface. The distance map is inverted as the watershed grows towards high intensity.

A Gaussian blur filter was applied to the distance transform map to smooth it and to minimize spurious peaks caused by the flat nature of the solids voxelated in the images. In order to minimize over-segmentation, a manual optimization yielded 0.35 and 5 as optimal values for the Gaussian blur filter and the Local Thickness filter, respectively.

Removal of the erroneous peaks that lead to over-segmentation was done individually and in an iterative way, using a minimal cubic structuring element during morphology dilation. The similarity between the maximized dilated peaks and the smooth distance map allows the recognition of new peaks. The comparison between the old peak(s) and the new peak(s) enables eliminating spurious errors. This is followed by applying the marker-based watershed transformation at the local minimum of a hypothetical 3D topography, from which basins are flooded to segment the images and separate the pores (Figure~\ref{fig:pnm_watershed_segmentation_workflow}F). It is possible to compute properties of pore-network elements, including diameters, coordinates, surface area, and volume, for pore bodies as well as diameters, perimeters, and lengths for pore throats. A detailed description of the extraction algorithm and the parameter definition of the PN elements is given in \cite{Gostick2017}. 

%===========================================================
\subsection{Pore-network construction}
\label{pnm_pore_network_construction}
\begin{figure}[hbtp]
\begin{center}
{
\small\includegraphics[width=0.54\textwidth]{NubianVisualizationb.pdf}} \llap{\parbox[b]{3.3in}{[A]\\\rule{0ex}{2.40in}}}\hfill
{\small\includegraphics[width=0.44\textwidth]{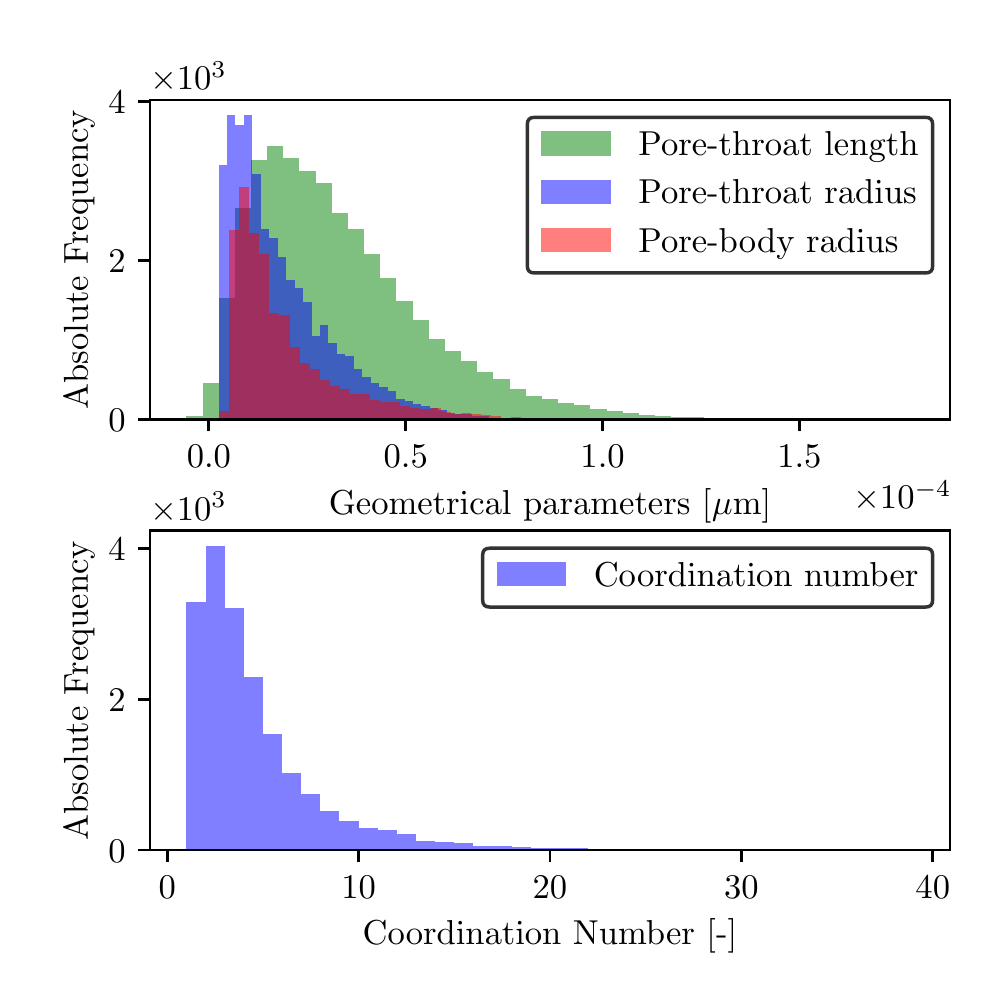}
{\llap{\parbox[b]{2.85in}{[B]\\\rule{0ex}{2.40in}}}}\hfill
{\llap{\parbox[b]{2.85in}{[C]\\\rule{0ex}{1.20in}}}}
}
\caption{\small{[A] 3D view of the Nubian Sandstone pore-network (domain size: $\!1.4\!\times\!1.4\!\times\!2.8 \;\mathrm{mm^3}$). The voxel SRXTM image shows the solid phase in gray, while the extracted pore elements are overlain. The pore bodies and pore throats are respectively rendered as red spheres and gray cylinders that are smaller than their actual sizes to improve visualization. [B] Histograms of pore-body and pore-throat radii as well as pore-throat lengths for the extracted pore network of the Nubian Sandstone. [C] The distribution of the coordination  number, ($z$), in the Nubian Sandstone pore network.} }
\label{fig:pnm_nubian_3d_visualization_statisitics}
\end{center}
\end{figure}

Since the pore morphology of Nubian Sandstone is quite complex (Figures~\ref{fig:pnm_nubian_sandstone_qemscan_srxtm}B and~\ref{fig:pnm_watershed_segmentation_workflow}A), its topological quantification is challenging. Its 3D intricate pore network was constructed using the discretized elements, resulting from the marker-based watershed segmentation method (see Section~\ref{pnm_image_processing_segmentation} and Figure~\ref{fig:pnm_watershed_segmentation_workflow}). In this
paper, the geometrical properties of the extracted pore network refer to the distributions of the pore-body and pore-throat radii, lengths, and volumes. Additionally, relevant topological traits include the spatial connectivity, which prescribes how pore bodies are connected via pore throats (i.e. coordination number, $z$) and a shape-describing parameter for each pore-network element that will be used during numerical fluid-flow modeling (Figure~\ref{fig:pnm_nubian_3d_visualization_statisitics}). Coordination numbers range from $\mathrm{1\!-\!39}$ for the whole extracted network, with a mean coordination number of $\overline{z} \approx 4.1$, which is close to the value found by \cite{Dong2009Pore-networkImages} for Fontainebleau and Berea sandstones. Isolated clusters of pores (i.e. pores with a coordination number of zero) were removed from the pore network.

A three-dimensional visualization of the extracted pore network is presented in Figure~\ref{fig:pnm_nubian_3d_visualization_statisitics}A, where the red spheres represent pore bodies and gray cylinders represent the pore throats, where both appear to be properly located within the rendered voxel image. The regularity of the pore-element shapes is of course an over-simplified representation of the natural porous medium, lacking information of high relevance for two-phase flow. Many studies have focused on idealizing the geometry of pore-network elements to mimic such structure complexities and angularities \citep{Mason1991, Patzek2001ShapeFlow, ren2003ReconstructionEffects}. These idealized pore elements will be discussed in more detail in Section~\ref{pnm_shape_factor}.  The distributions of inscribed radii for the extracted pore-network elements are shown in Figure~\ref{fig:pnm_nubian_3d_visualization_statisitics}, while Table~\ref{tab:statistics_pnm_nubian} provides the statistical information of the pore-network properties, used in the simulations.  
\input{pnm_statisitcs.tex}
%
%===========================================================
\subsection{Representative elementary volume}
\label{pnm_rev}
\begin{figure}[hbtp]
\begin{center}
{\small\includegraphics[width=0.87\textwidth]{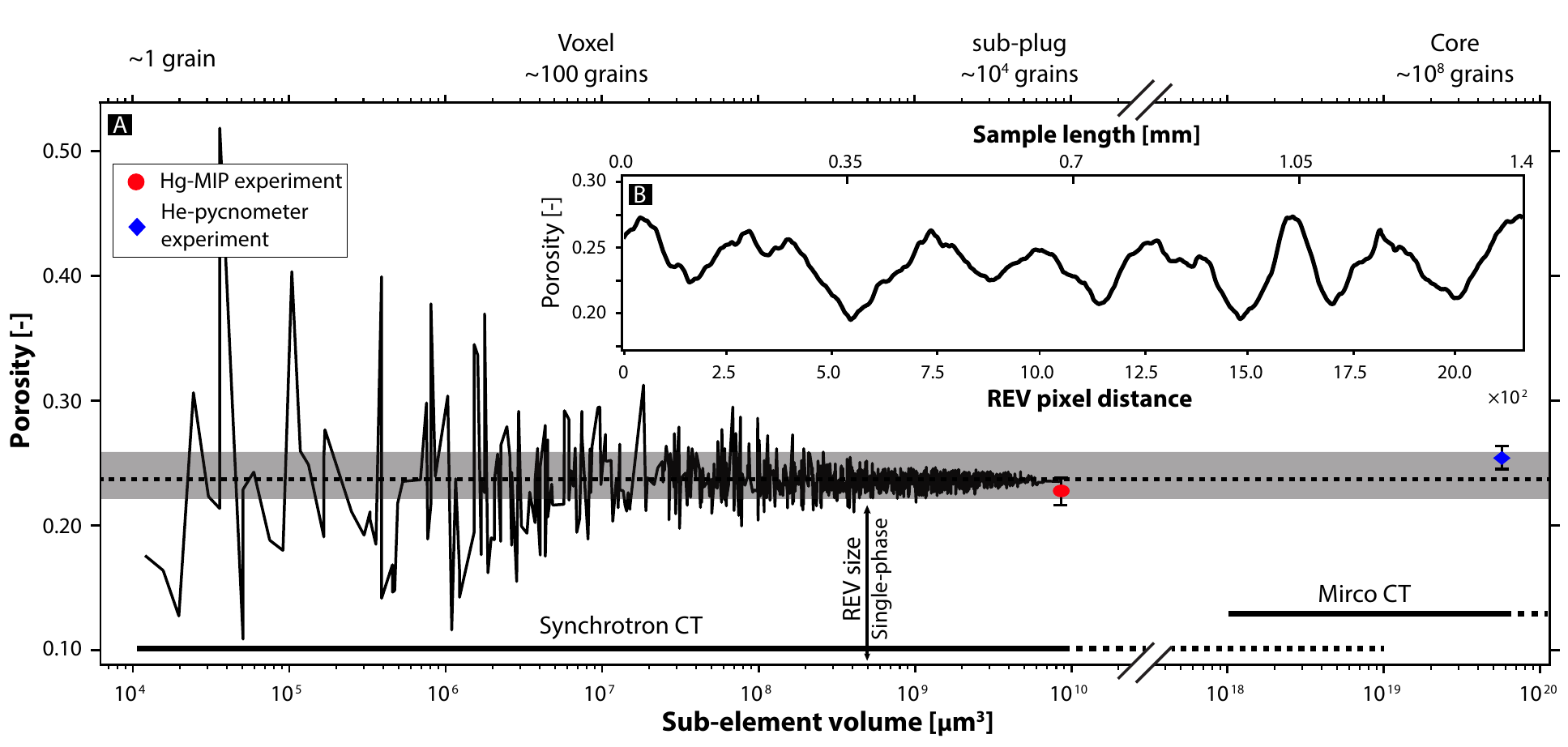}}
\caption{\small{Quantification of the porosity of the Nubian Sandstone retrieved from SRXTM images. [A] The porosity is calculated for an increasingly large cube, starting from a voxel centered in the pore space. Symbols are experimental results of the whole-plug sample. The measured MIP skeleton porosity is 0.263, yielding 0.032 Hg-inaccessible porosity. The gray bar represents the spatial variability zone of porosity. [B] Porosity profile across two opposite faces in the Nubian Sandstone sample.} }
\label{fig:pnm_nubian_rev}
\end{center}
\end{figure}

Before simulating two-phase fluid flow within the constructed pore network (PN), the existence of a representative elementary volume (REV) has to be confirmed by the spatial distribution of its components (e.g. solid--void spaces) to satisfy a lower and an upper bound of the fluid-flow field. These bounds are required to ensure that the REV is sufficiently larger than a single pore and smaller than the flow domain, such that the phase saturations are independent of the REV size and representative of the macroscopic domain. Figure~\ref{fig:pnm_nubian_rev}A shows that the physical properties (here the porosity) are highly fluctuating as the sub-element volume considered is small, and the values stabilize and approach the experiment values, measured via different techniques and for different sizes, as the averaging volume increases. 
The minimum size of the REV for Nubian Sandstone was found to be $\mathrm{\sim\!10^9 \;\micro m^3}$ (Figure~\ref{fig:pnm_nubian_rev}A).

In addition, Figure~\ref{fig:pnm_nubian_rev}B shows variations in the porosity profile as a function of distance across two opposite sides of the sample. These variations are very small and are thus neglected. The two sides correspond to the inlet and outlet boundaries in our PN simulations (Section~\ref{pnm_quasi_static_modelling}). Moreover, it is important to note that the anisotropy parameter of permeability, i.e. the ratio between the vertical and the horizontal permeability, is less than 5\%. Therefore, its significance, in relation to fluid flow effects can be neglected during our quasi-static simulations. We employ direct numerical simulations (DNS), based on the Lattice Boltzmann method (LBM), to determine the permeability anisotropy in the three orthogonal directions (Appendix~\ref{pnm_appendix_b}). In summary, we consider here that the Nubian Sandstone sample is mostly homogeneous in nature and that therefore the local fluid flow vectors always point in the opposite direction of the potential gradient during the simulations. The DNS results show no preferential flow direction, which, judging from the homogeneous nature of the samples (by visual inspection), is not surprising.  
% =========================================================
%\input{Sections/pnm_numerical_method.tex}
\section{Numerical methods}
\label{pnm_numerical_methods}
Over the past few decades, several studies have estimated fluid-flow properties from digital images of rocks \cite[e.g.][]{Valvatne2004PredictiveMedia,Blunt2013Pore-scaleModelling,Krevor2015,Rasmusson2016ATrapping,Rasmusson2018ResidualApproach,Qin2019AMedia}. In this paper, the laws of mass and volume conservation (incompressible and immiscible two-phase, laminar flow in a natural porous medium) are considered, where one fluid, supercritical CO$_{\text2}$ ($\mathrm{scCO_2}$), is the nonwetting phase, while the other (brine with a salinity of $\mathrm{\sim 40}$ \textperthousand) is the wetting phase. The thermophysical properties of these fluids are listed in Table~\ref{tab:pnm_thermophysics}. The salinity value refers to the surface water salinity in the Gulf of Suez after~\cite{Gavish1974GeochemistrySuez}. The assumption here is that seawater was injected into the  reservoir in the past to enhance the recovery of oil. We assume that initially, the brine fully filled the pore network of the Nubian Sandstone.

A quasi-static PNM simulator for $\mathrm{scCO_2}$ invasion after \cite{Qin2016Pore-NetworkCell} has been modified for this study, yielding capillary pressure and relative permeability curves:
\begin{itemize}
    \item The capillary pressure curve is achieved by simulating an invasion--percolation process. 
    \item Two-phase relative permeabilities are calculated at each saturation state of the two-phase quasi-static displacement process, once system equilibrium is reached.
\end{itemize}
The parameterization of the capillary pressure curve is done with the well-known van~Genuchten~(VG) model (Equation~\ref{eqn: pnm_van_genuchten_fitting}), given as \citep{vanGenuchten1980ASoils1}: 

\begin{equation}
    P_\mathrm{c} = \frac{1}{\alpha_{d,i}}\left( S_\mathrm{e}^{-1/m}-1\right)^{1/n}, \quad \mathrm{where} \; S_\mathrm{e} = \frac{S_\mathrm{w}-S_\mathrm{wr}}{1-S_\mathrm{wr}-S_\mathrm{nwr}}\,,
    \label{eqn: pnm_van_genuchten_fitting}
\end{equation}
In Equation \eqref{eqn: pnm_van_genuchten_fitting}, $S_\mathrm{e}$ is the effective saturation of the wetting phase, where $S_\mathrm{w}$, $S_\mathrm{wr}$, and $S_\mathrm{nwr}$ denote the wetting phase saturation attained by the primary drainage curve, the residual saturation of the wetting phase, and the residual saturation of the nonwetting phase, respectively, and $\alpha_\mathrm{d,i}$ $\left[\mathrm{Pa^{-1}}\right]$, $n$, and $m=1-1/n$ are the VG fitting parameters (Table~\ref{tab:pnm_result_sng_phase}). The subscripts `d' and `i' refer, respectively, to the drainage and imbibition process.

\input{pnm_thermophysics.tex}
%
%===========================================================
\subsection{Shape factor}
\label{pnm_shape_factor}
A 2D slice of Nubian Sandstone (Figure~\ref{fig:pnm_nubian_sandstone_qemscan_srxtm}B) shows highly irregular geometries of the pore elements. Even though it is not feasible to reproduce these geometries explicitly in the PNM, it is important to consider their effect on the displacement process. It is necessary, therefore, to replace this geometry by an equivalent idealized geometry which leads to mathematically tractable problems, while at the same time reproducing the main features of the pore space, pertinent to two-phase flow.

\begin{figure}[hbtp]
\begin{center}
{\small\includegraphics[width=0.55\textwidth]{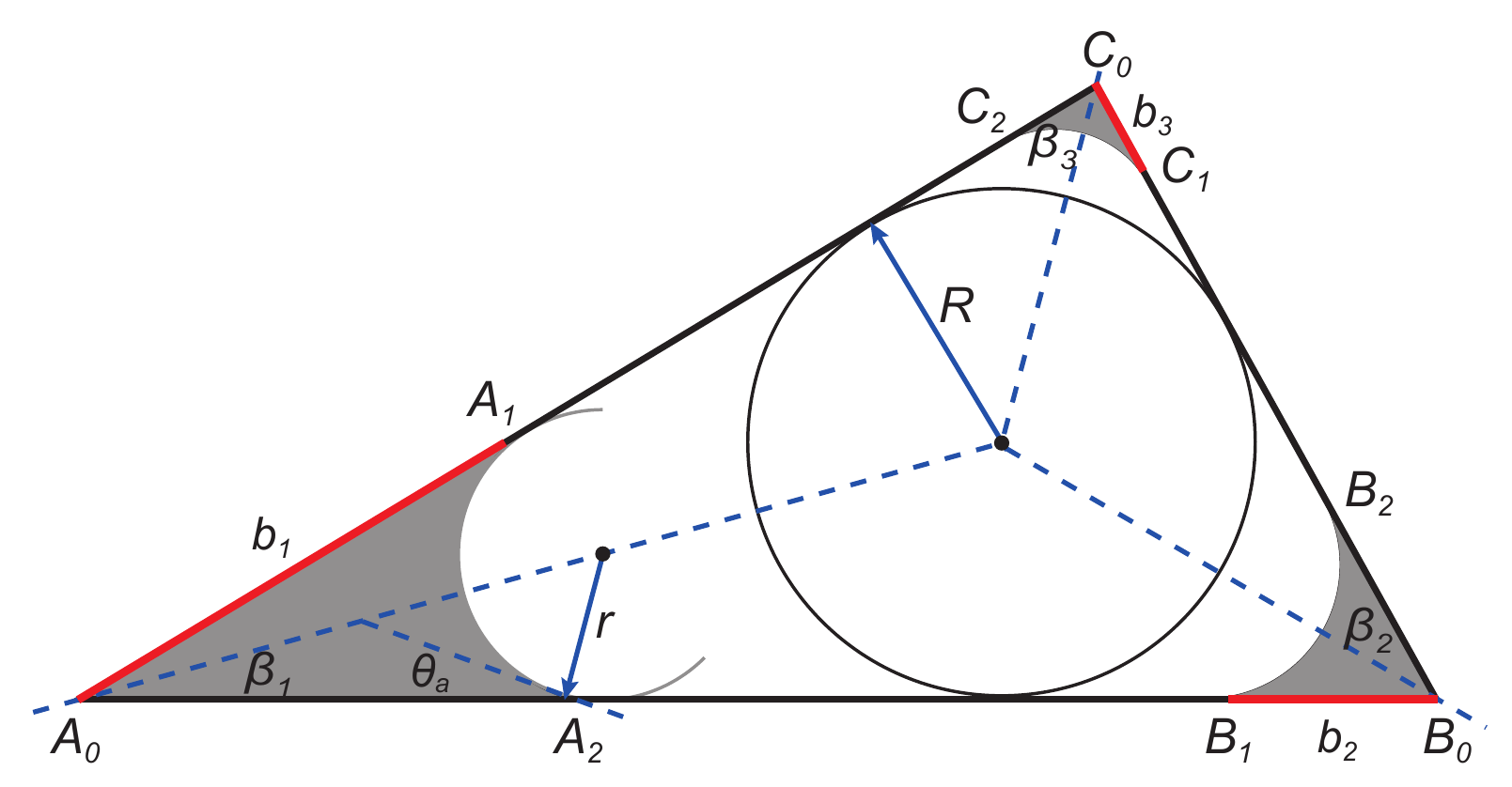}}
\caption{\small{Wetting-phase (gray) distribution in a pore element, which has a triangular cross-section. The three half-corner angles, $\beta_1$, $\beta_2$, and $\beta_3$, are $\mathrm{15^{\circ}}$, $\mathrm{30^{\circ}}$, and $\mathrm{45^{\circ}}$, respectively. $\theta_\text{a}$ is the static advancing contact angle, $R$ is the radius of the inscribed circle, $r$ is the radius of the nonwetting--wetting-interface curvature, and $b$ denotes the distance from each apex to the corresponding nonwetting--wetting interface along the solid wall (red).} }
\label{fig:pnm_triangle_cross_section_shape_factor}
\end{center}
\end{figure}

\cite{Mason1991} introduced the shape-factor concept to describe the geometry of pore-network elements (i.e. pore bodies and pore throats) and their angularity for reliable two-phase flow simulations, where the wetting phase occupies the angular corners and the nonwetting phase represents the bulk (Figure~\ref{fig:pnm_triangle_cross_section_shape_factor}). The most common approach is to choose the shape -- i.e. either a circle, a triangle, or a square $\mathrm{(C\!-\!T\!-\!S)}$ -- whose shape factor (as defined in Equation~\ref{eqn: pnm_shape_factor_1}) is closest to that of the real pore, as the idealized pore geometry, 
%---------------------------------Equation--------------------------------
\begin{equation}
    \label{eqn: pnm_shape_factor_1}
    G = \frac{A}{P^2} \,,
\end{equation}
%-------------------------------------------------------------------------
where $A$ represents the cross-sectional area of a pore or throat and $P$ is its corresponding perimeter.

Our marker-based watershed segmentation provided: (1) the region volume, $(V)$, by summing the number of voxels in that region; (2) the extended radius, $(R_\text{ext})$, given as half of the maximum value of the global distance map, lying within each pore region, which may lead to an overlap of two pore bodies; and (3) the inscribed radius, $(R_\text{ins})$, given as half of the maximum value of the local distance map (i.e. Euclidean distance for a single pore space), lying within each pore region. The length of an idealized pore element, $(L_\text{ipe})$, is the summation of the extended radius and the inscribed radius (i.e. Equation \ref{eqn: pnm_area_porespy}). The cross-sectional area of an idealized pore element, $(A_\text{ipe})$, and its perimeter, $(P_\text{ipe})$, can, therefore, be defined by Equations~\eqref{eqn: pnm_area_porespy} and \eqref{eqn: pnm_permeter_porespy}, respectively, given by:
%
%---------------------------------Equation--------------------------------
\begin{equation}
    \label{eqn: pnm_area_porespy}
    A_\text{ipe} = \frac{V}{R_\text{ins}+R_\text{ext}}, \quad  \mathrm{where} \; R_\text{ins}+R_\text{ext} = L_\text{ipe}\,,
\end{equation}
%-------------------------------------------------------------------------
%
%---------------------------------Equation--------------------------------
\begin{equation}
    \label{eqn: pnm_permeter_porespy}
    P_\text{ipe} = \frac{2V}{R_\text{ins}(R_\text{ins}+R_\text{ext})}\,. 
\end{equation}
%-------------------------------------------------------------------------

For pore throats, the marker-based watershed segmentation counts the number of voxels in the throat with a distance transform value of~1, then multiplies the result by the voxel length to provide the pore throat perimeter. Its inscribed radius can be found from the maximum of the global distance transform. With Equations~\eqref{eqn: pnm_area_porespy} and~\eqref{eqn: pnm_permeter_porespy}, and substitution with Equation~\eqref{eqn: pnm_shape_factor_1}, we obtain a new equation that describes the shape factor of three-dimensional idealized pore elements,
%
%---------------------------------Equation--------------------------------
\begin{equation}
    \label{eqn: pnm_shape_factor_new}
    G = \frac{R^2_\text{ins}}{4A_\text{ipe}}\,. 
\end{equation}
%-------------------------------------------------------------------------

From Equation~\eqref{eqn: pnm_shape_factor_new}, the shape factor values of $\sqrt{3}\big/36$, $1\big/4\pi$, and $1\big/16$ correspond to the 2D cross-sections of a scalene triangle, square, and circle, respectively. Statistical properties of pore-network elements using the newly reformulated shape factor (Equation~\ref{eqn: pnm_shape_factor_new}) are given in Table~\ref{tab:statistics_pnm_nubian}. The details of how to calculate the corner half-angles $\beta_1$, $\beta_2$, and $\beta_3$, with the convention of $[0 \leq  \beta_1 \leq \beta_2 \leq \beta_3 \leq \pi\big/2]$, as illustrated in Figure~\ref{fig:pnm_triangle_cross_section_shape_factor}, for a given shape factor, are given in \cite{Patzek2001VerificationImbibition}.

%===========================================================
\subsection{Quasi-static pore-network modelling}
\label{pnm_quasi_static_modelling}
The pore network (PN), initially fully filled with the wetting phase, is brought in contact with the nonwetting and wetting reservoirs at the inlet and outlet boundaries, respectively. The other boundaries are set to wall (no-flow) boundary conditions. Both reservoir pressure values are set to zero initially. When the primary drainage starts, we gradually increase the nonwetting reservoir pressure by small increments (less than~$\mathrm{5 \;Pa}$). After each increase, we check which pore bodies are invaded by the nonwetting phase. A pore body will be invaded by the nonwetting phase, once the following two conditions are both satisfied: 
\begin{itemize}
    \item The nonwetting reservoir pressure is larger than its entry pressure,
    \item Both nonwetting and wetting phases in the pore body have pathways to the inlet and outlet reservoirs, respectively.
\end{itemize}
Then, the phase occupation in both pore bodies and pore throats is updated under the equilibrium condition. The capillary-pressure relationships, with respect to the wetting saturation used, are given in Appendix~\ref{pnm_appendix_a}. The primary drainage process is terminated once a pre-set maximum drainage pressure is achieved. At this stage, the wetting phase is present in the tight pores and crevices of the network, due to pore-structure trapping, and resides in the corners of the surface roughness of the pores. We use an interfacial tension of $\sigma = 0.0202$~N/m.

For main imbibition, we decrease the nonwetting reservoir pressure step-wise, by a small deduction (less than $\mathrm{5 \;Pa}$), allowing the wetting phase gradually to fill the pore bodies in each step. After each decrease, we check the scenarios of cooperative pore-body filling ($cpf$) and snap-off ($so$) in pore bodies and pore throats. For the cooperative filling of a pore body, if the local capillary pressure is larger than the threshold capillary necessary to fill the pore body, and the nonwetting and wetting phases have pathways to the inlet and outlet reservoirs, the pore body will be fully filled with the wetting phase by spontaneous imbibition. When snap-off happens in a pore throat, the pore throat will be completely occupied by the wetting phase. When snap-off happens in a pore body, the nonwetting phase will be locally trapped. For the remaining pore bodies and pore throats, the phase occupation is updated by the piston-type movement of arc menisci (AMs) or the main terminal meniscus (MTM) filling \citep{Ma1996EffectTubes} under the equilibrium condition. The relationships for checking snap-off events and the threshold capillary pressure for the cooperative filling, used here, are given in Appendix~\ref{pnm_appendix_a}. 

Finally, in both drainage and imbibition processes, as long as the phase occupation needs to be updated, a search operation needs to be conducted to check if the nonwetting and wetting phases exhibit pathways that are connected to the inlet and outlet reservoirs, respectively. If both pathways can be found, the phase occupation will be updated to the new one. Otherwise, the update is aborted due to trapping.
%
%===========================================================
\subsection{Calculation of relative permeability}
\label{pnm_calculation_relative_permeability}
For a given stage in the drainage or imbibition process, the conducting pathways of the nonwetting and wetting phases are known. For either phase, its volumetric conservation in a pore body is given by:
%
%---------------------------------Equation---------------
\begin{equation}
    \label{eqn: pnm_volumetric_conservation_quasistatic }
     \sum_{j=1}^{N_i} K^{\kappa}_{ij}\left(P^{\kappa}_{i}-P^{\kappa}_{j} \right) = 0\,.
\end{equation}
%-----------------------------------------------------------
Here, $i$ is the pore-body index, $ij$ is the pore-throat index, $N_i$ is the coordination number, $\kappa$ is the phase indicator, $P\; \left[\mathrm{Pa}\right]$ is the pressure, and   $K^{\kappa}_{ij} \; \left[\mathrm{m^3/Pa/s}\right]$ is the conductivity, calculated by
%
%---------------------------------Equation------------------
\begin{equation}
    \label{eqn: pnm_relative_permeability_quasistatic}
    K^{\kappa}_{ij} = {1}\Bigg/{\left(\frac{l_{i}}{2g^{\kappa}_{i}} + \frac{l_{j}}{2g^{\kappa}_{j}}+\frac{l_{ij}}{g^{\kappa}_{ij}} \right)} \,,
\end{equation}
%-----------------------------------------------------------
where $l \; \mathrm{[m]}$ is the length, and $g \; \left[\mathrm{m^4/Pa/s}\right]$ is the conductance, given in Appendix~\ref{pnm_appendix_a}.
With proper inlet and outlet pressure boundary conditions, Equation~\eqref{eqn: pnm_volumetric_conservation_quasistatic } can be used to solve for the pressure field. Then, Darcy’s law is used to obtain the relative permeability for the given network saturation value.

% =========================================================
% \input{Sections/pnm_results.tex}
\section{Results}
\label{pnm_results}
In the following sections, the pore-space geometry of the Nubian Sandstone sample, as determined experimentally using the relevant data from the MIP experiment, serves as an input for the image-based PNM extraction approach. In all the simulations that follow, we run primary drainage--main imbibition cycles using the quasi-static pore-network model, while varying the saturation at the point of reversal from drainage to imbibition.
%
%===========================================================
\subsection{Pore-network calibration}
\label{pnm_result_pore_network_calibration}
We assume that the pore throats are volumeless elements of the pore network and account for resistance to flow, while pore bodies account for the pore volume but do not provide any resistance to flow. On a local level, however, the extracted PN of the Nubian Sandstone contains slate-shaped pores that have an aspect ratio (i.e. ratio of pore-body to pore-throat sizes) smaller than one. Only small spheres can be inscribed inside pore spaces with such pore-to-throat aspect ratios, but the throats could still be quite large. In this case, such a division into pore bodies and throats has little physical meaning. Therefore, we adjust the pore-throat radii until the measured capillary pressure--saturation relationship, obtained from the MIP experiment, is fulfilled. This is done by assuming that the aspect ratio of any pore-body/pore-throat pair must be greater than, or equal to, a given threshold value. The resulting minimum aspect ratio, $R_\text{asp}$, for the analyzed Nubian Sandstone sample in this work is 2.27, compared to 2.25 for Berea Sandstone \citep{ren2003ReconstructionEffects}. This calibration is done manually by trial-and-error. 

\begin{figure}[hbtp]
\begin{center}
{\small\includegraphics[width=0.80\textwidth]{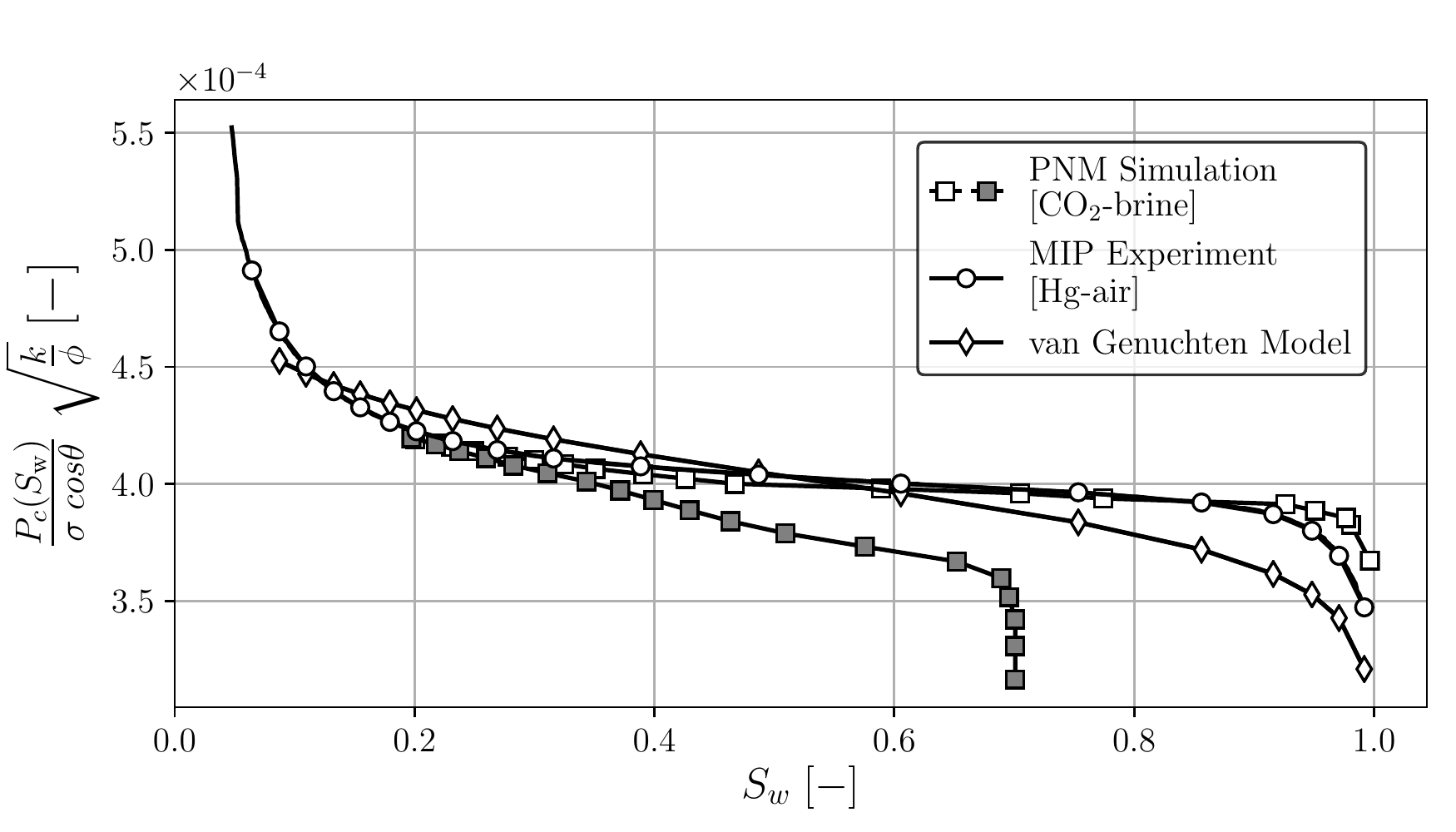}}
\caption{\small{Results of the PN model calibration for the Leverett-$J$ versus wetting-phase-saturation drainage curve, using the results from the MIP experiments. The van~Genuchten fit is also shown. The primary drainage curves are shown with unfilled symbols. The PNM is used to predict the main imbibition curve, shown here with gray-filled symbols.} }
\label{fig:pnm_nubian_pc_sw_mip_pnm_vg}
\end{center}
\end{figure}

The $P_\mathrm{c}\!-\!S_\mathrm{w}$ relationship used for the calibration is obtained by extending the results of the MIP experiments from the mercury--air fluid pair, using a Leverett-$J$ scaling \citep{Leverett1941} to the CO$_{\text2}$--brine system (Table~\ref{tab:pnm_result_sng_phase}). 
In addition, a van~Genuchten (VG) model was fitted to the $P_\mathrm{c}\!-\!S_\mathrm{w}$ curve as shown in Figure~\ref{fig:pnm_nubian_pc_sw_mip_pnm_vg}. 

\input{pnm_result_sng_phase2.tex}

Table~\ref{tab:pnm_result_sng_phase} summarizes the intrinsic hydraulic and capillary properties of the Nubian Sandstone, using experimental and computational methods. The results of the quasi-static single-phase flow simulation shows that the absolute permeability is in agreement with the value obtained from the MIP experiment. It can be seen that the absolute permeability of the network $\left(\mathrm{2.00\times\!10^{-12}}\; \mathrm{m^2}\right)$ is close to that of the plug sample $\left(\mathrm{MIP \;experiment;}\; \mathrm{1.70\times\!10^{-12}}\; \mathrm{m^2}\right)$. The permeability obtained from direct numerical simulation, using the Lattice Boltzmann method (our in-house code, LBHydra, \cite{Walsh2010InterpolatedKinetics,Walsh2010MacroscaleMedia, Walsh2013DevelopingUnits}) is with  $\mathrm{2.57\times\!10^{-12}}\; \mathrm{m^2}$ slightly higher.

%===========================================================

\subsection{Residual trapping in Nubian Sandstone}
\label{pnm_result_residual_trapping}
\begin{figure}[hbtp]
\begin{center}
{\small\includegraphics[width=0.80\textwidth]{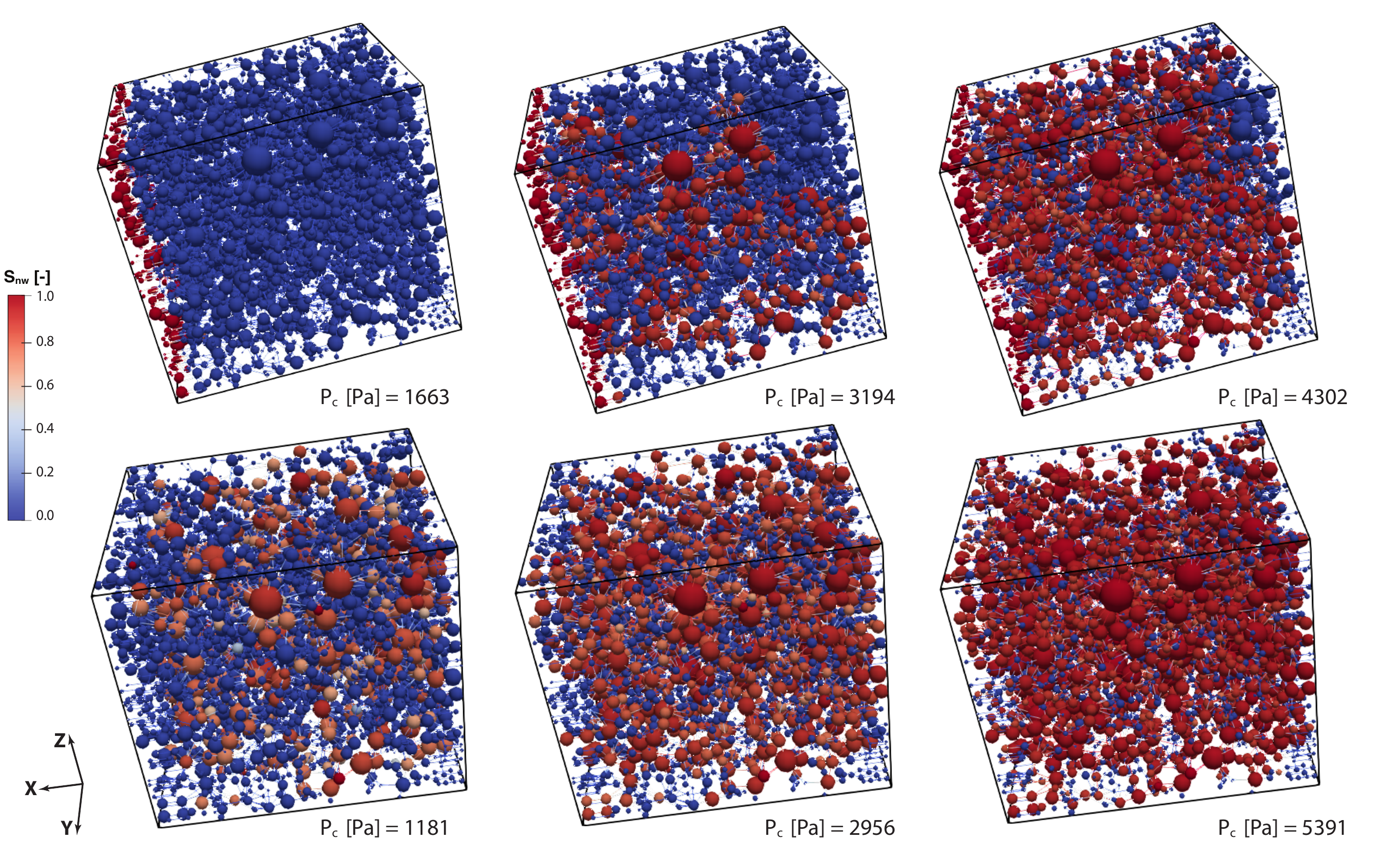}}
\caption{\small{Snapshots of the quasi-static simulations, with a wetting-phase (brine) saturation, at the point of reversal from drainage to imbibition, of $S_\mathrm{nw,i}=0.1$. The brine-filled pores are shown in blue, while red represents CO$_{\text2}$-filled pores. The upper panel shows a representative process of primary drainage at three different capillary pressures, while the lower panel shows imbibition. We use only half of the extracted pore network, shown in Figure~\ref{fig:pnm_nubian_3d_visualization_statisitics}A, as it is more appropriate for the simulations to have a cube-shaped domain.} }
\label{fig:pnm_nubian_quasi-static_ir01}
\end{center}
\end{figure}

In this section, we investigate pore-scale fluid-displacement and trapping mechanisms by simulating primary drainage--main imbibition cycles. The  simulations use the calibrated pore network of the Nubian Sandstone sample (Section~\ref{pnm_result_pore_network_calibration}). We assume water-wet conditions and that the contact angle is constant for each process, i.e. that the receding contact angle, $\theta_\mathrm{r}$, is 0$\mathrm{^{\circ}}$ (primary drainage), and the advancing contact angle, $\theta_\mathrm{a}$, is 35$\mathrm{^{\circ}}$ (main imbibition). The primary drainage simulations stop when the pore-space saturation of the non-wetting phase reaches a target value, $S_\mathrm{nw,i}$. This is the initial condition for the subsequent main imbibition process, where the wetting saturation increases till the non-wetting phase is no longer connected to its reservoir. A visualization of the primary drainage--main imbibition cycle, with $S_\mathrm{nw,i}=10\%$ and an advancing contact angle of $35 \mathrm{^{\circ}}$, is shown in Figure~\ref{fig:pnm_nubian_quasi-static_ir01}. The brine-filled pores are shown in blue, while red represents CO$_{\text2}$-filled pores. 
The figure shows clearly that most residual trapping takes place under low capillary pressures, i.e. towards the end of the imbibition cycle. 
In the simulations, $S_\mathrm{nw,i}$ is varied between~0.1 and~0.9, in order to assess the effect of initial saturation on the residual nonwetting saturation.  Further discussions on the relationship between $S_\mathrm{nw,i}$ and $S_\mathrm{nw,r}$ (i.e. the initial--residual relationship,~IR) are provided in Section~\ref{pnm_IR_discussion}.

\begin{figure}[hbtp]
\begin{center}
{\small\includegraphics[width=0.395\textwidth]{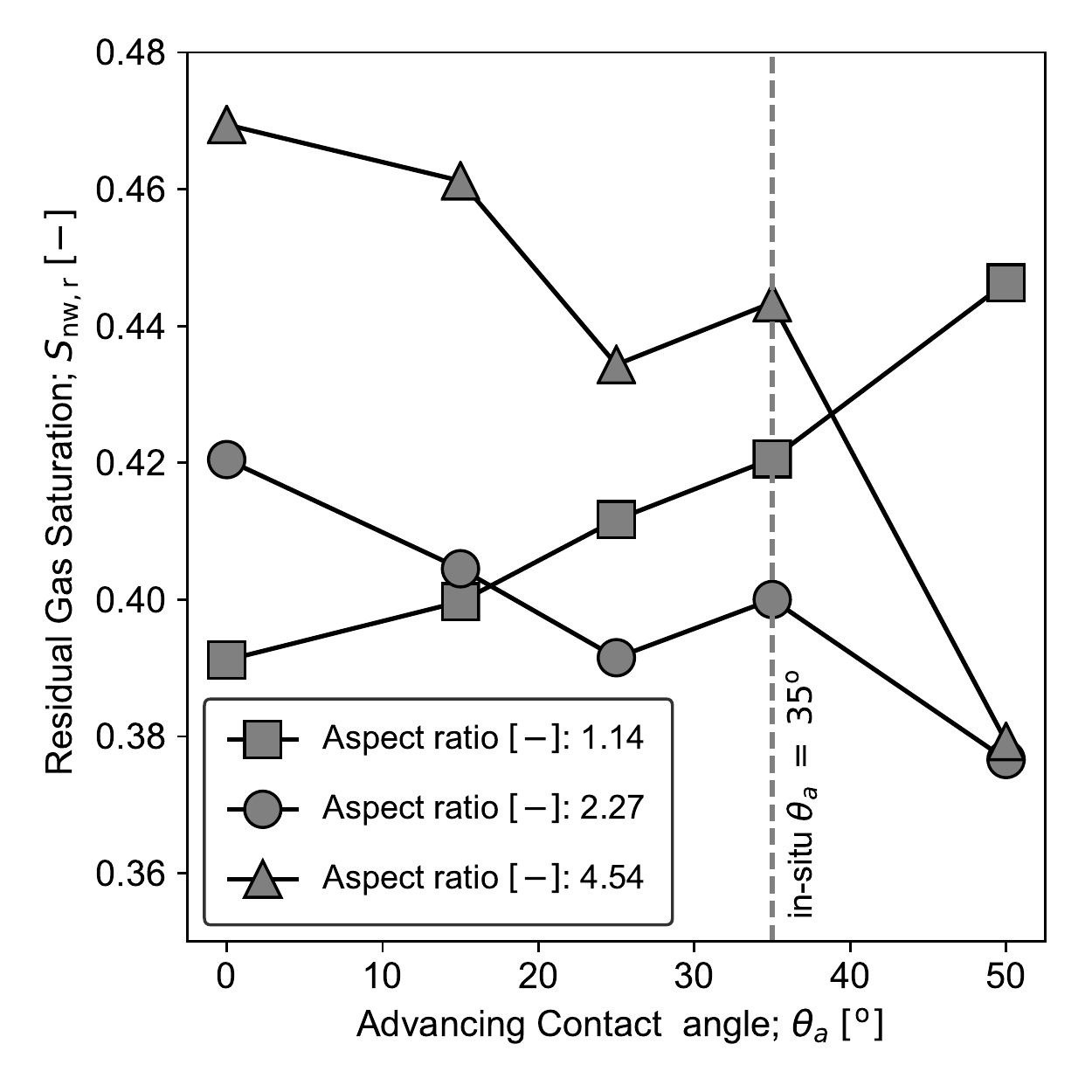}}
\caption{\small{Sensitivity of residual CO$_{\text2}$ trapping to different values of the advancing contact angle $\left(\theta_\mathrm{a}\;\left[ \mathrm{^{\circ}}\right]\right)$, as a function of the minimum aspect ratio of the pore network. The figure shows that an increase in the amount of trapped CO$_{\text2}$ corresponds to a decrease in the contact angle for the medium (2.27) and high (4.54) minimum aspect ratios. In contrast, the amount of residual CO$_{\text2}$ trapping increases with increasing advancing contact angle for the smallest minimum aspect ratio (1.14).} }
\label{fig:pnm_nubian_IR_CA_curve}
\end{center}
\end{figure}

Furthermore, we investigate two distinct processes, whose competition influences the residual trapping at the pore scale. The first is changing the system wettability from strongly water-wet (i.e. $\theta_\mathrm{a}= 0\mathrm{^{\circ}}$) to more neutrally water-wet (i.e. $\theta_\mathrm{a}= 50\mathrm{^{\circ}}$). The second is the sensitivity of residual trapping to the minimum aspect ratio, ${R_\text{asp}}$. In addition to ${R_\text{asp}}=2.27$, obtained during the pore-network calibration, we chose ${R_\text{asp}}$ values of~1.14 and~4.54 to analyze statistically the fluid distribution, the saturation map, and the trapping events. Figure~\ref{fig:pnm_nubian_IR_CA_curve} shows, how the amount of residually trapped CO$_{\text2}$ differs with contact angle and minimum aspect ratio, ${R_\text{asp}}$. Generally, more CO$_{\text2}$ trapping occurs at lower contact angles and higher minimum aspect ratios. However, this does not hold true for the lowest minimum aspect ratio of ${R_\text{asp}}=1.14$. In this case, $S_\text{nw,r}$ increases with increasing contact angle, and even surpasses those of the higher aspect ratios at high contact angles.  

%===========================================================
\subsection{Relative permeability}
\label{pnm_result_relative_permeability}
\begin{figure}[hbtp]
\begin{center}
{\small\includegraphics[width=0.87\textwidth]{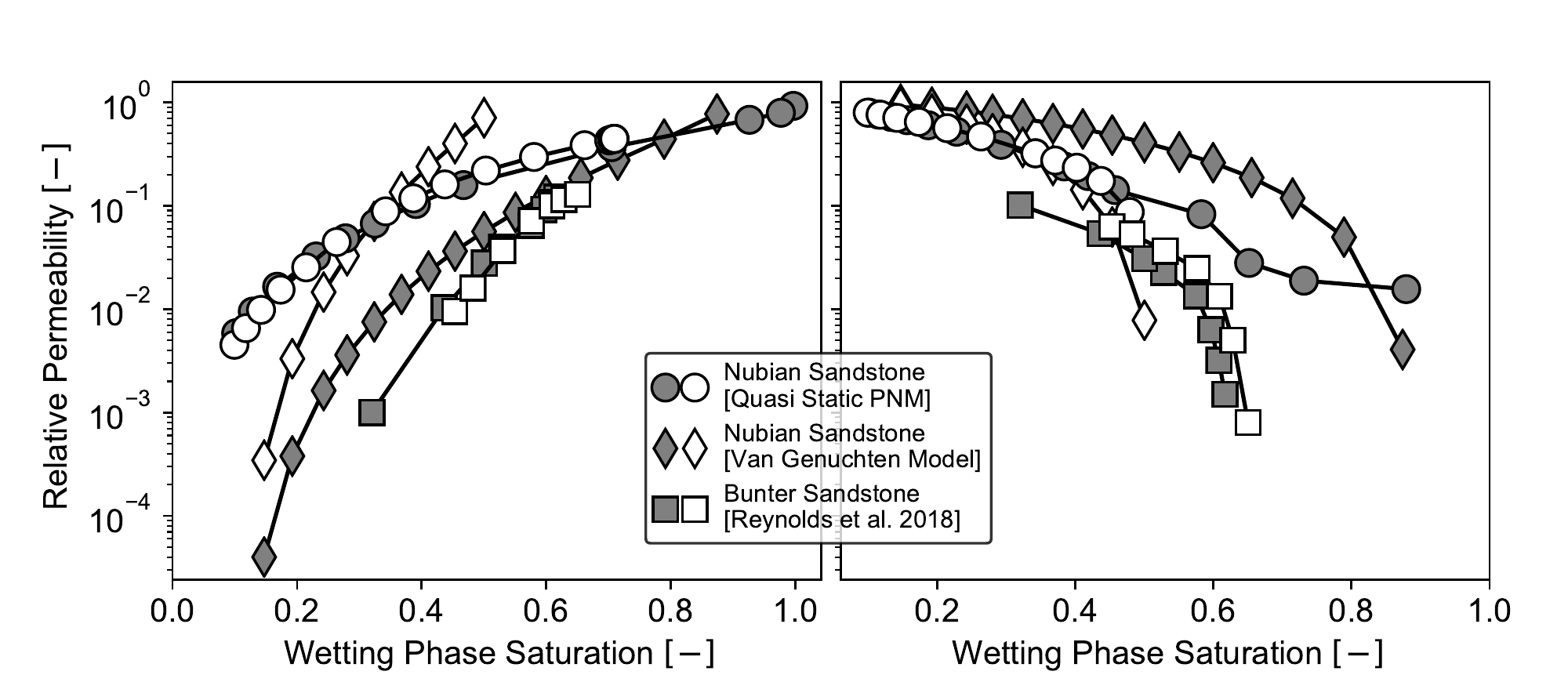}
\llap{\parbox[b]{4.85in}{[A: Brine]\\\rule{0ex}{1.9in}}}
\llap{\parbox[b]{0.95in}{[B: CO$_{\text2}$]\\\rule{0ex}{1.9in}}}
}
\caption{\small{Comparison between the predicted and the experiment-based relative permeability curves ($k_\mathrm{r}-S_\mathrm{w}$) for brine [A] and CO$_{\text2}$ [B] during the primary drainage and main imbibition processes, commencing at an initial wetting-phase saturation of~0.1 (i.e. $S_\mathrm{w,i}$, the saturation before the start of the main imbibition). Drainage is shown with filled symbols and imbibition is shown with open symbols. The comparison in relative permeability curves is between drainage--imbibition cycles of quasi-static pore-scale simulations and the fitted van~Genuchten model for Nubian Sandstone as well as experiment data for Bunter Sandstone after \cite{Reynolds2018MultiphaseKingdom}.} }
\label{fig:pnm_nubian_kr_mip_bunter_vg}
\end{center}
\end{figure}

In this study, the predicted values for the relative fluid permeabilities in the Nubian Sandstone sample are obtained via quasi-static pore-network modeling. These values are then compared to the two-phase experiment results, reported in \cite{Reynolds2018MultiphaseKingdom} for the Bunter Sandstone (UK) as well as to the VG functions, which were fitted to the capillary pressure curve, obtained from our MIP experiment (Figure~\ref{fig:pnm_nubian_kr_mip_bunter_vg}). Generally, there is a good agreement between the simulated relative permeabilities and the calculated values from the VG function (fitted with the capillary pressure curve). The Nubian Sandstone relative permeabilities tend to be higher than those observed experimentally for the Bunter Sandstone, but the same trend is seen. The end-point nonwetting-phase relative permeability, at the irreducible water saturation for the Nubian Sandstone sample, obtained from our PN simulations, is $k_\mathrm{r,nw}\left(S_\mathrm{w,ir} = 0.099 \right)\ = 0.794$.

We investigate the dependence of hysteresis, of the relative permeability curves, on the pore-scale phenomena described in Section~\ref{pnm_result_residual_trapping} (i.e. wettability and minimum aspect ratio), as shown in Figure~\ref{fig:pnm_nubian_kr_ca_ar_ir_curves}.  
Based on the observations from our pore-network study, we infer that the hysteresis in the $k_\mathrm{r,nw}-S_\mathrm{w}$ relationship is primarily caused by the dependency of the swelling of arc menisci in pore corners during imbibition. The lower the initial wetting saturation (i.e. the wetting saturation when drainage switches to imbibition), the more the nonwetting phase (here CO$_{\text2}$) in the pore bodies tends to get disconnected at the end of the imbibition process, thereby becoming more isolated and immobilized, in part via cooperative pore-filling. 

%========================================================

\subsection{Initial--residual saturation relationship for Nubian Sandstone}
\label{pnm_IR_discussion}
Two-phase modeling at the continuum scale can benefit significantly from realistic residual trapping (IR) constitutive relationships in order to quantify the magnitude of trapping \citep{Krevor2015}. Several empirical trapping models have been developed to describe residual trapping. The most widely used model for water-wet systems is that of \cite{Land1968CalculationProperties}. In this model, the residual saturation is calculated by
%---------------------------------Equation------------------
\begin{equation}
    \label{eqn: pnm_land_model}
    S_\mathrm{nw,r} = \frac{S_\mathrm{nw,i}}{1+CS_\mathrm{nw,i}}\, ,
\end{equation}
%-----------------------------------------------------------
where, $C$ is the Land trapping coefficient, which is found by fitting a curve to a plot of the residual saturation of the nonwetting phase ($S_\mathrm{{nw,r}}$) versus the initial saturation at the end of drainage and prior to imbibition ($S_\mathrm{{nw,i}}$). The higher $C$ is, the lower is the amount of trapped nonwetting fluid (here CO$_{\text2}$), and therefore the weaker the hysteresis between drainage and imbibition. We use the calibrated pore network (Section~\ref{pnm_result_pore_network_calibration}) to estimate the trapping curve for the $\mathrm{CO_2-brine}$ pair in Nubian Sandstone and compare it with the trapping curves that were experimentally determined for similar rock types by \cite{Akbarabadi2013RelativeConditions} and \cite{Reynolds2018MultiphaseKingdom}.

\begin{figure}[hbtp]
\begin{center}
{\small\includegraphics[width=0.65\textwidth]{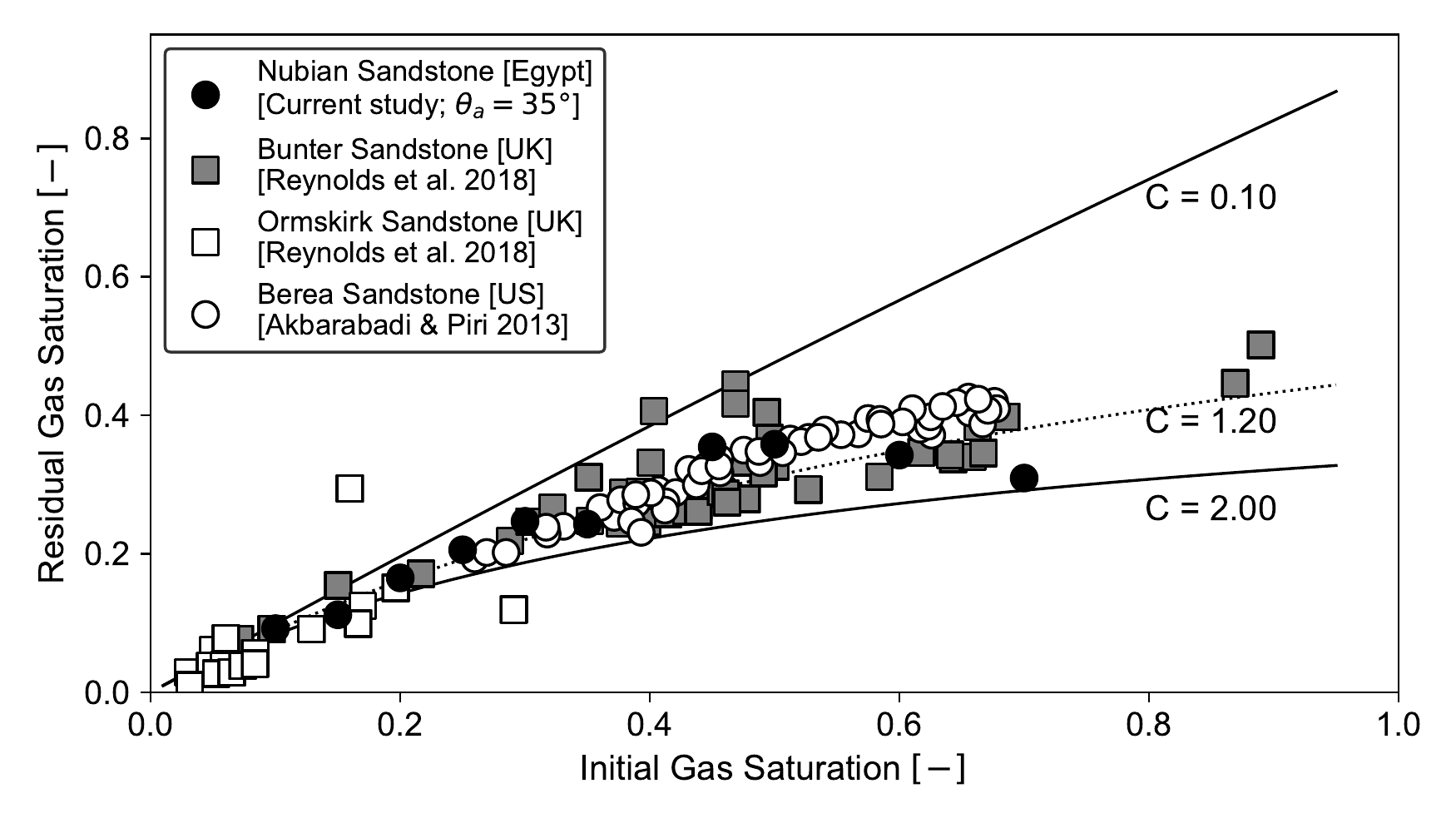}}%\llap{\parbox[b]{3.6in}{[B]\\\rule{0ex}{2.28in}}}}
\caption{\small{Predicted residual trapping of supercritical CO$_{\text2}$ in Nubian Sandstone, compared to experiment IR data for Berea and Ormskirk Sandstones. The predicted initial and residual saturations of CO$_{\text2}$ are obtained by quasi-static pore-network modelling. The solid lines represent Land model curves with the labeled coefficients, while the dashed line shows the fitted IR curve for PN-simulation results for the Nubian Sandstone.} }
\label{fig:pnm_nubian_IR_curve}
\end{center}
\end{figure}

For the Nubian Sandstone, with an advancing contact angle of 35$^\circ$, the best fit for the Land trapping coefficient (from Equation~\ref{eqn: pnm_land_model}) is~1.2, with~0.86 as the coefficient of determination. Nevertheless, $C$ exhibits a range of possible values between~0.1 \citep{Reynolds2018MultiphaseKingdom} and~1.6 \citep{Akbarabadi2013RelativeConditions} for Bunter and Berea Sandstones, respectively.
However, \cite{Krevor2015} reported a value of $C = 5$ for a wider range of rock types. Figure~\ref{fig:pnm_nubian_IR_curve} shows that the capillary trapping curves increase monotonically with initial saturation. At $S_\mathrm{nw,i}$ $\sim\!0.6$, the relationship is nearly linear. Interestingly, Nubian Sandstone has relatively good capillary trapping characteristics with up to 45\% of the CO$_{\text2}$ initially present remaining as a trapped fluid after imbibition. A similar residual trapping value has been observed by \cite{Spiteri2008ACharacteristics} for Berea Sandstone. It is worth noting, however, that in a mixed-wet system, residual trapping would be greatly reduced \citep{Spiteri2008ACharacteristics}. 

% =========================================================
%\input{Sections/pnm_discussion.tex}
\section{Discussion}
\label{pnm_discussion}
Overall, the Nubian Sandstone sample is relatively homogeneous, with only small variations in the porosity profile along its length (Figure~\ref{fig:pnm_nubian_rev}). The developed quasi-static pore network~(PN) simulator demonstrates the sensitivity of residual trapping in such a homogeneous sandstone to the key parameters wettability and minimum pore-body-to-throat aspect ratio. 
\subsection{Sensitivity analysis of Pore-Network modelling}
\paragraph{Displacement mechanisms:}
Generally, three different pore-scale displacement mechanisms are possible during imbibition, as explained in Section~\ref{pnm_capillarytrapping}. Piston-type displacement is always favored over cooperative pore-body filling and snap-off, wherever piston-type displacement is topologically possible. However, in order for nonwetting-phase trapping to occur, cooperative pore-body filling or snap-off is required. 

\begin{figure}[hbtp]
\begin{center}
{\small\includegraphics[width=0.9\textwidth]{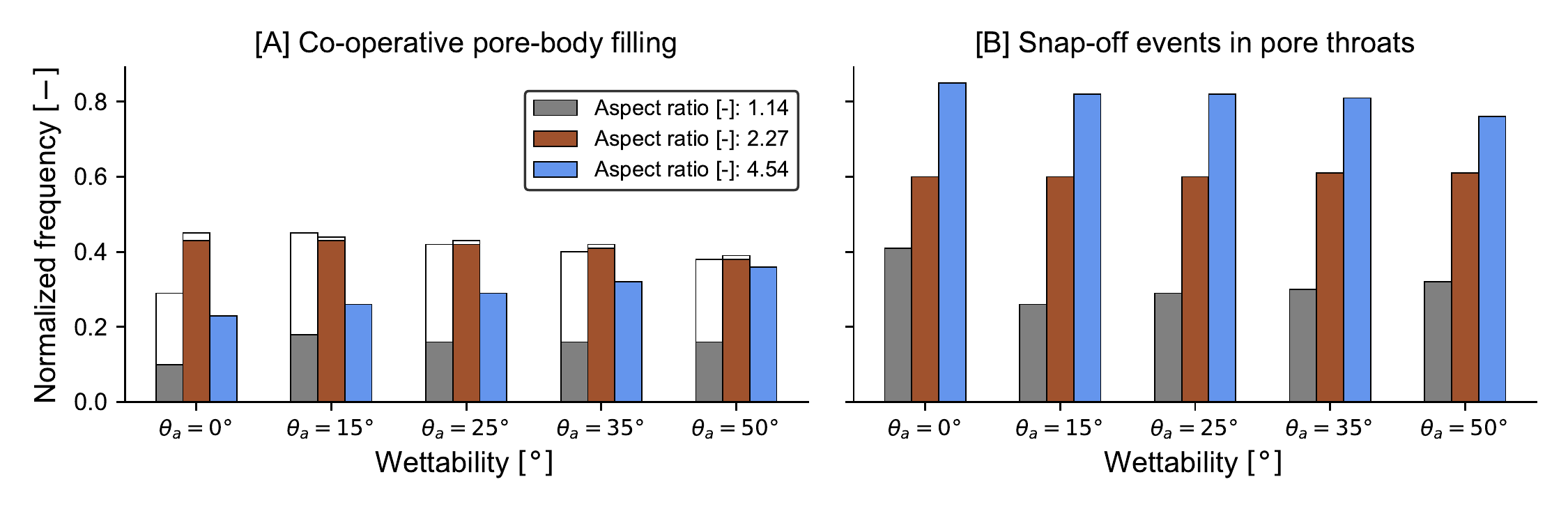}}
\caption{\small{Comparison between the fluid-displacement mechanisms in the pore-network model of the Nubian Sandstone as a function of wettability and minimum aspect ratio. We present here, the fluid-displacement processes that are responsible for (CO$_{\text2}$) trapping only: [A] cooperative pore-body filling ($cpf$) events and [B] snap-off events in pore throats. The white bars in the cooperative pore-body filling plot represent snap-off events in the pore body. The piston-like advance process, either in pore bodies [A] or pore throats [B], represent the remainder (up until a normalized frequency of~1) of each bar for each case.} }
\label{fig:pnm_nubian_pore_scale_displacment}
\end{center}
\end{figure}

Figures~\ref{fig:pnm_nubian_pore_scale_displacment}A and \ref{fig:pnm_nubian_pore_scale_displacment}B show two histograms of the number of occurrences of each displacement mechanism in the pore bodies and pore throats, respectively. One can see that, for the two higher minimum aspect ratios, co-operative pore-body filling ($cpf$) is the main mechanism responsible for the trapping of the nonwetting phase in the pore bodies. $cpf$ occurs most often for $R_\text{asp} = 2.27$. Snap-off ($so$) is all but absent in the pore bodies. For $R_\text{asp} = 1.14$, however, $so$ is slightly more common than $cpf$. In addition, the number of snap-off events in the pore throats increases with increasing minimum aspect ratio. Such behavior is expected, as it is easier for the wetting phase, present in the corners of the pore throats, to connect (leading to snap-off) in narrow throats.
\paragraph{Effect of wettability on residual CO$_{\text2}$ trapping:}
For $R_\text{asp} = 2.27$ and $R_\text{asp} = 4.54$, the relationship between the amount of residually trapped CO$_{\text2}$ and the contact angle is straightforward. The lower the contact angle, the higher the entry pressure required for a nonwetting CO$_{\text2}$ bubble in a pore body to invade a (narrow) pore throat. Hence, as shown in Figure~\ref{fig:pnm_nubian_IR_CA_curve}, $S_\text{nw,r}$ reduces with increasing contact angle. This behaviour is consistent with previous experiment results, using core-flooding computerized tomography on different sandstones \citep{Chatzis1983MagnitudeSaturation., Herring2013EffectSequestration,Geistlinger2014QuantificationMicrotomography, Krevor2015}.

\begin{figure}[hbtp]
\begin{center}
{\small\includegraphics[width=0.9\textwidth]{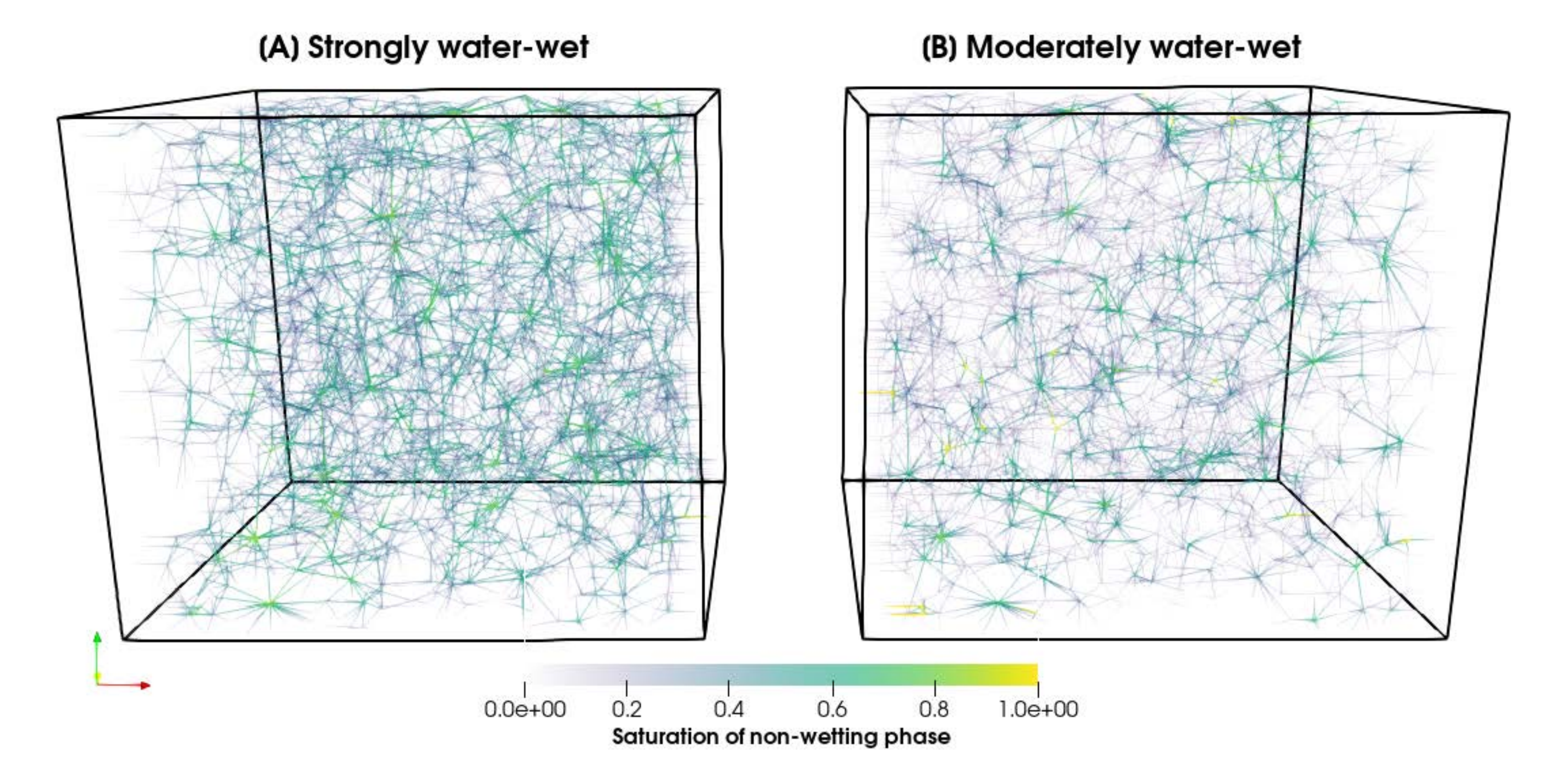}}

\caption{\small{Volume fraction (saturation) of the trapped nonwetting phase (CO$_{\text2}$) in the pore network, with $R_\text{asp}=1.14$, $S_\text{w,i}=10\,\%$, and $\theta_\mathrm{a}=0\,^\circ$ [A] and $\theta_\mathrm{a} =50\,^\circ$ [B] at the end of the imbibition process.} }
\label{fig:pnm_trapping_map_low_aspect_ratio}
\end{center}
\end{figure}

However, this tendency is inverted for $R_\text{asp}=1.14$. It has been shown in imbibition experiments with low-aspect-ratio capillary micromodels, that strongly water-wet displacement yields better displacement efficiencies (sweep) than mixed-wet systems \citep{Mattax1961}. Hence, lower contact angles result in smaller residual trapping values in this case. The reason for this effect is that systems with high capillarity tend to show fluid distributions that are more homogeneous \citep{Mattax1961}, while larger clusters of nonwetting fluid can persist when capillary forces are less relevant (Figure~\ref{fig:pnm_trapping_map_low_aspect_ratio}), yielding more residual trapping.
\paragraph{Effect of the pore-body-to-throat aspect ratio on residual trapping:}
For strongly water-wet systems, higher minimum aspect ratios yield more trapping (Figure~\ref{fig:pnm_nubian_IR_CA_curve}). This is again caused by the higher entry pressures that narrower pore throats have. As has been pointed out throughout this discussion, the case of $R_\text{asp} = 1.14$ behaves decidedly differently than the other two cases with larger aspect ratios. It should be noted though that such a low aspect ratio is uncommon in natural rocks. As such, the trapping behavior and trends depicted by the cases with $R_\text{asp} = 2.27$ and $R_\text{asp} = 4.54$ are more representative for CO$_{\text2}$ residual trapping in sandstone.

\subsection{Implications for CO$_{\text 2}$ storage in the Gulf of Suez Basin}
We explore the implications of this research, as a first attempt, for future implementation of CCS in the Egyptian context. Equation~\eqref{eqn: pnm_storage_capacity} is used to estimate the geologic CO$_{\text 2}$ storage capacity. It is consistent with methods used in previous studies to assess the prospective geologic storage of buoyant fluids in subsurface formations \citep{vanderMeer1995TheAquifers, Doughty2001,Kopp2009InvestigationsCoefficients,  Goodman2011U.S.Scale, NETL2015},
\begin{equation}
\label{eqn: pnm_storage_capacity}
    % M_\mathrm{CO_2}^\mathrm{eff} = \overline{\phi} V_\mathrm{res}^\mathrm{bulk} \rho_\mathrm{CO_2}  \zeta S_\mathrm{CO_2}
    M_\mathrm{CO_2}^\mathrm{eff} = \zeta_{\mathrm{eff}}  V_\mathrm{res}^\mathrm{bulk} \rho_\mathrm{CO_2} (T,p),
\end{equation}
where, \textit{M}$_\mathrm{CO_2}^\mathrm{eff}$ is the effective storage capacity $\mathrm{\left[kg\right]}$,  $V_\mathrm{res}^\mathrm{bulk}$ is the bulk reservoir volume  $\mathrm{\left[m^3\right]}$, and $\rho_\mathrm{CO_2}$ is the CO$_{\text 2}$ density $\mathrm{\left[kg.m^{-3}\right]}$ as a function of the corresponding reservoir temperature and pressure. The dimensionless CO$_{\text 2}$-storage efficiency factor, $\zeta_{\text{eff}} \; \mathrm{\left[-\right]}$, represents the fraction of the total pore volume that can be occupied by the injected CO$_{\text 2}$. \cite{Doughty2001} proposed a method in which $\zeta_{\text{eff}}$ is estimated by a combination of the parameters provided in Equation~\eqref{eqn: pnm_storage_capacity_efficiency_factor}. 
\begin{equation}
\label{eqn: pnm_storage_capacity_efficiency_factor}
    \zeta_{\mathrm{eff}}  = \overline{\phi}_{\text{avg}} \textit{C}_{\text g} \textit{C}_{\text h} C_\text{i},\quad \text{where} \; %\forall \; 
    \zeta_{\mathrm{eff}} \in \left[0, \overline{\phi}_{\text{avg}}\right]\, .
\end{equation}
Here, $\overline{\phi}_{\text{avg}}$ is the average formation porosity $\mathrm{\left[-\right]}$, $\textit{C}_{\text g}$ and $\textit{C}_{\text h}$ are coefficients for the geometric capacity $\mathrm{\left[-\right]}$ and the heterogeneity capacity $\mathrm{\left[-\right]}$, respectively. The coefficient $\textit{C}_{\text i}$ is intrinsic capacity $\mathrm{\left[-\right]}$, which is controlled by multi-phase flow and transport phenomena; Equation~\eqref{eqn: pnm_intrinsic_capacity_storage_ci}.
\begin{equation}
\label{eqn: pnm_intrinsic_capacity_storage_ci}
C_\text{i} = C_\text{ig}+C_\text{il}\, ,
\end{equation}
where
$C_{\text {ig}}$ and $\textit{C}_{\text {il}}$ are the intrinsic capacity $\mathrm{\left[-\right]}$ for the CO$_{\text 2}$ and brine phase, respectively. $\textit{C}_{\text {ig}}$ can be estimated using Equation~\eqref{eqn: pnm_storage_capacity_efficiency_factor_Cig}. 

\begin{equation}
\label{eqn: pnm_storage_capacity_efficiency_factor_Cig}
    \textit{C}_{\text {ig}}  \cong \textit{S}_\mathrm{g} , 
\end{equation}
where, the \textit{S}$_{\mathrm{g}}$ is the average saturation of CO$_{\text 2}$ within the CO$_{\text 2}$-plume. We use the residual CO$_{\text 2}$ saturation value of 0.4 obtained from the current pore-network analysis. Meanwhile, the \textit{C}$_{\mathrm{il}}$ coefficient can be estimated using Equation~\eqref{eqn: pnm_storage_capacity_efficiency_factor_Cil}. 
\begin{equation}
\label{eqn: pnm_storage_capacity_efficiency_factor_Cil}
    \textit{C}_{\text {il}}  \cong \textit{S}_\mathrm{l}\chi_\mathrm{l}^\mathrm{g} \frac{\rho_\mathrm{l}}{\rho_\mathrm{g}}, 
\end{equation}
where, \textit{S}$_\mathrm{l} = 1-\textit{S}_{\mathrm{g} }$ is the average brine saturation $\mathrm{\left[-\right]}$, $\chi_\mathrm{l}^\mathrm{g}$ is the average CO$_{\text 2}$ mass fraction dissolved in the brine phase $\left[\text{kg}/\text{kg}\right]$, and $\rho_\mathrm{l}/\rho_\mathrm{g}$ stands for the density ratio between the CO$_{\text 2}$ and brine phases $\mathrm{\left[-\right]}$. We employ the commonly used thermodynamic models to describe the mole
% \todo[inline]{Do you mean mole fracton or mass fraction?}
fraction (solubility) of CO$_{\text 2}$ for a CO$_{\text 2}$-brine system provided by \cite{Duan2003AnBar} and \cite{Duan2006}. 
The densities, as a function of the relevant pressure and temperature, are numerically obtained using the \cite{Span1996AMPa} equation of state for CO$_{\text 2}$ and \cite{Driesner2007TheXNaCl} for brine (considering the brine salinity discussed above).

In the Gulf of Suez basin, Nubian Sandstone is subdivided into three main zones with high reservoir qualities and interfingering of two relatively thin-shale aquitards between the sandstone aquifer zones. Such layer-type
heterogeneities tend to enhance sequestration capacity by counteracting gravitational forces \citep{Doughty2001}. Therefore, we choose a heterogeneity capacity coefficient $\textit{C}_{\text h}$ of 0.5 \citep{vanderMeer1995TheAquifers}. A range of geometrical capacity coefficients $\textit{C}_{\text g}$ is taken from \cite{Kopp2009InvestigationsCoefficients}: 0.19--0.63.

\begin{figure}[hbtp]
\begin{center}

{\small\includegraphics[width=0.4\textwidth]{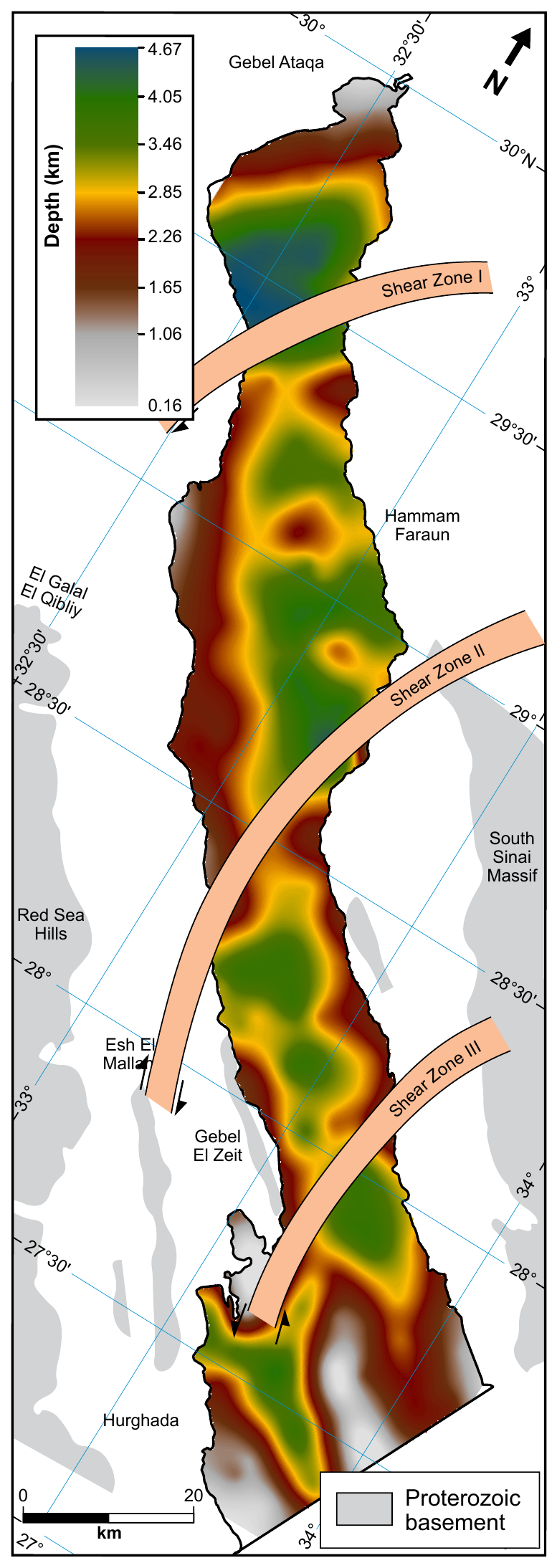}}
{\small\includegraphics[width=0.407\textwidth]{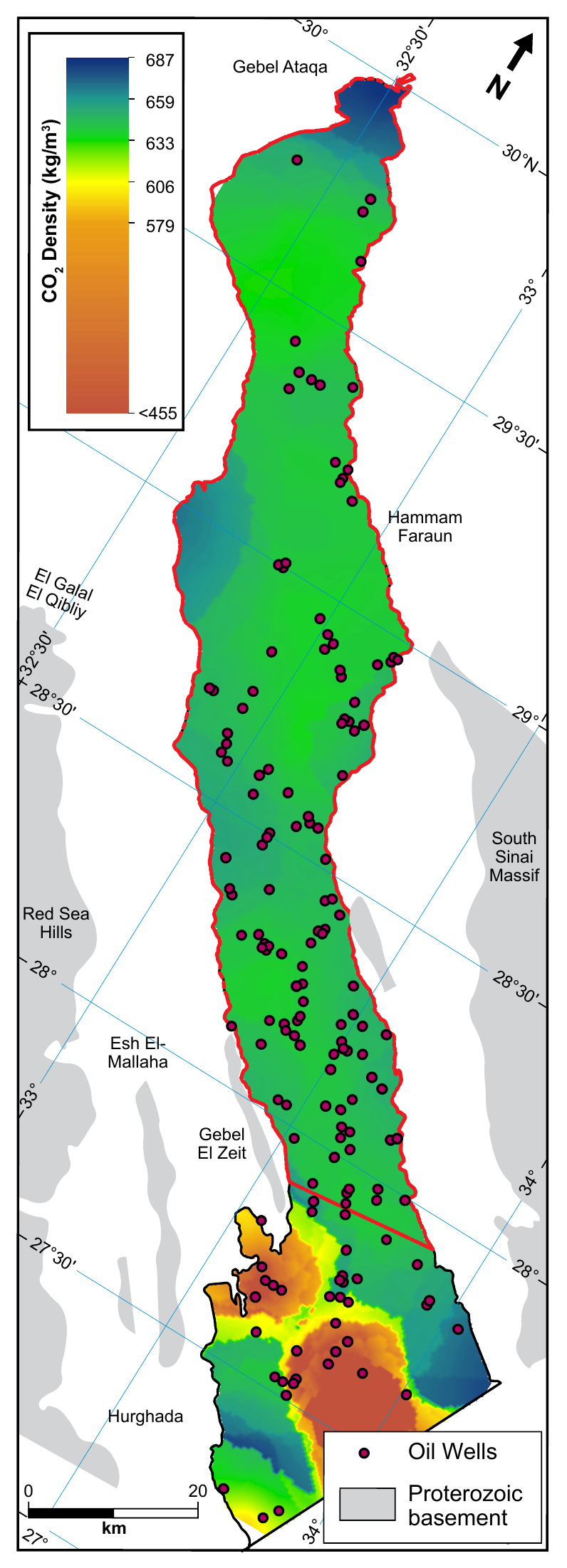}}
\llap{\parbox[b]{5.05in}{[\textbf{A}]\\\rule{0ex}{7.05in}}}
\llap{\parbox[b]{2.5in}{[\textbf{B}]\\\rule{0ex}{7.05in}}}
\caption{\small{[A] Depth to the top of the Nubian Sandstone sequence in the Gulf of Suez (Egypt). The map is constructed based on the interpretation of aeromagnetic and geological data by \cite{Mesheref1976} for basement rocks and modified after \cite{Farhoud2009}. [B] The distribution of in-situ CO$_{\text 2}$ densities across the Gulf of Suez. CO$_{\text 2}$ density is calculated at the equivalent temperature and pressure after \cite{Span1996AMPa}'s equation of state. The red outline  bounds the potential area (with 9\,950 km$^{\text{2}}$ footprint) in the Gulf of Suez for CO$_{\text 2}$ to be in the supercritical phase. }  }
\label{fig:pnm_nubian_sandstone_depth_gos}
\end{center}
\end{figure}

Stratigraphically, Nubian Sandstone overlies the basement complex throughout most of the Gulf of Suez basin with an average thickness of 0.45~km \citep{Alsharhan1997LithostratigraphyEgypt}. The depth map for the basement complex was interpreted from the aeromagnetic and the geological data by \cite{Mesheref1976, Farhoud2009}. Therefore, the depth to the top of the Nubian Sandstone sequence was constructed by adding a vertical offset of 0.45 km to the interpreted depth map of the basement complex. The resultant map is shown in Figure~\ref{fig:pnm_nubian_sandstone_depth_gos}A with a maximum depth of 4.67~km.

\input{pnm_capacity_estimation_gos}

Temperature and pressure generally increase with burial depth but have opposite effects on CO$_{\text 2}$ density. %CO$_{\text 2}$ density is numerically obtained using the \cite{Span1996AMPa}'s equation of state for CO$_{\text 2}$ as a function of the relevant pressure and temperature.
Following the geothermal gradient of 35.7 $^{\circ}$C/km and pressure gradient of 10.18 MPa/km after \cite{Hefny}, the range of CO$_{\text 2}$ density was found to be 640 -- 624 kg/m$^3$. 
\cite{Hefny2020} estimated $\chi_\mathrm{l}^\mathrm{g}\; \mathrm{\left[-\right]}$ and $\rho_\mathrm{l}/\rho_\mathrm{g}$ ratio $\mathrm{\left[-\right]}$ as 0.057 and 1.48, respectively, both are calculated for an average temperature (118.24~$^\circ$C) and pressure (34.7~MPa), and brine salinity (0.66 mol/kg; Table~\ref{tab:pnm_thermophysics}) found in the Gulf of Suez. All calculated coefficients and parameters used to estimate the CO$_{\text 2}$ storage capacity in Nubian Sandstone are provided in Table~\ref{tab:pnm_capacity_estimation_gos}. In conclusion, our $\zeta_{\mathrm{eff}}$ estimation for Nubian Sandstone was found to be in the range of 0.005 -- 0.017 $\mathrm{\left[-\right]}$. 

The \textit{M}$_\mathrm{CO_2}^\mathrm{eff}$ estimation covers a 9\,950 km$^{\text 2}$ surface area for the Gulf of Suez (marked by the red outline in Figure~\ref{fig:pnm_nubian_sandstone_depth_gos}B), where the subsurface conditions allow CO$_{\text 2}$ to be in the supercritical phase. We conclude that the Nubian Sandstone in the Gulf of Suez can store up to 14 -- 49 GtCO$_{\text 2}$, which alone can meet the Egyptian needs to store 250 MtCO$_{\text 2}$ emitted per annum \citep{Crippa2019} for the upcoming >227~years.

% =========================================================
\section{Conclusions}
\label{pnm_conclusions}
\begin{itemize}
    \item 
We investigate geologic CO$_{\text2}$ storage in a highly permeable formation, typical for Nubian Sandstone, Gulf of Suez (Egypt). This investigation consists of (1) constructing a realistic 3D pore network model that represents the characteristic features of Nubian Sandstone, (2) developing a quasi-static pore-network numerical simulator that mimics in-situ conditions that are similar to those prevailing at a CO$_{\text2}$-storage site, at the trailing edge of the CO$_{\text2}$ plume, and (3) predicting the two-phase flow characteristics.
    \item 
Two-phase constitutive relationships, including the capillary pressure and relative permeability curves, are computed for water-wet rocks at a low capillary number. We determine a Land trapping coefficient of $C=1.2$ for the Nubian sandstone sample. These relationships can be employed in field-scale numerical models to estimate the extent of CO$_{\text2}$-plume migration, at a representative geological scale.
    \item 
The estimated capillary-trapping curves for Nubian Sandstone are in good agreement with those observed experimentally for similar rocks. This confirms, that residual trapping due to hysteresis can be a key mechanism for long-term CO$_{\text2}$ storage in Nubian Sandstone. 
    \item 
One aspect, which has not been taken into account in this study, is contact-angle hysteresis, due to the lack of experiment measurements of the contact angle in Nubian Sandstone. We, therefore, consider uniform wettability in this work. Wettability hysteresis may be a topic of interest for future studies. 
    \item 
Our sensitivity study of the impact of wettability (i.e. advancing contact angle) on the magnitude of residual trapping shows that the trapped amount of CO$_{\text2}$ increases with decreasing contact angle for the cases, where the minimum pore-body-to-throat aspect ratio is~2.27 or~4.54. The inverse is the case when the minimum aspect ratio is only~1.14. 
    \item 
The pore-network model developed in this work improves our understanding of the different trapping mechanisms in Nubian Sandstone, as they pertain to CCS- and CPG-related applications. 
\item
The result of a conservative approach for a basin-scale estimation of CO$_{\text2}$ storage capacity shows an attractive potential to store 14 -- 49 GtCO$_{\text2}$ in Nubian Sandstone, Gulf of Suez basin. This assessment is based on a horizontal caprock and incorporates the arrangement of several injection wells accessing all fault-blocks with a compartment-like basin. The existence of many of these required wells, due to the locally active hydrocarbon industry, is another attractive factor, contributing to potentially making CCS and CPG techno-economically feasibile in the Gulf of Suez. 
\end{itemize}
% =========================================================
%\end{linenumbers}
\section*{Supplemental material}
\label{pnm_supplementary}
Supplemental material is provided to enable the reproducibility of our results. In addition, the SRXTM raw dataset of the Nubian Sandstone sample (in tiff format) has been made publicly accessible for the scientific community through the ETH~Zurich Research Collection repository [DOI: \href{https://doi.org/10.3929/ethz-b-000377881}{https://doi.org/10.3929/ethz-b-000377881}], which also contains descriptions of imaging conditions and characterizations of the rock samples through laboratory measurements \citep{Hefny2019DatasetDioxide}.

\section*{CRediT authorship contribution statement:}
\textbf{M. Hefny} conducted the experiments, analysed the data, prepared figures and wrote the main manuscript. \textbf{A. Ebigbo} supervised the whole work and wrote the main manuscript. \textbf{C.-Z. Qin} was responsible for coding pore-network model. \textbf{M.-O. Saar} secured the funding and the resources/infrastructure designed for the research work. All authors reviewed the manuscript.\par
% =========================================================
\section*{Acknowledgments}
\label{pnm_acknowledgments}

We gratefully acknowledge funding for this work by the Government of Egypt and the Werner Siemens Foundation (Werner Siemens-Stiftung) for their support of the Geothermal Energy and Geofluids (GEG.ethz.ch) Group at ETH~Zürich, Switzerland. We appreciate the contributions made by C.~Schlepütz and M.~Stampanoni from TOMCAT’s Paul Scherrer Institute in acquiring the SRXTM dataset used in this work. Thanks also go to the ClayLab ETH~Zürich, and in particular to M.~Plötz for his support with the Mercury Intrusion Porosimetry analysis. We are grateful to the Rock Physics and Mechanics Laboratory at ETH~Zürich for the permeability measurement support, employing experiments and numerical simulators. We further thank A.~Moscariello for allowing us to use the QEMSCAN analysis equipment at the University of Geneva. We conducted Pore-Network Extraction and IP Simulations, using the High-Performance Computing Cluster of the Geothermal Energy and Geofluid (GEG.ethz.ch) group at ETH~Zürich.
% =========================================================
% \input{Sections/pnm_ORCiDs.tex}
\section*{ORCID}
M. Hefny:  https://orcid.org/0000-0003-2568-6902 \\
C.-Z. Qin:  https://orcid.org/0000-0003-2793-3161 \\
M.O. Saar: https://orcid.org/0000-0002-4869-6452 \\
A. Ebigbo: https://orcid.org/0000-0003-3972-3786 \\

% =========================================================
\section*{Conflict of interest}
The authors declare that they have no conflict of financial interest or personal relationships that could have appeared to influ-ence the work reported in this paper.
% =========================================================
\markboth{Bibliography}{}
\bibliographystyle{apalike}
\bibliography{pnm_bib}
% =========================================================
\newpage
\appendix

\renewcommand\thefigure{\thesection.\arabic{figure}} 
\section{Appendices}
\label{pnm_Appendices}

% =========================================================
\subsection{Pore-network modeling}
\label{pnm_appendix_a}
% =========================================================
\renewcommand{\thefigure}{A.\arabic{figure}}
\setcounter{figure}{0}

\renewcommand{\thetable}{A.\arabic{table}}
\setcounter{table}{0}

\renewcommand{\theequation}{A.\arabic{equation}}
\setcounter{equation}{0}

% =========================================================

\subsubsection{Capillary pressure }
\begin{figure}[H]
\begin{center}
{\small\includegraphics[width=0.70\textwidth]{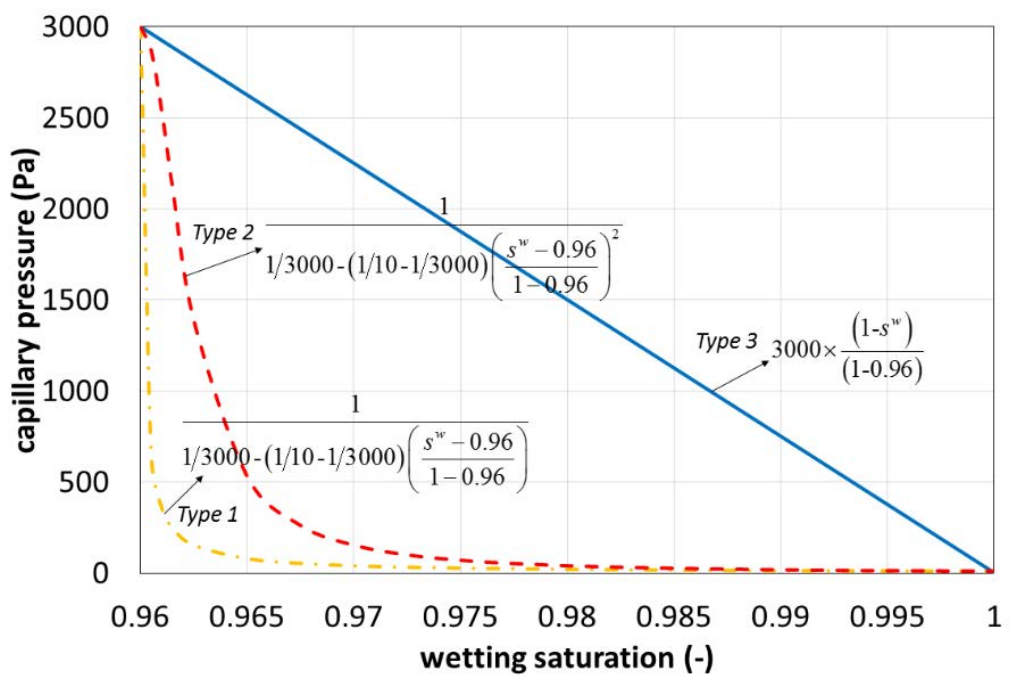}}
\caption{\small{Three types of transitional functions used to force the capillary pressure ($3000\; \left[ \mathrm{Pa} \right]$) to fast approach zero at the end of the MTM filling.} }
\label{fig:pnm_transitional_functions}
\end{center}
\end{figure}

There are two imbibition mechanisms for a pore body, namely, filling from arc menisci (AMs) and the main terminal meniscus (MTM). For the MTM filling, the capillary pressure (i.e., the entry pressure) will keep constant as the increase of wetting saturation until a pre-set value of saturation is reached. In this work, we set it to 0.96. Then, a transitional function of the wetting saturation is used to force the capillary pressure to fast approach zero. Figure \ref{fig:pnm_transitional_functions} shows three types of functions. The type 1 is primarily used in this work.	

The capillary entry pressure for a pore body or a pore throat is calculated by the following nonlinear equations based on the MS-P theory \citep{Patzek2001VerificationImbibition}:

\begin{equation}
    \label{eqn: pnm_radius}
    r = \frac{A_{\mathrm{eff}}}{L_\mathrm{{nw}}+L_\mathrm{{ns}} \cos \theta}
\end{equation}

\begin{equation}
    \label{eqn: pnm_effective_area}
    A_{\mathrm{eff}} = A-R^2 \sum_{i=1}^{n}\Bigg[ \frac{\cos \theta \cos(\theta+\beta_i)}{\sin \beta_i} + \theta + \beta_i- \frac{\pi}{2} \Bigg]
\end{equation}

\begin{equation}
    \label{eqn: pnm_ns_length}
    L_{ns} = \frac{R}{2G}-2\sum_{i=1}^{n} b_i
\end{equation}

\begin{equation}
    \label{eqn: pnm_nw_length}
    L_{nw} = 2r \sum_{i=1}^{n} a\sin \Bigg( \frac{b_i \sin \beta_i}{r} \Bigg)
\end{equation}

\begin{equation}
    \label{eqn: pnm_bi}
    b_i = r \frac{\mathrm{max}\big(0, \cos \big(\theta+\beta_i \big) \big)}{\sin \beta_i}
\end{equation}
where, $A_{\mathrm{eff}}$ is the effective cross-sectional area occupied by the nonwetting phase, $A$ is the cross-sectional area of the pore body, $L_{\mathrm{ns}}$ and $L_{\mathrm{nw}}$ denote the length of nonwetting-wetting interface and the length of the nonwetting-wall interface, respectively, and for the remaining notations, one can refer to Figure \ref{fig:pnm_triangle_cross_section_shape_factor}.   \\

\begin{figure}[H]
\begin{center}
{\small\includegraphics[width=0.49\textwidth]{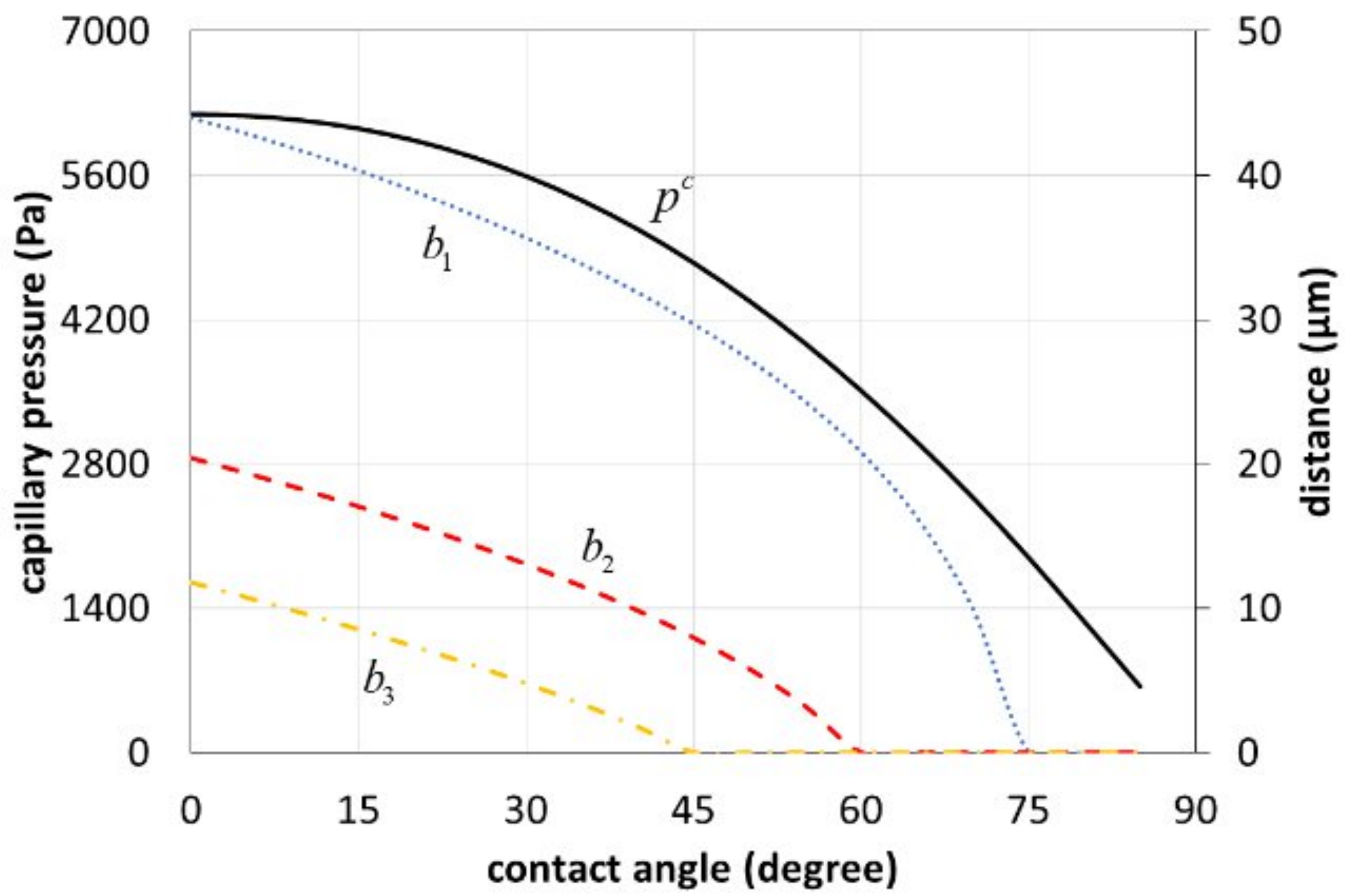}
\llap{\parbox[b]{0.60in}{[A]\\\rule{0ex}{1.8in}}}}\hfill
{\small\includegraphics[width=0.49\textwidth]{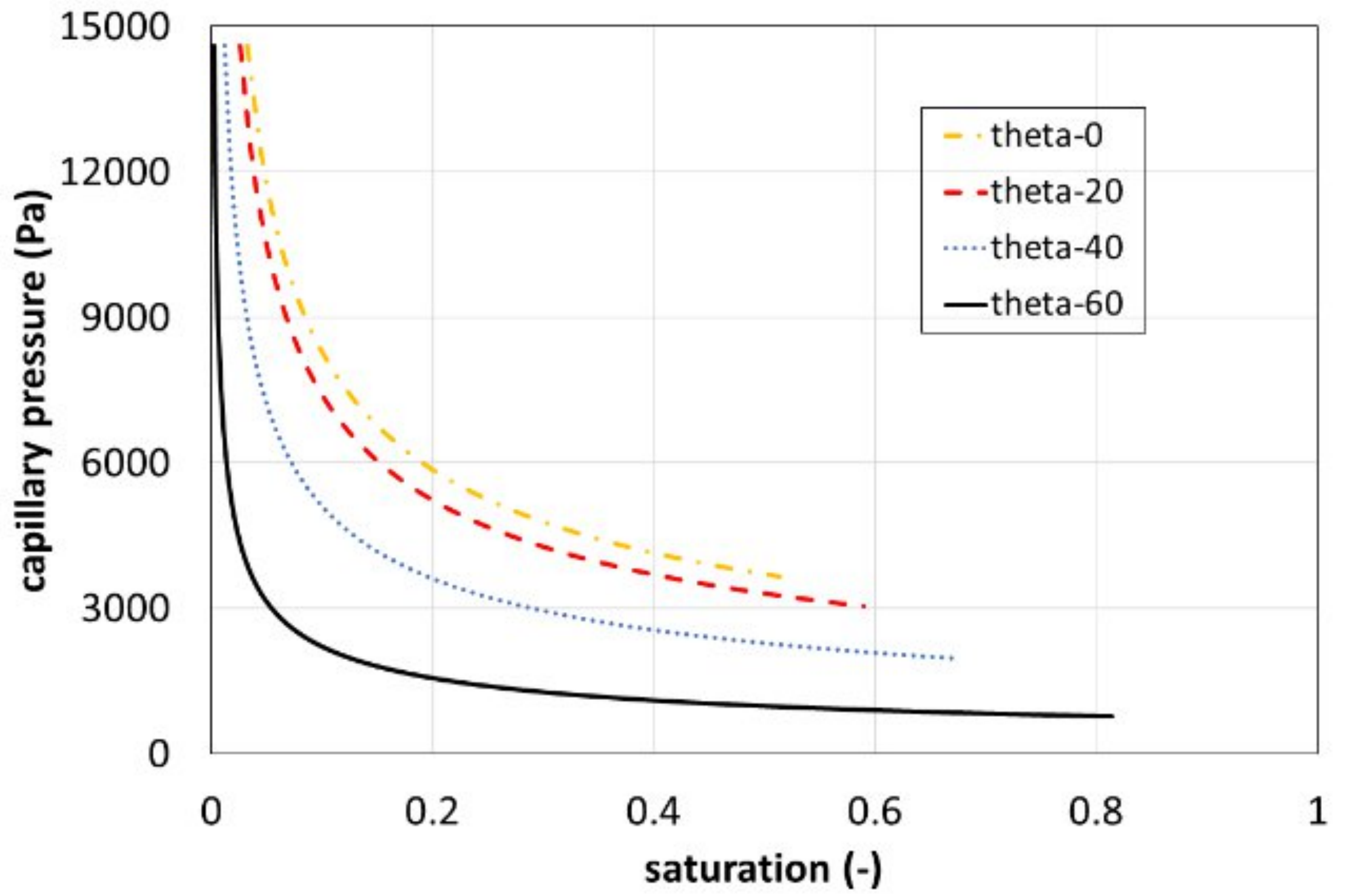}\llap{\parbox[b]{0.25in}{[B]\\\rule{0ex}{1.80in}}}}
\caption{\small{[A] imbibition entry capillary pressure versus the static contact angle, and the corresponding distances, $b_1$, $b_2$, and $b_3$; [B] the capillary pressure versus the wetting saturation at four different contact angle values for the AMs filling. The cross section is the right triangle, the inscribed circle radius is $20\; \mathrm{\left[\micro m\right]}$ , and the surface tension is $0.073\;  \mathrm{\left[N/m\right]}$.}} 
\label{fig:pnm_pc_contact_angle_saturation}
\end{center}
\end{figure}

For a pore body with the cross section of the right triangle described in Figure \ref{fig:pnm_triangle_cross_section_shape_factor}, assuming  $R= 20 \;\left[\mathrm{\micro m}\right]$  and the surface tension ($\sigma$) of 0.073 $\left[\mathrm{N/m}\right]$, the obtained entry capillary pressure versus the static contact angle is plotted in Figure \ref{fig:pnm_pc_contact_angle_saturation}A. Also, the corresponding distances $b_1$, $b_2$, and $b_3$ are plotted. The distances go to zero if corner flow cannot be formed.  \\

For the imbibition filling from AMs, taking the parameters used in Figure \ref{fig:pnm_pc_contact_angle_saturation}A, the capillary pressure versus saturation curves are plotted in Figure \ref{fig:pnm_pc_contact_angle_saturation}B for four contact angle values. Notice that each curve is truncated at its snap-off saturation. The snap-off event is triggered once more than one AM touch each other. For a triangle cross-section, we check the following three inequalities:

\begin{equation}
    \label{eqn: pnm_b1_b3}
    b_1 + b_3 < I_{A_0C_0}
\end{equation}

\begin{equation}
    \label{eqn: pnm_b1_b2}
    b_1 + b_2 < I_{A_0B_0}
\end{equation}

\begin{equation}
    \label{eqn: pnm_b2_b3}
    b_2 + b_3 < I_{B_0C_0}
\end{equation}
where  $I$ denotes the side length. If one or more are false, the snap-off is triggered. When the snap-off occurs, the nonwetting phase is trapped in the pore body. As mentioned in Section \ref{pnm_result_pore_network_calibration}, no volume is assigned to pore throats. The capillary pressure in a pore throat is approximated by the capillary pressure in its connected upstream pore body.

% =========================================================
\subsubsection{Phase conductance}
Relations of capillary pressure and phase conductivities for idealized pore elements are shown in Figure \ref{fig:pnm_triangle_cross_section_shape_factor}. Flow resistance in both pore bodies and pore throats is taken into account. If a pore is fully filled by either nonwetting or wetting phase, the single-phase conductance is given by \cite{Patzek2001ShapeFlow}:

\begin{equation} 
\label{eqn: pnm_phase_conductance}
    g = 
    \begin{cases}
\begin{tabular}{@{}cl@{}}
   $0.6GS^2$ & $\rightarrow$  triangle\\
    $0.5632GS^2 $&  $\rightarrow$  square\\
    $0.5GS^2$& $\rightarrow$ circle
\end{tabular}
\end{cases}
\end{equation}
where, $g$  is the phase conductance. \\

\begin{figure}[H]
\begin{center}
{\small\includegraphics[width=0.60\textwidth]{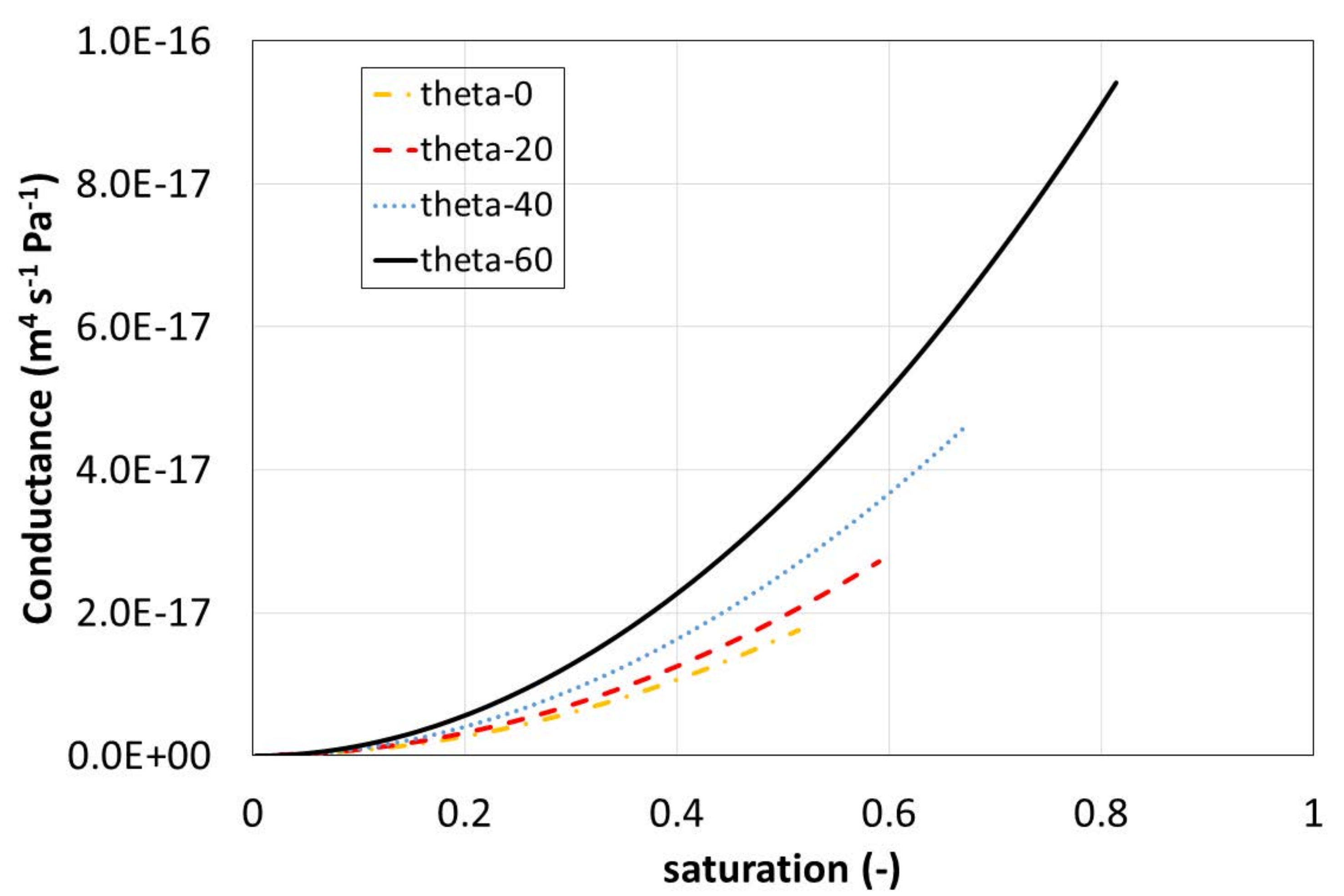}}
\caption{\small{Wetting-phase Conductance versus the wetting saturation for a two-phase occupied pore at four different contact angle values. The cross section is the right triangle, the inscribed circle radius is $20\; \mathrm{\left[\micro m\right]}$, and the surface tension is $0.073\;  \mathrm{\left[N/m\right]}$.} }
\label{fig:pnm_conductance_saturation}
\end{center}
\end{figure}

If a pore is occupied by both nonwetting and wetting phases, the conductance for the wetting phase is calculated as follows. First, for an AM (see Figure \ref{fig:pnm_triangle_cross_section_shape_factor}, with $b=1$, the dimensionless corner wetting area is given as proposed by \cite{Patzek2001ShapeFlow}:

\begin{equation}
    \label{eqn: pnm_corner_wetting_area}
    \tilde{A}_\mathrm{w} = \Bigg( \frac{\sin \beta}{cos(\theta+\beta)}\Bigg)^2 \Bigg( \frac{\cos \theta \cos(\theta+\beta)}{\sin \beta} + \theta + \beta -\frac{\pi}{2} \Bigg)
\end{equation}
Further, we define the shape factor for the corner wetting as:
\begin{equation}
    \label{eqn: pnm_shape_factor_corner_wetting}
    \tilde{G}_\mathrm{w}=\frac{\tilde{A}_\mathrm{w} }{4\big[1- (\theta + \beta - \pi \big/2) \sin \beta \big/ \cos \big( \theta + \beta \big) \big]}
\end{equation}
Then, the semi-empirical dimensionless conductance with the perfect slip boundary condition between wetting and nonwetting phases is calculated by \cite{Patzek2001ShapeFlow}:

\begin{equation}
    \label{eqn: pnm_dimensionless_conductance}
    \tilde{\textit{g}}_{\text{w}}=\text{exp}\left[\left[ \text{ -18.2066} \tilde{\textit{G}}_{\text{w}}^{\text{2}} + \text{5.88287} \tilde{\textit{G}}_{\text{w}} - \text{0.351809}+ \text{0.02}\text{sin}\Big( \beta - \pi \big/\text{6}\Big)  \right] \bigg/ \Big(\text{1}\big/\text{4}\big/\pi-\tilde{\textit{G}}_{\text{w}}  \Big) + \text{2}\ln\textit{S}_{\text{w}}\right]
\end{equation}
Finally, the dimensional one is obtained as $\tilde{g}_w=\big( b^4\big/\mu\big)\tilde{g}_w$. Taking the parameters used in Figure \ref{fig:pnm_pc_contact_angle_saturation}. Figure \ref{fig:pnm_conductance_saturation} shows the curves of conductance versus saturation for four contact angle values. Obviously, for the same saturation value, a larger contact angle gives larger wetting-phase conductance along the corners. For the nonwetting-phase conductance, it is approximated by \cite{Qin2015WaterModeling}:

\begin{equation}
    \label{eqn: pnm_nonwetting_phase_conductance}
        \tilde{g}_\mathrm{{nw}} = 
    \begin{cases}
\begin{tabular}{@{}cl@{}}
   $0.6GS \big(1-A_\mathrm{w} \big)$ & $\rightarrow$  triangle\\
    $0.5632GS \big(1-A_\mathrm{w} \big)$&  $\rightarrow$  square
\end{tabular}
\end{cases}
\end{equation}

\subsubsection{Relative Permeability hysteresis}
\begin{figure}[hbtp]
\begin{center}
{\small\includegraphics[width=0.85\textwidth]{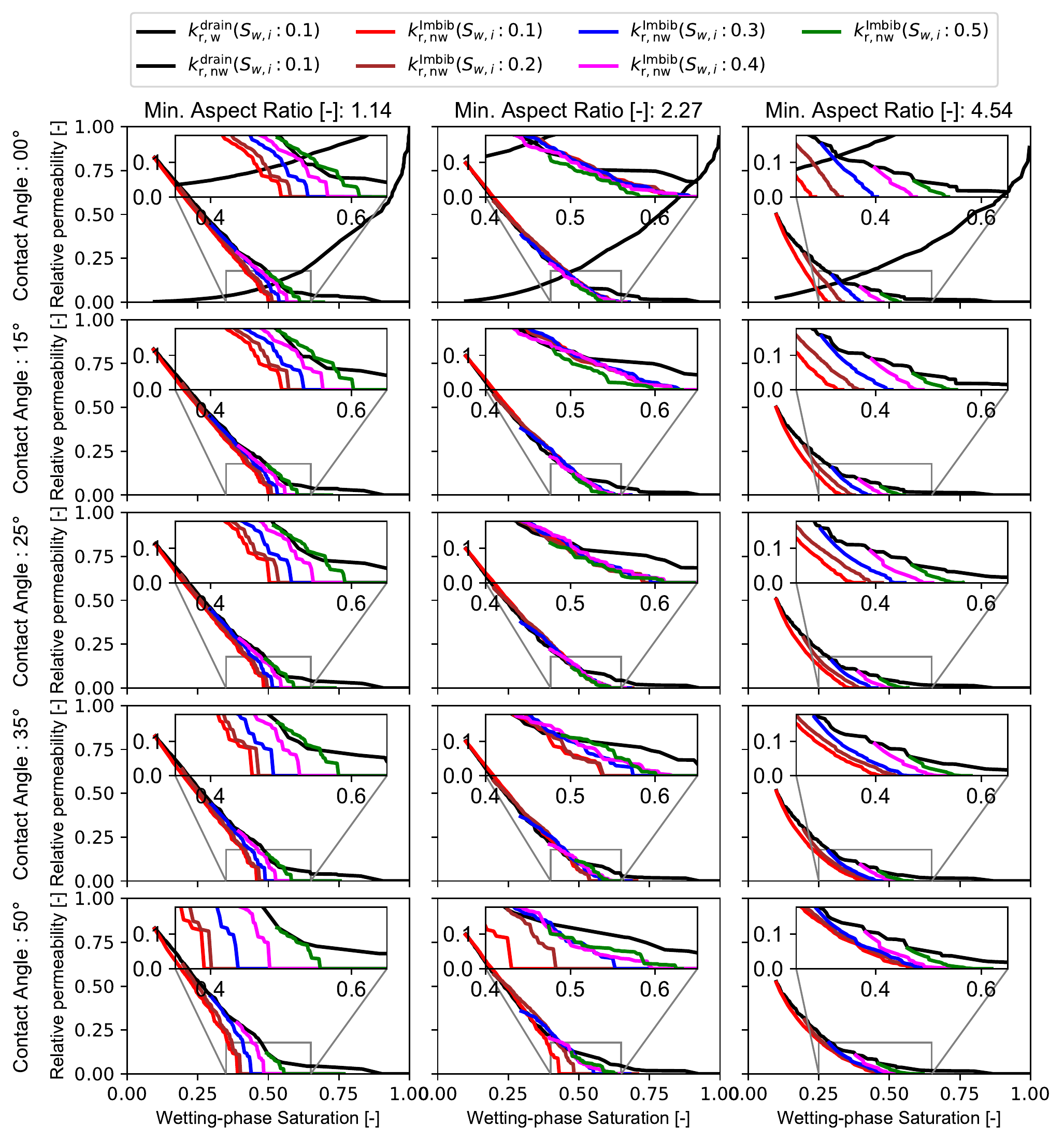}}%\llap{\parbox[b]{3.6in}{[B]\\\rule{0ex}{2.28in}}}}
\caption{\small{Relative permeability scanning curves during the main imbibition process as a function of wettability [rows], minimum pore network aspect ratio [columns], and initial wetting-phase (brine) saturation [curves].} }
\label{fig:pnm_nubian_kr_ca_ar_ir_curves}
\end{center}
\end{figure}

The hysteresis patterns appear in relative permeability curves are shown in Figure~\ref{fig:pnm_nubian_kr_ca_ar_ir_curves}. These simulated $k_{\text r}-S_{\text w}$ curves are presented as a function of initial wetting-phase saturation, wettability and aspect ratios.

% =========================================================
\subsection{Direct Numerical Simulation (DNS); Absolute Permeability}
\label{pnm_appendix_b}

\renewcommand{\thefigure}{B.\arabic{figure}}
\setcounter{figure}{0}

\renewcommand{\thetable}{B.\arabic{table}}
\setcounter{table}{0}

\renewcommand{\theequation}{B.\arabic{equation}}
\setcounter{equation}{0}
% =========================================================

At the pore-scale, the permeability and flow field has been determined using Lattice Boltzmann Method (LBM) to solve the conservation equations in the complex geometry of Nubian Sandstone. A representative geometry of the Nubian Sandstone was reconstructed from a high-resolution Synchrotron Radiation X-ray Tomographic Microscopy (SRXTM) binarized volume of $\mathrm{1.4\!\times\!1.4\!\times\!1.3}$ $\mathrm{mm^3}$. For the stability of LBM simulation, we performed a voxel-averaging in order to down-scaling the resolution SRXTM's images to $270\!\times\!270\!\times\!270 \; \mathrm{pixel}$ while keeping the 3D internal geometrical structures of unchanged. We used ImageJ open-source image analysis tool to lowering the voxel size to the value of $\mathrm{5.13}\; \mathrm{\micro m^3}$. See the supplementary section for accessing the raw SRXTM grey images (original pixel size). 

\begin{figure}[H]
\begin{center}
{\small\includegraphics[width=0.37\textwidth]{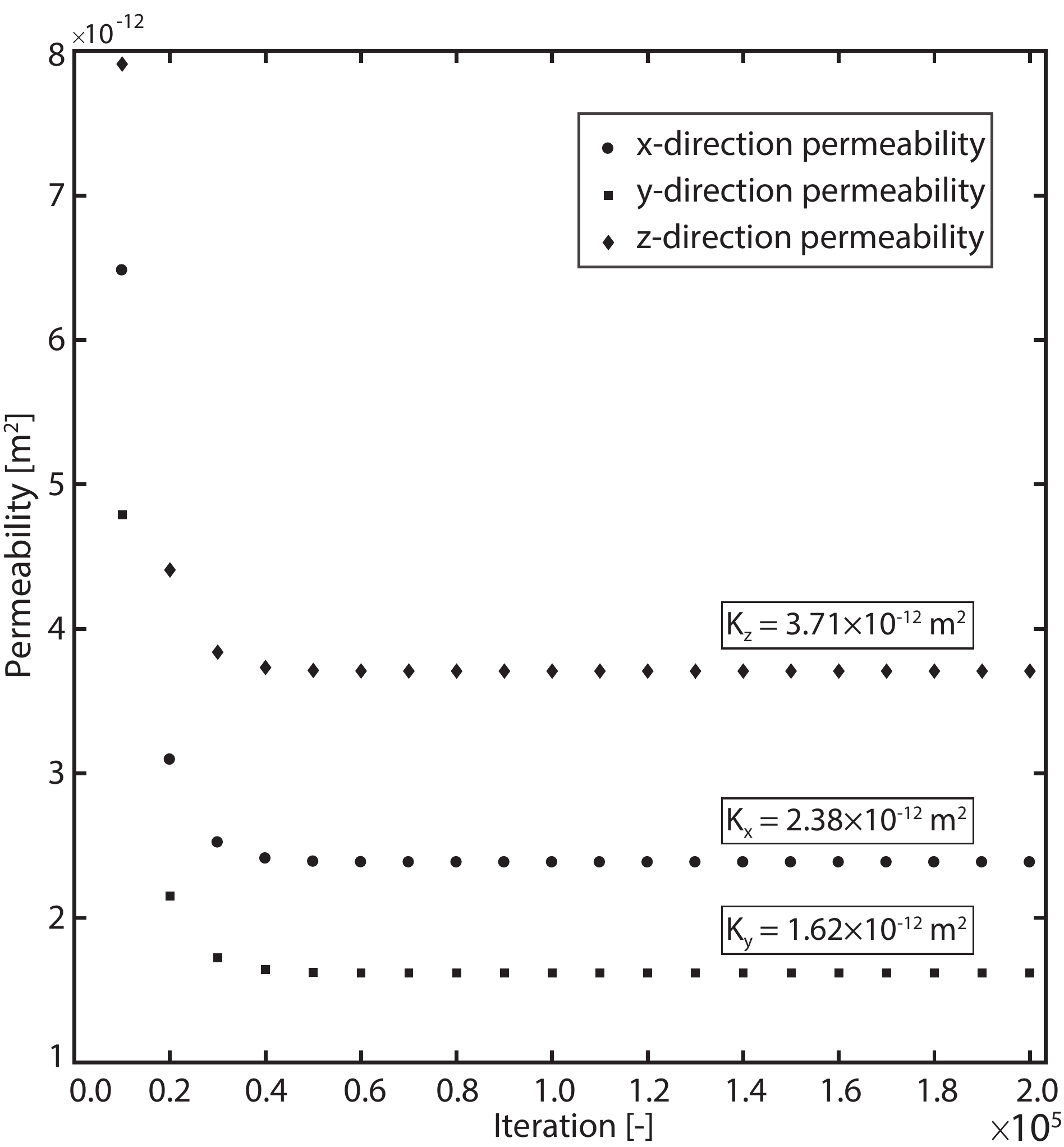}
\llap{\parbox[b]{2.5in}{[A]\\\rule{0ex}{2.23in}}}}
{\small\includegraphics[width=0.47\textwidth]{StreamTracer_NubianSandstone270_MH.pdf}
\llap{\parbox[b]{0.4in}{[B]\\\rule{0ex}{2.23in}}}}

\caption{\small{[A] The estimated permeability once flow was deemed to have reached steady-state conditions. [B] A 3-D visualization of the simulated stream tracer distribution inside the 3D rendering of SRXTM images (domain size: $\mathrm{1.4\!\times\!1.4\!\times\!1.3 \;mm^3}$, with voxel size $\mathrm{5.13\; \micro m^3}$). }} \label{fig:pnm_dns_permeability_nubian}
\end{center}
\end{figure}

Fluid flow inside the reconstructed Nubian Sandstone is performed simulating a single phase flows using an in-house LBM numerical permeameter \cite[LBHydra;][]{Walsh2010InterpolatedKinetics,Walsh2010MacroscaleMedia, Walsh2013DevelopingUnits}, to validate the estimated permeability of pore-network modeling. The D3Q19 model is implemented in the solution of 3D incompressible fluid flows in the LBM. At the solid boundary of the pore space, no-slip and no-flow BCs can typically be utilized by “bounce back” implementation. During the LBM simulations, once flow was deemed to have reached steady state (Figure  \ref{fig:pnm_dns_permeability_nubian}A), the permeability of the medium was calculated based on the Darcy’s law and after the conversion from lattice units into physical units be achieved.  Figure \ref{fig:pnm_dns_permeability_nubian}B displays the 3D distribution of streamlines of velocity in z-direction of Nubian Sandstone. The pressure difference between the inlet and outlet, which is expressed by the density difference in the simulation, is 0.0005 in lattice units.

As shown in Figure~\ref{fig:pnm_dns_permeability_nubian}, the streamlines represents the continuous pathways from the inlet to the outlet are quite straight, resulting in quite high permeability of Sandstone. The permeability is determined in the three flow directions at the same scanned data of Nubian Sandstone sample to investigate its permeability's anisotropy (Figure~\ref{fig:pnm_dns_permeability_nubian}B and Table~\ref{tab:pnm_appednix_LBHydra_nubianl}). Based on the coupled LBM method, the predicted permeability results are $\mathrm{2.38\times10^{-12}}$, $\mathrm{1.62\times10^{-12}}$, and $\mathrm{3.71\times10^{-12}}$ that corresponding, respectively, to permeability $\left[\mathrm{m^2}\right]$ in x-, y-, and z-directions. The mean value of the permeability is $\mathrm{2.57\times10^{-12}}\; \mathrm{m^2}$. The standard deviation is $\mathrm{1.06\times10^{-12}}\; \mathrm{m^2}$. Its anisotropy, the ratio of the horizontal to the vertical permeability, to be $\mathrm{<1}$ confirming the homogeneity of Nubian sandstone.

\input{pnm_lbm_k_nubian2.tex}

% =========================================================
\includepdf[pages=-]{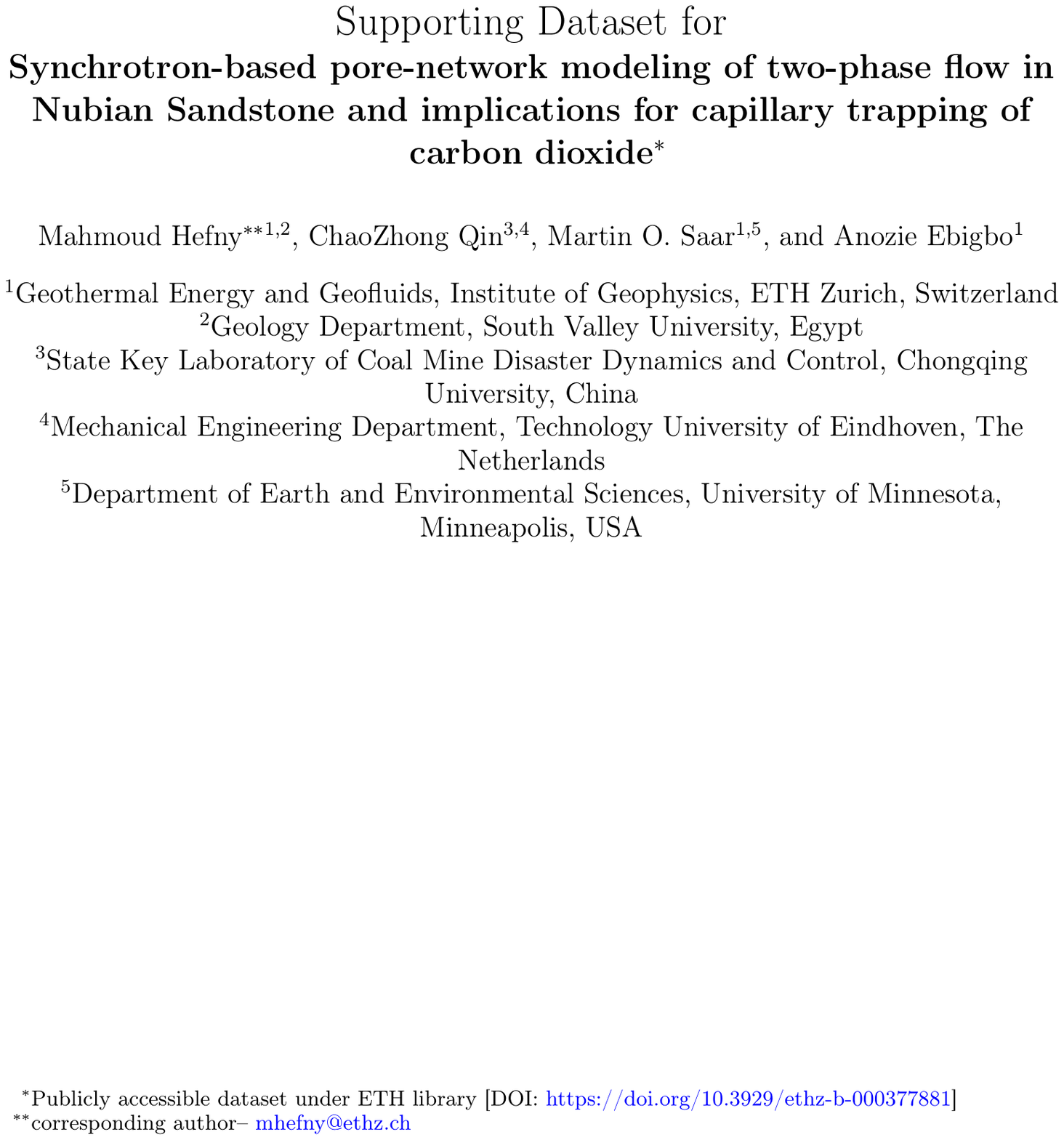}
\end{document}

%% file: pnm_statisitcs.tex
\begin{table}[htbp]
    \small
    \centering
    \caption{Parameter statistics of the pore-network structure for the Nubian Sandstone sample, extracted by the stochastic pore-network extractor, and used in quasi-static PN modeling. Network size (domain) $\mathrm{\left[mm^3\right]}$ is $\mathrm{1.41\!\times\!1.41\!\times\!2.82}$.}
  
    \begin{threeparttable}

        \begin{tabular}{p{4.5cm}cc}
            \Xhline{3\arrayrulewidth}
            \textbf{Pore-network parameters} & \textbf{Pore bodies} & \textbf{Pore throats} \\
            \hline
            Total number $\mathrm{\left[-\right]}$ & $19 692$ & $38 933$ \\
            
            \multirow{2}{*}{Extended diameter; $\mathrm{D_{ext} \; \left[\mu m\right]}$ \tnote{a}} & $\mathrm{8.1 < D_{ext}^{pb} < 192.28;}$  & $\mathrm{2.59\; < D_{ext}^{pt} <\; 129.47;}$  \\
            &  $\mathrm{\overline{D}_{ext}^{pb}}$: 49.73  &   $\mathrm{\overline{D}_{ext}^{pt}}$: 26.83 \\
            
            \multirow{2}{*}{Inscribed diameter; $\mathrm{D_{ins} \; \left[\mu m\right]}$ \tnote{b}} & $\mathrm{5.18 < D_{ins}^{pb} < 115.6;}$  & $\mathrm{4.38 < D_{ins}^{pt} < 91.03;}$ \\
            &  $\mathrm{\overline{D}_{ins}^{pb}}$: 28.71 &   $\mathrm{\overline{D}_{ins}^{pt}}$: 20.83  \\
            
            Cross-sectional fraction &  \multirow{2}{*}{$\mathrm{[0.915,\;  0.070, \; 0.015]}$} &  \multirow{ 2}{*}{$\mathrm{[0.988,\; 0.016,\;  0.001]}$}\\
            
            [Triangle, Square, Circle]  \tnote{c}  & \\

            \Xhline{3\arrayrulewidth}
        \end{tabular}
        
        \begin{tablenotes}
                %\small{
                \item[a] Extended diameter of spherical pore element is defined as the maximum value of the global distance map lying within each pore region.
                \item[b] Inscribed diameter of spherical pore element is obtained in the same manner as the extended diameter, but a local distance map of just the pore region is used. 
                \item[c] Cross-sectional fraction was calculated after determination of the  dimensionless shape factor ($G$) \citep[derived from ][]{Mason1991}; see Section \ref{pnm_shape_factor} for more details. The shape-factor concept is used to describe the geometry of the pore-network elements and to relate it to the displacement processes being simulated.
                
        \end{tablenotes}
        
    \end{threeparttable}
\label{tab:statistics_pnm_nubian}  
\end{table}

%% file: pnm_thermophysics.tex
\begin{table}[htbp]
    \small
    \centering
    \caption{The thermophysical properties of the fluids (brine and \scco) used for simulating drainage--imbibition invasion processes in the Nubian Sandstone samples. The $\mathrm{P\!-\!T}$ conditions are representative of conditions at a depth of \SI{4}{\km} after \cite{Hefny} (pressure: \SI{40e6}{\pascal}, and temperature: \SI{142}{\degreeCelsius}). }
    
    \begin{threeparttable}
        \begin{tabular}{lccc}
            \toprule
            \textbf{Physical Parameter} & \textbf{Value} & \textbf{Unit} & \textbf{Reference} \\
            \midrule
            Interfacial tension between \scco $\;$and brine, $\mathrm{\sigma}$  & $\mathrm{0.0202}$    & $\left[\si{\newton\per\meter}\right]$ & \cite{Bachu2009InterfacialMgL-1} \\
    
            Intrinsic contact angle for brine-\scco-quartz, $\mathrm{\theta}$ \tnote{a} & $\mathrm{35}$  & $\left[\si{\degree}\right]$   & \cite{Jung2012SupercriticalMeasurements} \\
    
            Dynamic viscosity of wetting fluid, $\mathrm{\eta_{brine}}$ & $\mathrm{6.55\times\!10^{-04}}$ & $\left[\si{\pascal\second}\right] $  & \cite{Sharqawy2010ThermophysicalData} \\
    
            Dynamic viscosity of non-wetting fluid, $\mathrm{\eta_{scCO_2}}$ \tnote{b}& $\mathrm{5.21\times\!10^{-05}}$ & $\left[\si{\pascal\second}\right] $ & \cite{Ouyang2011NewDioxide} \\
    
            Densities of brine, \scco, $\mathrm{\Delta\rho}$ \tnote{c} & $\mathrm{960,\; 627,\; 442}$  & $ \left[\si{\kilogram\per\metre\cubed}\right] $& \cite{Span1996AMPa}  \\
    
            Brine salinity  & $\mathrm{4\times\!10^{4}}$ & $\left[\si{\mg\per\liter} \right]$ & \cite{Gavish1974GeochemistrySuez}  \\
            \bottomrule
        \end{tabular}
        
        \begin{tablenotes}
                %\small{
                \item[a] Value was measured through the aqueous phase.
                
                \item[b] Value was calculated as a function of salinity, temperature and pressure (Equation 20; \cite{Sharqawy2010ThermophysicalData}). 
                
                \item[c] Density was obtained after \cite{Span1996AMPa}'s equation of state.
                
        \end{tablenotes}
        
    \end{threeparttable}
\label{tab:pnm_thermophysics}
\end{table}

%% file: pnm_result_sng_phase2.tex
\begin{table}[htbp]
\small
\centering
\caption{The average values of hydraulic and capillarity properties for Nubian Sandstone, determined by employing laboratory experiments and pore-network simulations. Subscripts `d' and `i' refer, respectively, to the drainage and imbibition process.}
\begin{threeparttable}
    \begin{tabular}{llcc}
%    
    %\toprule
    \specialrule{.15em}{.05em}{.05em}
    \multicolumn{2}{l}{\textbf{Parameter}} & \textbf{Nubian Sandstone} & \textbf{Unit} \\
    %\midrule
    \specialrule{.1em}{.1em}{0em}
    \multicolumn{4}{l}{\textbf{Hydraulic Properties:}} \\
    %\midrule
    \specialrule{.1em}{.1em}{0em}
    \multirow{2}[1]{*}{Porosity; $ \mathrm{\phi_{eff}}$} & Laboratory \tnote{(1)} &   $\mathrm{0.269}$    & \multirow{2}[1]{*}{$\left[\mathrm{-}\right]$} \\
          & SRXTM \tnote{(2)} &  $\mathrm{0.254}$     &  \\
    \midrule
    \multirow{3}[1]{2.6cm}{Permeability; $k_\mathrm{eff}$} & Laboratory \tnote{(3)} &  $\mathrm{1.70\times\!10^{-12}}$     & \multirow{3}[1]{*}{$\left[\mathrm{m^2}\right]$} \\
          & PNM \tnote{(4)}  &  $\mathrm{2.00\times\!10^{-12}}$     &  \\
          & LBM \tnote{(5)}  &  $\mathrm{2.57\times\!10^{-12}}$     &  \\
    %\midrule
    \specialrule{.1em}{.1em}{0em}
    \multicolumn{4}{l}{\textbf{Capillarity Properties:}} \\
    %\midrule
    \specialrule{.1em}{.1em}{0em}
    \multirow{3}[1]{*}{Entry Pressure; $P_\mathrm{e}$} & MIP ($\mathrm{Hg}\!-\!\mathrm{air}$) &   $\mathrm{25347.6}$     & {$\left[\mathrm{Pa}\right]$} \\
          
          & MIP (Leverett-$J_\mathrm{d}$) &  $3.91\times\!10^{-4}$   & \multirow{2}[1]{*}{$\left[\mathrm{-}\right]$} \\
          & PNM (Leverett-$J_\mathrm{i}$) &  $3.59\times\!10^{-4}$     &  \\
   \midrule
   
 \multirow{2}[1]{*}{Irreducible Saturation; $ S_\mathrm{w,r}$\tnote{(6)}} & MIP (Leverett-$J_\mathrm{d}$) &   $\mathrm{0.089}$    & \multirow{2}[1]{*}{$\left[\mathrm{-}\right]$} \\
          & PNM &  $\mathrm{0.099}$     &  \\
    \midrule
   
    \multicolumn{1}{l}{\multirow{4}[1]{2.6cm}{van Genuchten parameters }} & $m_\mathrm{d}$    &  $0.466$     & $\left[\mathrm{-}\right]$ \\
          & $\alpha_\mathrm{d}$ &    $\mathrm{3.72\times\!10^{-3}}$   & $\left[\mathrm{MPa^{-1}}\right]$  \\
          & $m_\mathrm{i}$  &     $0.580$  & $\left[\mathrm{-}\right]$  \\
          & $\alpha_\mathrm{i}$ &  $\mathrm{1.85\times\!10^{-3}}$     & $\left[\mathrm{MPa^{-1}}\right]$ \\
    %\bottomrule
    \specialrule{.15em}{.05em}{.05em}
    \end{tabular}
    \begin{tablenotes}
        \item[(1)] Measured via He pycnometry on a cylindrical plug of the same rock (25.4 $\mathrm{mm}$ diameter cylindrical sample). This value is based on averaging three plugs. 
        \item[(2)] Calculated from a cylindrical subplug $\left(\mathrm{1.4\!\times\!1.4\!\times\!2.8\; mm^3}\right)$, using binarized Synchrotron-based X-ray computed tomography images.
        \item[(3)] The permeability was determined, based on \citep[][Equation (08)]{Kuijpers2017SorptionImaging} and using the MIP-processed data, including porosity, average pore-diameter, and the pore space tortuosity  factor, $\tau= 1.969$.
        \item[(4)] The permeability was estimated, based on steady-state, single-phase flow across a constant pressure difference between the inlet and the outlet of the pore network.
        \item[(5)] The permeability represents the average of the permeabilities calculated in the three main (x, y, z) directions, employing direct numerical simulations, using our in-house Lattice Boltzmann simulator, LBHydra  \cite{ Walsh2010InterpolatedKinetics,Walsh2010MacroscaleMedia, Walsh2013DevelopingUnits} as a numerical permeameter (see Appendix~\ref{pnm_appendix_b} for more details). The standard deviation is $\mathrm{1.06\times\!10^{-12}\; m^2}$.
        \item[(6)] The irreducible saturation, $S_\mathrm{{w,r}} [-]$, is required by the van Genuchten fitting model.
    \end{tablenotes} 
\end{threeparttable}    
\label{tab:pnm_result_sng_phase}  
\end{table}

%% file: pnm_capacity_estimation_gos.tex
\begin{table}[htbp]
    \small
    \centering
    \caption{The physical parameters used to calculate the storage capacity of Nubian Sandstone in Gulf of Suez basin (Egypt). Intrinsic capacity is defined as $\textit{C}_{\text {i}} = \textit{C}_{\text {ig}}+\textit{C}_{\text {il}}$, where the coefficient \textit{C}$_{\text {ig}}$ equals to the residual CO$_{\text{2}}$ saturation \textit{S}$_{\text{g}}$. }
    
    \begin{threeparttable}
        \begin{tabular}{lccp{1.5cm}}
            \toprule
            \textbf{Parameter} & \textbf{Value} & \textbf{Model} \\
            \midrule
            \textit{S}$_{\text{g}}$ $\mathrm{\left[-\right]}$ & 0.4    &  current study \\
            \textit{S}$_{\text{l}}$ $\mathrm{\left[-\right]}$ &  $1-S_{\text{g}}$  &  -- \\
            $\chi_\mathrm{l}^\mathrm{g}$ $\mathrm{\left[kg/kg\right]}$&  0.057 \tnote{a}  & \cite{Duan2006}  \\
            $\rho_\mathrm{g}$ $\mathrm{\left[kg/m^3\right]}$ &  647.7 \tnote{a}  &  \cite{Span1996AMPa} \\
            $\rho_\mathrm{l}$ $\mathrm{\left[kg/m^3\right]}$&   959.47 \tnote{a} &   \cite{Driesner2007TheXNaCl} \\

            \textit{C}$_{\text {il}}$ $\mathrm{\left[-\right]}$ &  0.05  & --  \\
            $\overline{\phi}_{\text{avg}}$ $\mathrm{\left[-\right]}$  & 0.12 %$\pm0.04$
            &  \cite{Hefny} \\
            \midrule
            \textit{C}$_{\text {i}}$ $\mathrm{\left[-\right]}$ &  0.45  &  -- \\
            \textit{C}$_{\text {g}}$ $\mathrm{\left[-\right]}$ & 0.19--0.63   & \cite{Kopp2009InvestigationsCoefficients}  \\
            \textit{C}$_{\text {h}}$ $\mathrm{\left[-\right]}$ &  0.5  & \cite{vanderMeer1995TheAquifers}  \\
             \midrule
            $\zeta_{\mathrm{eff}}$ $\mathrm{\left[\%\right]}$ & 0.5 -- 1.7 & --\\
            Surface area $\mathrm{\left[km^2\right]}$ &  9\,950 \tnote{b}  &  -- \\
            \textit{V}$_{\text{res}}^{\text{bulk}}$ $\mathrm{\left[km^3\right]}$ &  4\,477.5 \tnote{c}  &  -- \\

            \bottomrule
        \end{tabular}
        
        \begin{tablenotes}
                %\small{
                \item[a] These values are calculated as a function of brine salinity, and reservoir temperature and pressure \citep{Hefny2020}.
                
                \item[b] The area estimation (marked with the red outline in Figure~\ref{fig:pnm_nubian_sandstone_depth_gos}B) bounds the potential area in the Gulf of Suez where the depth to Nubian sandstone will be > 1.5 km and the CO$_{\text 2}$ will be in the supercritical phase. The depth map is constructed based on the interpretation of aeromagnetic and geological data by \cite{Mesheref1976} for the basement rocks and modified after \cite{Farhoud2009}.
                \item[c] Volume calculation is based on an average thickness of 0.45 km after \cite{Alsharhan1997LithostratigraphyEgypt}.
                
        \end{tablenotes}
        
    \end{threeparttable}
\label{tab:pnm_capacity_estimation_gos}
\end{table}

%% file: pnm_lbm_k_nubian2.tex
\begin{table}[htbp]
\small
  \centering
  \caption{Permeability calculations from LBM simulation in different directions.}
    \begin{tabular}{ccccccc}
    \toprule
    \textbf{Iteration } & \multicolumn{2}{c}{\textbf{x-direction}} & \multicolumn{2}{c}{\textbf{y-direction}} & \multicolumn{2}{c}{\textbf{z-direction}} \\
    \cmidrule{2-7}    $\mathrm{[\times 10^4]}$  & $k_\mathrm{{xx}} \; \mathrm{[L]}$ & $k_\mathrm{xx} \; \mathrm{[\times10^{-12} \; m^2]}$ & $k_\mathrm{{yy}} \; \mathrm{[L]}$ & $k_\mathrm{yy} \; \mathrm{[\times10^{-12} \; m^2]}$& $k_\mathrm{{zz}} \; \mathrm{[L]}$ & $k_\mathrm{zz} \; \mathrm{[\times10^{-12} \; m^2]}$\\
    \midrule
    1     & 0.2412 & 6.48  & 0.1783 & 4.79  & 0.2943 & 7.91 \\
    2     & 0.1152 & 3.1   & 0.0801 & 2.15  & 0.164  & 4.41 \\
    3     & 0.0938 & 2.52  & 0.0642 & 1.73  & 0.143  & 3.84 \\
    4     & 0.0897 & 2.41  & 0.0611 & 1.64  & 0.139  & 3.73 \\
    5     & 0.0889 & 2.39  & 0.0604 & 1.62  & 0.139  & 3.71 \\
    6     & 0.0887 & 2.38  & 0.0603 & 1.62  & 0.138  & 3.71 \\
    \vdots& \vdots & \vdots  & \vdots & \vdots  & \vdots  & \vdots \\
    20    & 0.0887 & 2.38  & 0.0603 & 1.62  & 0.138  & 3.71 \\
    \bottomrule
    \end{tabular}
  \label{tab:pnm_appednix_LBHydra_nubianl}
\end{table}

%% file: main.bbl
\begin{thebibliography}{}

\bibitem[Adams et~al., 2014]{Adams2014OnSystems}
Adams, B.~M., Kuehn, T.~H., Bielicki, J.~M., Randolph, J.~B., and Saar, M.~O.
  (2014).
\newblock {On the importance of the thermosiphon effect in CPG (CO$_\text2$
  plume geothermal) power systems}.
\newblock {\em Energy}, 69:409--418.

\bibitem[Adams et~al., 2015]{Adams2015AConditions}
Adams, B.~M., Kuehn, T.~H., Bielicki, J.~M., Randolph, J.~B., and Saar, M.~O.
  (2015).
\newblock {A comparison of electric power output of CO$_\text2$ Plume
  Geothermal (CPG) and brine geothermal systems for varying reservoir
  conditions}.
\newblock {\em Applied Energy}, 140:365--377.

\bibitem[Akbarabadi and Piri, 2013]{Akbarabadi2013RelativeConditions}
Akbarabadi, M. and Piri, M. (2013).
\newblock {Relative permeability hysteresis and capillary trapping
  characteristics of supercritical CO$_\text2$/brine systems: An experimental
  study at reservoir conditions}.
\newblock {\em Advances in Water Resources}, 52:190--206.

\bibitem[Alsharhan and Salah, 1997]{Alsharhan1997LithostratigraphyEgypt}
Alsharhan, A.~S. and Salah, M.~G. (1997).
\newblock {Lithostratigraphy, sedimentology and hydrocarbon habitat of the
  pre-Cenomanian Nubian sandstone in the Gulf of Suez oil province, Egypt}.
\newblock {\em GeoArabia}.

\bibitem[Andrew et~al., 2014]{Andrew2014Pore-scaleMicrotomography}
Andrew, M., Bijeljic, B., and Blunt, M.~J. (2014).
\newblock {Pore-scale contact angle measurements at reservoir conditions using
  X-ray microtomography}.
\newblock {\em Advances in Water Resources}, 68:24--31.

\bibitem[Andrew et~al., 2015]{Andrew2015TheConditions}
Andrew, M., Menke, H., Blunt, M.~J., and Bijeljic, B. (2015).
\newblock {The Imaging of Dynamic Multiphase Fluid Flow Using Synchrotron-Based
  X-ray Microtomography at Reservoir Conditions}.
\newblock {\em Transport in Porous Media}, 110(1):1--24.

\bibitem[Bachu and Bennion, 2009]{Bachu2009InterfacialMgL-1}
Bachu, S. and Bennion, D.~B. (2009).
\newblock {Interfacial Tension between CO$_\text2$, Freshwater, and Brine in
  the Range of Pressure from (2 to 27) MPa, Temperature from (20 to 125)
  {${}^\circ$}C, and Water Salinity from (0 to 334 000)
  mg{\textperiodcentered}L{\{}{\$}{\{}{\}}{\^{}}{\{}-1{\}}{\$}{\}}}.
\newblock {\em Journal of Chemical {\&} Engineering Data}, 54(3):765--775.

\bibitem[Benson and Cole, 2008]{Benson2008COsub2/subFormations}
Benson, S.~M. and Cole, D.~R. (2008).
\newblock {CO$_\text2$ sequestration in deep sedimentary formations}.
\newblock {\em Elements}, 4(5):325--331.

\bibitem[Berg et~al., 2016]{Berg2016ConnectedImbibition}
Berg, S., R{\"{u}}cker, M., Ott, H., Georgiadis, A., van~der Linde, H.,
  Enzmann, F., Kersten, M., Armstrong, R.~T., de~With, S., Becker, J., and
  Wiegmann, A. (2016).
\newblock {Connected pathway relative permeability from pore-scale imaging of
  imbibition}.
\newblock {\em Advances in Water Resources}, 90:24--35.

\bibitem[Blunt et~al., 2013]{Blunt2013Pore-scaleModelling}
Blunt, M.~J., Bijeljic, B., Dong, H., Gharbi, O., Iglauer, S., Mostaghimi, P.,
  Paluszny, A., and Pentland, C. (2013).
\newblock {Pore-scale imaging and modelling}.
\newblock {\em Advances in Water Resources}, 51:197--216.

\bibitem[Chatzis et~al., 1983]{Chatzis1983MagnitudeSaturation.}
Chatzis, I., Morrow, N.~R., and Lim, H.~T. (1983).
\newblock {Magnitude and detailed structure of residual oil saturation.}
\newblock {\em Society of Petroleum Engineers journal}, 23(2):311--326.

\bibitem[Crippa et~al., 2019]{Crippa2019}
Crippa, M., Oreggioni, G., Guizzardi, D., Muntean, M., Schaaf, E., Lo~Vullo,
  E., Solazzo, E., Monforti-Ferrario, F., Olivier, J. G.~J., and Vignati, E.
  (2019).
\newblock {Fossil CO$_\text2$ and GHG emissions of all world countries}.
\newblock {\em Luxemburg: Publication Office of the European Union}.

\bibitem[Dong and Blunt, 2009]{Dong2009Pore-networkImages}
Dong, H. and Blunt, M.~J. (2009).
\newblock {Pore-network extraction from micro-computerized-tomography images}.
\newblock {\em Physical Review E}, 80(3):036307.

\bibitem[Donoghue et~al., 2006]{Donoghue2006SynchrotronEmbryos}
Donoghue, P. C.~J., Bengtson, S., Dong, X.-p., Gostling, N.~J., Huldtgren, T.,
  Cunningham, J.~A., Yin, C., Yue, Z., Peng, F., and Stampanoni, M. (2006).
\newblock {Synchrotron X-ray tomographic microscopy of fossil embryos}.
\newblock {\em Nature}, 442(7103):680--683.

\bibitem[Doughty et~al., 2001]{Doughty2001}
Doughty, C., Pruess, K., Benson, S.~M., Hovorka, S.~D., Knox, P.~R., and Green,
  C.~T. (2001).
\newblock {Capacity investigation of brine-bearing sands of the Frio Formation
  for geologic sequestration of CO$_\text2$}.
\newblock In {\em First National Conference on Carbon Sequestration, May
  14–17,}, Washington, D.C., sponsored by National Energy Technology
  Laboratory, CD-ROM. GCCC Digital Publication Series {\#}01-03.

\bibitem[Driesner, 2007]{Driesner2007TheXNaCl}
Driesner, T. (2007).
\newblock {The system H$_\text2$O-NaCl. Part II: Correlations for molar volume,
  enthalpy, and isobaric heat capacity from 0 to 1000 {${}^\circ$}C, 1 to 5000
  bar, and 0 to 1 XNaCl}.
\newblock {\em Geochimica et Cosmochimica Acta}, 71(20):4902--4919.

\bibitem[Duan and Sun, 2003]{Duan2003AnBar}
Duan, Z. and Sun, R. (2003).
\newblock {An improved model calculating CO$_\text2$ solubility in pure water
  and aqueous NaCl solutions from 273 to 533 K and from 0 to 2000 bar}.
\newblock {\em Chemical Geology}, 193(3-4):257--271.

\bibitem[Duan et~al., 2006]{Duan2006}
Duan, Z., Sun, R., Zhu, C., and Chou, I.~M. (2006).
\newblock {An improved model for the calculation of CO$_\text2$ solubility in
  aqueous solutions containing Na$^\text{+}$, K$^\text{+}$, Ca$^\text{2+}$,
  Mg$^\text{2+}$, Cl$^\text{-}$, and SO$_\text{4}^\text{-2}$}.
\newblock {\em Marine Chemistry}, 98(2-4):131--139.

\bibitem[Ezekiel et~al., 2020]{Ezekiel2020CombiningGeneration}
Ezekiel, J., Ebigbo, A., Adams, B.~M., and Saar, M.~O. (2020).
\newblock {Combining natural gas recovery and CO$_\text2$-based geothermal
  energy extraction for electric power generation}.
\newblock {\em Applied Energy}, In Press.

\bibitem[Farhoud, 2009]{Farhoud2009}
Farhoud, K. (2009).
\newblock {Accommodation zones and tectono-stratigraphy of the Gulf of Suez,
  Egypt: A contribution from aeromagnetic analysis}.
\newblock {\em GeoArabia}, 14(4):139--162.

\bibitem[Fasihi et~al., 2019]{Fasihi2019Techno-economicPlants}
Fasihi, M., Efimova, O., and Breyer, C. (2019).
\newblock {Techno-economic assessment of CO$_\text2$ direct air capture
  plants}.
\newblock {\em Journal of Cleaner Production}, 224:957--980.

\bibitem[Garapati et~al., 2015]{Garapati2015BrineWell}
Garapati, N., Randolph, J.~B., and Saar, M.~O. (2015).
\newblock {Brine displacement by CO$_\text2$, energy extraction rates, and
  lifespan of a CO$_\text2$-limited CO$_\text2$-Plume Geothermal (CPG) system
  with a horizontal production well}.
\newblock {\em Geothermics}, 55:182--194.

\bibitem[Gavish, 1974]{Gavish1974GeochemistrySuez}
Gavish, E. (1974).
\newblock {Geochemistry and mineralogy of a recent sabkha along the coast of
  Sinai, Gulf of Suez}.
\newblock {\em Sedimentology}, 21(3):397--414.

\bibitem[Geistlinger et~al.,
  2014]{Geistlinger2014QuantificationMicrotomography}
Geistlinger, H., Mohammadian, S., Schlueter, S., and Vogel, H.~J. (2014).
\newblock {Quantification of capillary trapping of gas clusters using X-ray
  microtomography}.
\newblock {\em Water Resources Research}, 50(5):4514--4529.

\bibitem[Goodman et~al., 2011]{Goodman2011U.S.Scale}
Goodman, A., Hakala, A., Bromhal, G., Deel, D., Rodosta, T., Frailey, S.,
  Small, M., Allen, D., Romanov, V., Fazio, J., Huerta, N., McIntyre, D.,
  Kutchko, B., and Guthrie, G. (2011).
\newblock {U.S. DOE methodology for the development of geologic storage
  potential for carbon dioxide at the national and regional scale}.
\newblock {\em International Journal of Greenhouse Gas Control}, 5(4):952--965.

\bibitem[Gostick, 2017]{Gostick2017}
Gostick, J.~T. (2017).
\newblock {Versatile and efficient pore network extraction method using
  marker-based watershed segmentation}.
\newblock {\em Physical Review E}, 96(2).

\bibitem[Hartfield et~al., 2018]{Hartfield2018State2017}
Hartfield, G., Blunden, J., Arndt, D.~S., Hartfield, G., Blunden, J., and
  Arndt, D.~S. (2018).
\newblock {State of the Climate in 2017}.
\newblock {\em Bulletin of the American Meteorological Society},
  99(8):Si--S310.

\bibitem[Hefny, 2019]{Hefny2019DatasetDioxide}
Hefny, M. (2019).
\newblock {\em {Dataset for "Synchrotron-based pore-network modeling of
  two-phase flow in Nubian Sandstone and implications for capillary trapping of
  carbon dioxide"}}.
\newblock ETH Zurich.

\bibitem[Hefny, 2020]{Hefny2020}
Hefny, M. (2020).
\newblock {\em {Rock physics and heterogeneities characterization controlling
  fluid flow in reservoir rocks}}.
\newblock PhD thesis, ETH Zurich.

\bibitem[Hefny et~al., 2019]{Hefny}
Hefny, M., Ebigbo, A., Adams, B.~M., Hassan, Z., Saar, M.~O., and Hammed, M.
  (2019).
\newblock {Assessing the feasibility of CO$_\text2$-plume geothermal in a
  hydrocarbon-depleted Nubian Sandstone}.
\newblock {\em under prepartion}.

\bibitem[Herring et~al., 2013]{Herring2013EffectSequestration}
Herring, A.~L., Harper, E.~J., Andersson, L., Sheppard, A., Bay, B.~K., and
  Wildenschild, D. (2013).
\newblock {Effect of fluid topology on residual nonwetting phase trapping:
  Implications for geologic CO$_\text2$ sequestration}.
\newblock {\em Advances in Water Resources}, 62:47--58.

\bibitem[{IPCC}, 2005]{IPCC2005CarbonStorage}
{IPCC} (2005).
\newblock {Carbon Dioxide Capture and Storage}.
\newblock Technical report, Cambridge University Press.

\bibitem[{IPCC}, 2014]{IPCC2014ClimateChange.}
{IPCC} (2014).
\newblock {\em {Climate Change 2014 Mitigation of Climate Change: Working Group
  III Contribution to the IPCC Fifth Assessment Report of the Intergovernmental
  Panel on Climate Change.}}
\newblock Cambridge University Press, Cambridge.

\bibitem[Jung and Wan, 2012]{Jung2012SupercriticalMeasurements}
Jung, J.-W. and Wan, J. (2012).
\newblock {Supercritical CO$_\text2$ and Ionic Strength Effects on Wettability
  of Silica Surfaces: Equilibrium Contact Angle Measurements}.
\newblock {\em Energy {\&} Fuels}, 26(9):6053--6059.

\bibitem[K{\"{a}}rger, 2011]{Karger2011FlowRock}
K{\"{a}}rger, J. (2011).
\newblock {Flow and Transport in Porous Media and Fractured Rock}.
\newblock {\em Zeitschrift f{\"{u}}r Physikalische Chemie},
  194(Part{\_}1):135--136.

\bibitem[Kopp et~al., 2009]{Kopp2009InvestigationsCoefficients}
Kopp, A., Class, H., and Helmig, R. (2009).
\newblock {Investigations on CO$_\text2$ storage capacity in saline
  aquifers-Part 2: Estimation of storage capacity coefficients}.
\newblock {\em International Journal of Greenhouse Gas Control}, 3(3):277--287.

\bibitem[Krevor et~al., 2015]{Krevor2015}
Krevor, S., Blunt, M.~J., Benson, S.~M., Pentland, C., Reynolds, C.,
  Al-Menhali, A., and Niu, B. (2015).
\newblock {Capillary trapping for geologic carbon dioxide storage – From pore
  scale physics to field scale implications}.
\newblock {\em International Journal of Greenhouse Gas Control}, 40:221--237.

\bibitem[Krevor et~al., 2012]{Krevor2012RelativeConditions}
Krevor, S., Pini, R., Zuo, L., and Benson, S.~M. (2012).
\newblock {Relative permeability and trapping of CO$_\text2$ and water in
  sandstone rocks at reservoir conditions}.
\newblock {\em Water Resources Research}, 48(2).

\bibitem[Kuijpers et~al., 2017]{Kuijpers2017SorptionImaging}
Kuijpers, C., Huinink, H., Tomozeiu, N., Erich, S., and Adan, O. (2017).
\newblock {Sorption of water-glycerol mixtures in porous Al$_\text2$O$_\text3$
  studied with NMR imaging}.
\newblock {\em Chemical Engineering Science}, 173:218--229.

\bibitem[Land, 1968]{Land1968CalculationProperties}
Land, C.~S. (1968).
\newblock {Calculation of Imbibition Relative Permeability for Two- and
  Three-Phase Flow From Rock Properties}.
\newblock {\em Society of Petroleum Engineers Journal}, 8(02):149--156.

\bibitem[Lenormand et~al., 1988]{Lenormand1988NumericalMedia}
Lenormand, R., Touboul, E., and Zarcone, C. (1988).
\newblock {Numerical models and experiments on immiscible displacements in
  porous media}.
\newblock {\em Journal of Fluid Mechanics}, 189:165--187.

\bibitem[Lenormand et~al., 1983]{Lenormand1983MechanismsDucts}
Lenormand, R., Zarcone, C., and Sarr, A. (1983).
\newblock {Mechanisms of the displacement of one fluid by another in a network
  of capillary ducts}.
\newblock {\em Journal of Fluid Mechanics}, 135(-1):337.

\bibitem[Leverett, 1941]{Leverett1941}
Leverett, M. (1941).
\newblock {Capillary behavior in porous solids}.
\newblock {\em Transactions of the AIME}, 142(01):152--169.

\bibitem[Ma et~al., 1996]{Ma1996EffectTubes}
Ma, S., Mason, G., and Morrow, N.~R. (1996).
\newblock {Effect of contact angle on drainage and imbibition in regular
  polygonal tubes}.
\newblock {\em Colloids and Surfaces A: Physicochemical and Engineering
  Aspects}, 117(3):273--291.

\bibitem[Marone and Stampanoni, 2012]{Marone2012RegriddingImaging.}
Marone, F. and Stampanoni, M. (2012).
\newblock {Regridding reconstruction algorithm for real-time tomographic
  imaging.}
\newblock {\em Journal of synchrotron radiation}, 19(Pt 6):1029--37.

\bibitem[Mason and Morrow, 1991]{Mason1991}
Mason, G. and Morrow, N.~R. (1991).
\newblock {Capillary behavior of a perfectly wetting liquid in irregular
  triangular tubes}.
\newblock {\em Journal of Colloid and Interface Science}, 141(1):262--274.

\bibitem[Mattax and Kyte, 1961]{Mattax1961}
Mattax, C.~C. and Kyte, J.~R. (1961).
\newblock {Ever see a water flood}.
\newblock {\em Oil Gas J}, 59(42):115--128.

\bibitem[Mesheref et~al., 1976]{Mesheref1976}
Mesheref, W.~M., Rafel, E.~M., and Abdel~Baki, S.~H. (1976).
\newblock {Structural interpretation of the Gulf of Suez and its oil
  potentialities. 3rd EGPC}.
\newblock In {\em Exploration Seminar, Cairo, Egypt}.

\bibitem[Michael et~al., 2010]{Michael2010GeologicalOperations}
Michael, K., Golab, A., Shulakova, V., Ennis-King, J., Allinson, G., Sharma,
  S., and Aiken, T. (2010).
\newblock {Geological storage of CO$_\text2$ in saline aquifers—A review of
  the experience from existing storage operations}.
\newblock {\em International Journal of Greenhouse Gas Control}, 4(4):659--667.

\bibitem[{NETL}, 2015]{NETL2015}
{NETL} (2015).
\newblock {Carbon Storage Atlas, Fifth Edition, USA}.
\newblock Technical report, National Energy Technology Laboratory.

\bibitem[{\O}ren and Bakke, 2003]{ren2003ReconstructionEffects}
{\O}ren, P.~E. and Bakke, S. (2003).
\newblock {Reconstruction of Berea sandstone and pore-scale modelling of
  wettability effects}.
\newblock {\em Journal of Petroleum Science and Engineering}, 39(3-4):177--199.

\bibitem[Oren et~al., 1998]{Oren1998ExtendingModels}
Oren, P.-E., Bakke, S., and Arntzen, O. (1998).
\newblock {Extending Predictive Capabilities to Network Models}.
\newblock {\em SPE Journal}, 3(04):324--336.

\bibitem[Otsu, 1979]{Otsu1979AHistograms}
Otsu, N. (1979).
\newblock {A Threshold Selection Method from Gray-Level Histograms}.
\newblock {\em IEEE Transactions on Systems, Man, and Cybernetics},
  9(1):62--66.

\bibitem[Ouyang, 2011]{Ouyang2011NewDioxide}
Ouyang, L.-B. (2011).
\newblock {New Correlations for Predicting the Density and Viscosity of
  Supercritical Carbon Dioxide}.
\newblock {\em The Open Petroleum Engineering Journal}, 4:13--21.

\bibitem[Pacala and Socolow, 2004]{Pacala2004StabilizationTechnologies.}
Pacala, S. and Socolow, R. (2004).
\newblock {Stabilization wedges: solving the climate problem for the next 50
  years with current technologies.}
\newblock {\em Science (New York, N.Y.)}, 305(5686):968--72.

\bibitem[Patzek, 2001]{Patzek2001VerificationImbibition}
Patzek, T.~W. (2001).
\newblock {Verification of a Complete Pore Network Simulator of Drainage and
  Imbibition}.
\newblock {\em SPE Journal}, 6(02):144--156.

\bibitem[Patzek and Kristensen, 2001]{Patzek2001ShapeFlow}
Patzek, T.~W. and Kristensen, J. (2001).
\newblock {Shape Factor Correlations of Hydraulic Conductance in Noncircular
  Capillaries: II. Two-Phase Creeping Flow}.
\newblock {\em Journal of Colloid and Interface Science}, 236(2):305--317.

\bibitem[Qin, 2015]{Qin2015WaterModeling}
Qin, C.-Z. (2015).
\newblock {Water Transport in the Gas Diffusion Layer of a Polymer Electrolyte
  Fuel Cell: Dynamic Pore-Network Modeling}.
\newblock {\em Journal of The Electrochemical Society}, 162(9):F1036--F1046.

\bibitem[Qin et~al., 2016]{Qin2016Pore-NetworkCell}
Qin, C.-Z., Hassanizadeh, S.~M., and Van~Oosterhout, L. (2016).
\newblock {Pore-Network Modeling of Water and Vapor Transport in the Micro
  Porous Layer and Gas Diffusion Layer of a Polymer Electrolyte Fuel Cell}.
\newblock {\em Computation}, 4(2):21.

\bibitem[Qin and van Brummelen, 2019]{Qin2019AMedia}
Qin, C.-Z. and van Brummelen, H. (2019).
\newblock {A dynamic pore-network model for spontaneous imbibition in porous
  media}.
\newblock {\em Advances in Water Resources}, 133:103420.

\bibitem[Randolph and Saar, 2011a]{Randolph2011CombiningSequestration}
Randolph, J.~B. and Saar, M.~O. (2011a).
\newblock {Combining geothermal energy capture with geologic carbon dioxide
  sequestration}.
\newblock {\em Geophysical Research Letters}, 38(10):L10401.

\bibitem[Randolph and Saar, 2011b]{Randolph2011}
Randolph, J.~B. and Saar, M.~O. (2011b).
\newblock {Coupling carbon dioxide sequestration with geothermal energy capture
  in naturally permeable, porous geologic formations: Implications for
  CO$_\text2$ sequestration}.
\newblock {\em Energy Procedia}, 4:2206--2213.

\bibitem[Rasmusson et~al., 2018]{Rasmusson2018ResidualApproach}
Rasmusson, K., Rasmusson, M., Tsang, Y., Benson, S.~M., Hingerl, F., Fagerlund,
  F., and Niemi, A. (2018).
\newblock {Residual trapping of carbon dioxide during geological
  storage—Insight gained through a pore-network modeling approach}.
\newblock {\em International Journal of Greenhouse Gas Control}, 74:62--78.

\bibitem[Rasmusson et~al., 2016]{Rasmusson2016ATrapping}
Rasmusson, K., Rasmusson, M., Tsang, Y., and Niemi, A. (2016).
\newblock {A simulation study of the effect of trapping model, geological
  heterogeneity and injection strategies on CO$_\text2$ trapping}.
\newblock {\em International Journal of Greenhouse Gas Control}, 52:52--72.

\bibitem[Reynolds et~al., 2018]{Reynolds2018MultiphaseKingdom}
Reynolds, C.~A., Blunt, M.~J., and Krevor, S. (2018).
\newblock {Multiphase Flow Characteristics of Heterogeneous Rocks From
  CO$_\text2$ Storage Reservoirs in the United Kingdom}.
\newblock {\em Water Resources Research}, 54(2):729--745.

\bibitem[Sharqawy et~al., 2010]{Sharqawy2010ThermophysicalData}
Sharqawy, M.~H., Lienhard~V, J.~H., and Zubair, S.~M. (2010).
\newblock {Thermophysical properties of seawater: a review of existing
  correlations and data}.
\newblock {\em Desalination and Water Treatment}, 16:354--380.

\bibitem[Soille, 2014]{Soille2014MorphologicalApplications}
Soille, P. (2014).
\newblock {Morphological Image Analysis: Principles and Applications}.
\newblock {\em Sensor Review}, 20(3):391.

\bibitem[Span and Wagner, 1996]{Span1996AMPa}
Span, R. and Wagner, W. (1996).
\newblock {A New Equation of State for Carbon Dioxide Covering the Fluid Region
  from the Triple‐Point Temperature to 1100 K at Pressures up to 800 MPa}.
\newblock {\em Journal of Physical and Chemical Reference Data},
  25(6):1509--1596.

\bibitem[Spiteri et~al., 2008]{Spiteri2008ACharacteristics}
Spiteri, E.~J., Juanes, R., Blunt, M.~J., and Orr, F.~M. (2008).
\newblock {A new model of trapping and relative permeability hysteresis for all
  wettability characteristics}.
\newblock {\em SPE Journal}, 13(3):277--288.

\bibitem[Valvatne and Blunt, 2004]{Valvatne2004PredictiveMedia}
Valvatne, P.~H. and Blunt, M.~J. (2004).
\newblock {Predictive pore-scale modeling of two-phase flow in mixed wet
  media}.
\newblock {\em Water Resources Research}, 40(7).

\bibitem[van~der Meer, 1995]{vanderMeer1995TheAquifers}
van~der Meer, L.~G. (1995).
\newblock {The CO$_\text2$ storage efficiency of aquifers}.
\newblock {\em Energy Conversion and Management}, 36(6-9):513--518.

\bibitem[van Genuchten, 1980]{vanGenuchten1980ASoils1}
van Genuchten, M.~T. (1980).
\newblock {A Closed-form Equation for Predicting the Hydraulic Conductivity of
  Unsaturated Soils1}.
\newblock {\em Soil Science Society of America Journal}, 44(5):892.

\bibitem[Walsh and Saar, 2010a]{Walsh2010InterpolatedKinetics}
Walsh, S.~D. and Saar, M.~O. (2010a).
\newblock {Interpolated lattice Boltzmann boundary conditions for surface
  reaction kinetics}.
\newblock {\em Physical Review E - Statistical, Nonlinear, and Soft Matter
  Physics}, 82(6).

\bibitem[Walsh and Saar, 2010b]{Walsh2010MacroscaleMedia}
Walsh, S. D.~C. and Saar, M.~O. (2010b).
\newblock {Macroscale lattice-Boltzmann methods for low Peclet number solute
  and heat transport in heterogeneous porous media}.
\newblock {\em Water Resources Research}, 46(7):W07517.

\bibitem[Walsh and Saar, 2013]{Walsh2013DevelopingUnits}
Walsh, S. D.~C. and Saar, M.~O. (2013).
\newblock {Developing Extensible Lattice-Boltzmann Simulators for
  General-Purpose Graphics-Processing Units}.
\newblock {\em Communications in Computational Physics}, 13(3):867--879.

\bibitem[Zuo and Benson, 2014]{Zuo2014Process-dependentSandstone}
Zuo, L. and Benson, S.~M. (2014).
\newblock {Process-dependent residual trapping of CO$_\text2$ in sandstone}.
\newblock {\em Geophysical Research Letters}, 41(8):2820--2826.

\end{thebibliography}
